\documentclass[11pt]{article}

\usepackage{amsmath}
\usepackage{amsfonts}
\usepackage[T1]{fontenc}
\usepackage[utf8]{inputenc}
\usepackage{graphicx}
\usepackage{epstopdf}
\usepackage{pgfplots}
\usepackage{setspace}
\usepackage[noadjust]{cite}
\usepackage{filecontents}
\usepackage{subcaption}
\usepackage{float}
\usepackage{wrapfig}
\usepackage{caption}
\usepackage{subcaption}
\usepackage{indentfirst}

\usepackage[toc,page]{appendix}
\usepackage{url}
\usepackage{braket}
\usepackage{hyperref}
\usepackage[all]{hypcap}
\usepackage[hypcap]{caption}
\hypersetup{
	colorlinks=true,
	linkcolor=blue,
	linktoc=all,
	filecolor=black,      
	urlcolor=blue,
	citecolor=blue,
	pdfstartview=FitB,
}

\title{Well-posedness of cubic Horndeski theories}

\author{\'Aron D. Kov\'acs\\
\\
DAMTP, Centre for Mathematical Sciences, University of Cambridge,\\
Wilberforce Road, Cambridge CB3 0WA, UK}

\begin{document}
	
	\maketitle
	
	\begin{abstract}
		We study the local well-posedness of the initial value problem for cubic Horndeski theories. Three different strongly hyperbolic modifications of the ADM formulation of the Einstein equations are extended to cubic Horndeski theories in the weak field regime. In the first one, the equations of motion are rewritten as a coupled elliptic-hyperbolic system of partial differential equations. The second one is based on the BSSN formulation with a generalised Bona-Mass\'o slicing (covering the 1+log slicing) and non-dynamical shift vector. The third one is an extension of the CCZ4 formulation with a generalised Bona-Mass\'o slicing (also covering the 1+log slicing) and a gamma driver shift condition. This final formulation may be particularly suitable for applications in non-linear numerical simulations.
	\end{abstract}
	
	\section{Introduction}

	In the past few decades there has been a growing interest in modifying general relativity (GR) and exploring various properties of these modified theories. The most common reason for this is that the cosmological constant problem led some people to believe that GR may not be the correct theory of gravity, even at large distance scales and low energies. Investigating the properties of extensions of GR is also relevant from an effective field theory (EFT) point of view: since GR is non-renormalizable, one expects that it is only an EFT valid up to some energy scale. When describing strong field phenomena, other operators (besides the Einstein-Hilbert term) might become relevant in the action of the underlying UV complete theory. Finally, testing the rigidity of the predictions and mathematical properties of GR to small deformations can also help us better understand GR itself and discover new techniques.
	
	Trying to modify GR with the purpose to cure one of its shortcomings, however, typically introduces new unwanted pathologies elsewhere. For example, theories with equations of motion containing higher than second derivatives generically suffer from the so-called Ostrogradsky instabilities\footnote{However, in the EFT context it may be argued that such a “runaway” behaviour can be discarded as the mass of the ghost is usually around the cutoff and solutions of this type fall beyond the range of validity of the EFT (see e.g. \cite{eft}).}. To avoid such instabilities, investigations are usually restricted to theories with second order equations of motion. Examples of such theories include Horndeski theories \cite{horndeski}. Horndeski theories are the most general diffeomorphism-invariant scalar-tensor theories with an action principle that has second order equations of motion. The action for general Horndeski theory is
	
	\begin{equation}
	S=\frac{1}{16\pi G}\int \mathrm{d}^4 x\sqrt{-g} \left(\mathcal{L}_1+\mathcal{L}_2+\mathcal{L}_3+\mathcal{L}_4+\mathcal{L}_5\right)
	\end{equation}
	
	\noindent
	with
	
	\begin{align*}
	\mathcal{L}_1&=\mathcal{R}+X\\
	\mathcal{L}_2&=G_2(\phi,X)\\
	\mathcal{L}_3&=G_3(\phi,X)\Box \phi\\
	\mathcal{L}_4&=G_4(\phi,X)\mathcal{R}+\partial_X G_4(\phi,X)\delta_{cd}^{ab}\nabla_a\nabla^c\phi\nabla_b\nabla^d\phi\\
	\mathcal{L}_5&=G_5(\phi,X)\mathcal{G}_{ab}\nabla^a\nabla^b\phi-\frac16 \partial_X G_5(\phi,X)\delta_{def}^{abc}\nabla_a\nabla^d\phi\nabla_b\nabla^e\phi\nabla_c\nabla^f\phi
	\end{align*}
	
	\noindent
	and $X\equiv -\frac12 (\partial\phi)^2$, $G_i$ $(i=2,3,4,5)$ are freely specifiable functions, $\mathcal{R}$ and $\mathcal{G}_{ab}$ denote the Ricci scalar and the Einstein tensor, respectively (corresponding to the spacetime metric $g$) . 
	
	Even a theory with second order equations of motion cannot automatically be considered as a viable physical theory, unless it possesses a well-posed initial value formulation (Cauchy problem). This means that the theory can be formulated in terms of a system of time evolution equations that satisfies the following two physically reasonable properties. Given suitable initial data (that satisfies certain constraints), then i) a unique solution must exist to the evolution equations and ii) the solution must depend continuously on the initial data (in a suitable norm). Apart from being a crucial mathematical criterion, the problem of well-posedness is also interesting from a numerical and experimental point of view. Observations of gravitational wave signatures from coalescing black holes provide new possibilities to test general relativity and its alternatives. In order to do this, however, one needs a stable numerical scheme to solve the equations of motion which is not possible without a well-posed initial value formulation of the theory. Ill-posed systems are unsuitable for numerical computer simulations because initial numerical errors tend to grow drastically during the evolution.
	
	In general, the equations of motion in these theories is a nonlinear system of PDEs. A generic feature of these types of equations is that there may not exist global in time solutions for all data, solutions may blow up in a finite time. The best one can hope for (at least for generic initial data) is to establish local well-posedness, that is to say, the above two criteria are only required to hold for a finite (but strictly non-zero) time. To establish local well-posedness of the non-linear equations, it is sufficient to study the properties of the highest derivative (principal) terms in the linearized equations of motion in a \textit{generic} background.
	
	The set of algebraic conditions on the principal terms in a system of PDEs that are relevant for well-posedness is generally called \textit{hyperbolicity}. There exist multiple notions of hyperbolicity in the literature such as weak, strong, strict and symmetric hyperbolicity. It can be shown \cite{taylor} that \textit{strongly hyperbolic} equations possess a locally well-posed initial value formulation, whereas weaker notions of hyperbolicity do not guarantee that (see definitions and more precise statements in Appendix \ref{sec:back}) For this reason, strong hyperbolicity is considered to be a minimal requirement to perform numerical simulations and in this paper, we shall be primarily concerned with strong hyperbolicity.
	
	The problem of well-posedness in theories of gravity is exacerbated by the fact that most of these theories are diffeomorphism-covariant. This implies that solutions to the equations of motion are never unique in a mathematical sense. This problem is usually solved by fixing the gauge. This involves imposing a condition on the components of the metric and/or its derivatives, and modifying the equations of motion by terms that vanish when the gauge condition holds. In general relativity the simplest gauge condition leading to a well-posed formulation is provided by the harmonic gauge \cite{ycb,bruhat} which reduces the Einstein equations to a system of quasilinear wave equations on each component $g_{\mu\nu}$.
	
	The question of well-posedness in modified theories is far from settled. In fact, for non-trivial extensions of GR, it had not been discussed until recently. In \cite{gp_hsr, gp} Papallo and Reall studied the linearized equations of motion of Lovelock \cite{lovelock} and Horndeski theories in a generic weak field background and in (generalized) harmonic gauge. The restriction to a weak field background is due to the fact that in these theories well-posedness is known to break down when the background fields are strong \cite{gp_hsr,r_t_w,strong1,strong2,strong3}. However, one may still hope that the theory does not lose its predictive power when it is used to describe small deviations from GR. It turns out that the only subclass of these theories that has well-posed Cauchy problem in a generalized harmonic gauge is the so-called $k$-essence-type theories ($G_3=\partial_XG_4=G_5=0$). It should be noted that the \textit{linearized} equations of motion in theories with non-trivial $G_3$ were shown to be strongly hyperbolic in a specific choice of generalized harmonic gauge. However, this result does not extend to the \textit{non-linear} case in a generic background.

	Despite this result, one cannot immediately conclude that more general theories are useless. There exists several different well-posed formulations of GR so one might hope that some other formulation and a different choice of gauge could be suitably extended to more general Horndeski theories. In this paper, we focus on the cubic subclass $(G_4=G_5=0)$ of Horndeski theories\footnote{This is related to the theory given by $G_2\neq 0$, $G_3=b(\phi)X$, $G_4=G_4(\phi)$ and $G_5=0$ by a field redefinition.} and study its initial value formulation in more detail. Recently, this class of theories has received some attention among cosmologists due to the fact that this model can describe a non-singular bouncing universe \cite{bounce1,bounce2,bounce3,bounce4}. Furthermore, cubic Horndeski theories naturally arise as certain low energy limits of massive gravity theories \cite{massive1,massive2}.
	
	The main result of this paper is that cubic Horndeski theories \textit{do} possess a well-posed initial value formulation (at least in the weak field regime) and we provide three examples of strongly hyperbolic formulations. Since readers with different backgrounds may find different parts of this paper interesting, we intend to organise our results accordingly.
	
	We begin with a general discussion of the ADM formulation of cubic Horndeski theories in Section \ref{sec:setup}. More specifically, we present the standard ADM evolution and constraint equations of cubic Horndeski theories and show that a suitable linear combination of these equations give a scalar evolution equation which contains no second derivatives of the spacetime metric. This observation has already been made in \cite{bounce1} but we emphasize this fact here again, since it is a key step to obtain well-posed formulations. The section is concluded by a preliminary discussion of constraint propagation.
	
	Section \ref{sec:ell_hyp} is mathematically more involved\footnote{Some additional information is provided on pseudodifferential calculus and its applications to hyperbolic and elliptic PDEs in Appendix \ref{sec:back}.} so readers interested in numerical applications may jump straight to Section \ref{sec:bssn}. In Section \ref{sec:ell_hyp} we present an elliptic-hyperbolic formulation of cubic Horndeski theories, using ideas put forward by Andersson and Moncrief in \cite{andersson} for vacuum GR. After briefly reviewing \cite{andersson}, we show how a suitable modification of the constant (or arbitrarily prescribed) mean curvature and spatial harmonic gauge conditions lead to second order elliptic equations for the lapse function and the shift vector. In the weak field regime and on slices with negative Ricci curvature, existence and uniqueness of solutions to these elliptic equations is guaranteed. It is finally shown that under these assumptions, the strong well-posedness result of Andersson and Moncrief for GR extends to cubic Horndeski theories.
	
	In Section \ref{sec:bssn} we consider a version of the Baumgarte-Shapiro-Shibata-Nakamura (BSSN) formulation \cite{bssn1,bssn2} with a generalized Bona-Mass\'o slicing condition \cite{bona_masso} and non-dynamical (i.e. arbitrary but a priori fixed) shift vector. This formulation contains $2$ free parameters: one that describes the slicing condition and one that describes how we modify the evolution system by the momentum constraint. It is shown that when these parameters obey a lower bound then the system of equations is strongly hyperbolic in the weak field regime.
	
	Finally, Section \ref{sec:ccz4} is the most relevant to those with interests in non-linear numerical computer simulations. Here we review the so-called covariant conformal Z4 (CCZ4) formulation \cite{ccz4} which was constructed to enhance the accuracy of numerical simulations in GR. This was achieved by an appropriate modification of Einstein’s equation so that constraint violations are damped away during the evolution. Together with a $2$-parameter family of dynamical gauge conditions (generalized Bona-Mass\'o slicing and “gamma driver” conditions), a straightforward generalization of the CCZ4 system to cubic Horndeski theories constitutes a strongly hyperbolic system of PDEs whenever a simple lower bound on these parameters is imposed and the fields are sufficiently weak. In particular, the slicing conditions selected by strong hyperbolicity include the 1+log slicing which is used in many numerical applications. We also comment on the issue of constraint damping in cubic Horndeski theories.

	\section{Setting up the problem}\label{sec:setup}
	
	\subsection{Equations of motion}
	
	In this section we provide the ideas that all three formulations (presented in the subsequent sections) share.
	
	We adapt the following notation. We are going to use the Latin letters $(a,b,c,...)$ for abstract indices and Greek letters $(\mu,\nu,\rho,...)$ for coordinate indices. The Latin letters $(i, j, k, ...)$ will be used for spatial indices. As mentioned before, we use calligraphic letters ($\mathcal{R}$, $\mathcal{R}_{ab}$, $\mathcal{G}_{ab}$, etc.) for spacetime curvature tensors, whereas curvature tensors defined on spatial slices are denoted by regular ($R$, $R_{ab}$, etc.) letters.  For a metric $m$, $|\cdot|_m$ denotes the pointwise norm with respect to $m$ (e.g. for a vector field $v^a$ we have $|v|_m=m_{ab}v^av^b$). Our convention on the metric signature is $(-,+,+,+)$.
	
	As mentioned in the Introduction, the class of theories under consideration can be described by the action ($X\equiv -\frac12(\partial\phi)^2$)
	
	\begin{equation}\label{cubic_act}
	S=\frac{1}{16\pi G}\int \mathrm{d}^4 x\sqrt{-g} \left(\mathcal{R}+X+G_2(\phi,X)+G_3(\phi,X)\Box\phi\right).
	\end{equation}
	
	The reason for separating out $X$ in this action is that we are going to view Horndeski theories as small deformations of Einstein’s theory with a minimally coupled scalar field (referred to as the Einstein-scalar-field theory later, similarly to \cite{gp_hsr}). 
	
	\noindent
	Varying the action (\ref{cubic_act}) with respect to the metric yields the equation of motion \cite{gp}
	
	\begin{align}\label{eom_g3}\nonumber
	E_{ab}\equiv~ & \mathcal{G}_{ab}-\frac12 \left(X+G_2+2X\partial_\phi G_3\right)g_{ab}-\frac12\left(1+\partial_XG_2+2\partial_\phi G_3\right)\nabla_a\phi\nabla_b\phi \\
	&+\frac12\partial_XG_3\left(-\Box\phi\nabla_a\phi\nabla_b\phi+ 2\nabla_{(a}\phi\nabla_{b)}\nabla_c\phi\nabla^c\phi-\nabla_c\nabla_d\phi\nabla^c\phi\nabla^d\phi g_{ab}\right)=0.
	\end{align}
	
	In the ADM-type formulations of general relativity, it is often beneficial to use the linear combination $E_{ab}-\frac12 E g_{ab}=\mathcal{R}_{ab}=0$ as equation of motion, rather than $E_{ab}=\mathcal{G}_{ab}=0$. In fact, it turns out that it is useful to consider the same combination of the gravitational equations of motion in Horndeski theories:
	
	\begin{align}\label{eom_g3_2}\nonumber
	E_{ab}-\frac12 E g_{ab}=& ~\mathcal{R}_{ab}+\frac12 \left(G_2-X\partial_XG_2-X\partial_X G_3\Box\phi\right)g_{ab}-\frac12\left(1+\partial_XG_2+2\partial_\phi G_3\right)\nabla_a\phi\nabla_b\phi \\
	&+\frac12\partial_XG_3\left(-\Box\phi\nabla_a\phi\nabla_b\phi+ 2\nabla_{(a}\phi\nabla_{b)}\nabla_c\phi\nabla^c\phi\right)=0.
	\end{align}
	
	\noindent
	In the scalar evolution equation
	
	\begin{align}\label{original_sc}\nonumber
	E_\phi\equiv&-\left(1+\partial_XG_2+2X\partial_X^2G_2+2\partial_\phi G_3+2X\partial_{X\phi}^2G_3\right)\Box\phi+\partial_X G_3 \mathcal{R}_{ab}\nabla^a\phi \nabla^b\phi\\ \nonumber
	&-\left(\partial_X^2G_2+2\partial_{X\phi}^2G_3\right)\left((\partial\phi)^2\Box\phi-\nabla^a\phi\nabla^b\phi\nabla_a\nabla_b\phi\right)-\partial_XG_3\left((\Box\phi)^2-\nabla_a\nabla_b\phi\nabla^a\nabla^b\phi\right)\\
	&+\partial_X^2G_3\nabla^a\phi\nabla^b\phi\left(\Box\phi\nabla_a\nabla_b\phi-\nabla_a\nabla^c\phi\nabla_c\nabla_b\phi\right)+2X\left(\partial_\phi^2G_3+\partial_{X\phi}^2G_2\right)-\partial_\phi G_2=0
	\end{align}
	
	\noindent
	(obtained by varying the action (\ref{cubic_act}) with respect to $\phi$) the only term involving second derivatives of the metric is $\mathcal{R}_{ab}\nabla^a\phi\nabla^b\phi$. We will see that it is useful to express this from $\left(E_{cd}-\frac12Eg_{cd}\right)\nabla^c\phi\nabla^d\phi$. In other words, instead of the equation $E_\phi=0$, we are going to use
	
	\begin{align}\label{ev_sc}\nonumber
	\tilde E_{\phi}\equiv &~E_\phi-\partial_X G_3\left(E_{cd}-\frac12Eg_{cd}\right)\nabla^c\phi\nabla^d\phi=	-\left(1+\partial_XG_2+2X\partial_X^2G_2+2\partial_\phi G_3+2X\partial_{X\phi}^2G_3\right)\Box\phi\\ \nonumber
	&-\left(\partial_X^2G_2+2\partial_{X\phi}^2G_3\right)\left((\partial\phi)^2\Box\phi-\nabla^a\phi\nabla^b\phi\nabla_a\nabla_b\phi\right)-\partial_XG_3\left((\Box\phi)^2-\nabla_a\nabla_b\phi\nabla^a\nabla^b\phi\right)\\ \nonumber
	&+\partial_X^2G_3\nabla^a\phi\nabla^b\phi\left(\Box\phi\nabla_a\nabla_b\phi-\nabla_a\nabla^c\phi\nabla_c\nabla_b\phi\right)+2X\left(\partial_\phi^2G_3+\partial_{X\phi}^2G_2\right)-\partial_\phi G_2\\ 
	&+\partial_XG_3\left(2X^2+XG_2+X^2\partial_XG_2+X^2\partial_XG_3\Box\phi+4X^2\partial_\phi G_3+2X\partial_XG_3\nabla^a\phi\nabla^b\phi\nabla_a\nabla_b\phi\right)=0
	\end{align}
	
	\noindent
	as the scalar evolution equation. The reason for this is that this equation contains derivatives of the scalar field up to second order and derivatives of the metric only up to first order. (Some benefits of the use of this particular linear combination were also noticed in \cite{bounce1}.)
	
	Now we assume that the spacetime manifold $(M,g)$ is globally hyperbolic $M=\mathbb{R}\times \Sigma$ and $h_{ab}$ is the spatial metric induced on the spacelike Cauchy surfaces $\Sigma_t$. Let $n^a$ be the future directed unit normal to $\Sigma_t$. The lapse function $N$ and the shift vector $N^a$ are then defined by
	
	\begin{equation}
	\left(\frac{\partial}{\partial t}\right)^a=N n^a+N^a.
	\end{equation}
	
	\noindent
	The convention on the extrinsic curvature used here is
	
	\begin{equation}\label{extr_curv_def}
	K_{ab}=-\frac12\mathcal{L}_nh_{ab}=-\frac{1}{2N}\left(\partial_t-\mathcal{L}_N\right)h_{ab}.
	\end{equation}
	
	\noindent
	We also need ADM variables for the derivatives of the scalar field, let
	
	\begin{equation}\label{def_A}
	A\equiv n^a\nabla_a\phi=\frac{1}{N}\left(\partial_t-\mathcal{L}_N\right)\phi
	\end{equation}
	
	\noindent
	and 
	
	\begin{equation}
	A_a\equiv h_a^b\nabla_b\phi.
	\end{equation}

	For convenience, we also introduce a fixed, smooth background metric on the spatial slices $\mathring{h}$ and denote the corresponding covariant derivative and Christoffel symbol by $\mathring{D}$ and $\mathring{ \Gamma}^i_{jk}$, respectively.
	
	Now we are ready to provide the standard ADM-type equations of motion in cubic Horndeski theories. Taking the spatial projection of (\ref{eom_g3_2}) in both indices yields the tensor evolution equation (some helpful formulas for carrying out ADM decompositions are provided in Appendix \ref{app:adm})
	
	\begin{align}\label{g3_ev} \nonumber
	&\left(\partial_t-\mathcal{L}_N\right)K_{ij}-\frac12\partial_XG_3(Xh_{ij}+A_iA_j)\left(\partial_t-\mathcal{L}_N\right)A=-D_i D_jN +N\biggl\{R_{ij}+KK_{ij}-2K_{ik}K^k_j \\ \nonumber
	&+\frac12h_{ij}\left(G_2-X\partial_XG_2-X\partial_XG_3\left(D^kA_k+AK+A^kD_k\ln N\right)\right)-\frac12\left(1+\partial_XG_2+2\partial_\phi G_3\right)A_iA_j \\ 
	&+\frac12\partial_XG_3\left(A^kD_k\left(A_iA_j\right)-A_iA_jD^kA_k-A_iA_jAK-A_iA_jA^kD_k\ln N-2AA_{(i}D_{j)}A\right)\biggr\}.
	\end{align}
	
	\noindent
	Note that
	
	\begin{equation}
	X=\frac12\left(A^2-A^kA_k\right).
	\end{equation}
	
	\noindent
	Similarly to general relativity, the projections $E_{ab}n^an^b$ and $E_{cb}h^c_an^b$ in Horndeski theories yield constraint equations: the Hamiltonian constraint is
	
	\begin{align}\label{ham_g3}\nonumber
	2\mathbf{H}\equiv 2E_{\mu\nu}n^\mu n^\nu=&~ R+K^2-K_{ij}K^{ij}-\left(\tfrac12 A_iA^i+ \tfrac12 A^2-G_2+A^2\partial_XG_2+(A^2+A_iA^i)\partial_\phi G_3\right)\\
	&-\partial_X G_3\left(A^3K+A^2D^iA_i-K_{ij}A^iA^jA-A^iA^jD_iA_j\right)=0,
	\end{align}
	
	\noindent
	while the momentum constraint reads as
	
	\begin{align}\label{mom_g3}\nonumber
	\mathbf{M}_i\equiv E_{\mu i}n^\mu =&~ D_iK-D^jK_{ij}-\frac12\left(1+\partial_XG_2+2\partial_\phi G_3\right)A A_i\\
	&-\frac12\partial_X G_3\left(A^2A_iK+A A_iD^kA_k-K_{kl}A^kA^lA_i-AA^jD_jA_i-A_iA^kD_kA+A^2D_iA\right)=0.
	\end{align}
	
	Even though we are not going to use the explicit form of the scalar evolution equation (\ref{ev_sc}), only some of its properties, we rewrite it in terms of the ADM variables, for reference. One obtains\footnote{The \textit{Mathematica} package \textit{xAct} \cite{xact} was of great help in the derivation of the equations.}
	
	\begin{align}\label{ev_sc_adm}
	  E_\phi\equiv{} &\Phi \biggl(1
+ \partial_XG_2 + A^2 \partial_{{X}}^2G_2
+ 2 \partial_{\phi}G_3
+  \partial_{X\phi}^2G_3 ( A^2
+   {A}_{i}  {A}_{j}  {h}^{ij} )
+ 2  {\Phi}_{ij} \partial_XG_3  {h}^{ij} \nonumber\\
& + \tfrac{1}{4} \bigl(\partial_XG_3\bigr)^2 \Bigl(3 A^4
-  2  {A}_{i}  {A}_{j} A^2  {h}^{ij} 
-   {A}_{i}  {A}_{j}  {A}_{k}  {A}_{l}  {h}^{ik}  {h}^{jl} \Bigr)\nonumber\\
& +  \partial_{{X}}^2G_3 \Bigl( A^2  {\Phi}_{ij}  {h}^{ij} 
-    {A}_{k}  {A}_{l}  {\Phi}_{ij}  {h}^{ik}  {h}^{jl} \Bigr)\biggr)\nonumber\\
&- \partial_{\phi}G_2
-   {\Phi}_{ij}  {h}^{ij} 
- 2  {\Phi}_{ij} \partial_{\phi}G_3  {h}^{ij} 
+  \partial_{X\phi}^2G_2 (A^2 -    {A}_{i}  {A}_{j}  {h}^{ij} )\nonumber\\
& +  \partial_{\phi}^2G_3 ( A^2 -    {A}_{i}  {A}_{j}  {h}^{ij} )
+ \tfrac{1}{2} G_2 \partial_XG_3 (A^2 -    {A}_{i}  {A}_{j}  {h}^{ij} )\nonumber\\
& +  \partial_{{X}}^2G_2 (- 2 {A}_{i} A  {\Phi}_{j}  {h}^{ij} 
+   {A}_{k}  {A}_{l}  {\Phi}_{ij}  {h}^{ik}  {h}^{jl} )
\nonumber\\
&  + \partial_XG_2 \Bigl(-  {\Phi}_{ij}  {h}^{ij} + \tfrac{1}{4} \partial_XG_3 \bigl(A^4
-  2  {A}_{i}  {A}_{j} A^2  {h}^{ij} 
+  {A}_{i}  {A}_{j}  {A}_{k}  {A}_{l} {h}^{ik}  {h}^{jl} \bigr)\Bigr)
\nonumber\\
& +  \partial_{X\phi}^2G_3 \Bigl(- 4  {A}_{i} A  {\Phi}_{j}  {h}^{ij} 
+ A^2  {\Phi}_{ij}  {h}^{ij} 
+2  {A}_{k}  {A}_{l}  {\Phi}_{ij}  {h}^{ik}  {h}^{jl} -   {A}_{i}  {A}_{j}  {\Phi}_{kl}  {h}^{ij}  {h}^{kl} \Bigr)\nonumber\\
&  + \partial_XG_3 \Bigl( \partial_{\phi}G_3 \bigl( A^4
-  2  {A}_{i}  {A}_{j} A^2  {h}^{ij} +   {A}_{i}  {A}_{j}  {A}_{k} {A}_{l}  {h}^{ik}  {h}^{jl} \bigr)+ A^4
-  2  {\Phi}_{i}  {\Phi}_{j} \
 {h}^{ij} 
\nonumber\\
&  -  2  {A}_{i}  {A}_{j} A^2  {h}^{ij} 
+   {A}_{i}  {A}_{j}  {A}_{k} \
 {A}_{l}  {h}^{ik}  {h}^{jl} 
-    {\Phi}_{ij}  {\Phi}_{kl}  {h}^{ij} \
 {h}^{kl} 
+   {\Phi}_{ij}  {\Phi}_{kl}  {h}^{ik} \
 {h}^{jl} \Bigr)\nonumber\\
& +  \partial_{{X}}^2G_3 \Bigl(-  A^2  {\Phi}_{i}  {\Phi}_{j}  {h}^{ij} 
-  2  {A}_{i} A  {\Phi}_{jl}  {\Phi}_{k} \
 {h}^{ik}  {h}^{jl} +  {A}_{i}  {A}_{j}  {\Phi}_{k} \
 {\Phi}_{l}  {h}^{ik}  {h}^{jl} \nonumber\\
& 
+ 2  {A}_{k} A  {\Phi}_{ij} \
 {\Phi}_{l}  {h}^{ik}  {h}^{jl} -    {A}_{l}  {A}_{m}  {\Phi}_{in} \
 {\Phi}_{jk}  {h}^{il}  {h}^{jm}  {h}^{kn} 
+   {A}_{m}  {A}_{n}  {\Phi}_{ij} \
 {\Phi}_{kl}  {h}^{im}  {h}^{jn} \
 {h}^{kl} \Bigr)\nonumber\\
&+\tfrac14  \bigl(\partial_XG_3\bigr)^2 \Bigl(- 8  {A}_{i} A^3  {\Phi}_{j}  {h}^{ij} 
+ A^4  {\Phi}_{ij}  {h}^{ij} 
+ 4  {A}_{k}  {A}_{l} A^2  {\Phi}_{ij} \
 {h}^{ik}  {h}^{jl} \nonumber\\
& 
{}-  2  {A}_{i}  {A}_{k} A^2  {\Phi}_{jl}  {h}^{ik}  {h}^{jl}  +
 {A}_{i}  {A}_{j}  {A}_{k}  {A}_{l} \
 {\Phi}_{mn}  {h}^{ik}  {h}^{jl} \
 {h}^{mn} \nonumber\\
& {}-  4  {A}_{i}  {A}_{k}  {A}_{l} \
 {h}^{ik}  {h}^{jl} \bigl(-2 A  {\Phi}_{j} + \
 {A}_{m}  {\Phi}_{jn}  {h}^{mn} \bigr) \Bigr)=0.
\end{align}

\noindent
where we used the following auxiliary variables

\begin{align}\label{Phi}
{\Phi}\equiv{}&\frac{1}{N}\left(\partial_t-\mathcal{L}_N\right)A-{A}^i{D}_i\ln N, \\
{\Phi}_i\equiv{}&{D}_iA+K_{ij}{A}^j,\\
{\Phi}_{ij}\equiv{}&{D}_i{A}_j+AK_{ij}.
\end{align}

	In the modifications of the ADM formulation considered in this paper, the system of evolution equations takes the general form

\begin{subequations}
\begin{equation}\label{gen_def}
\partial_t u=\mathcal{L}_N u+N v
\end{equation}

\begin{equation}\label{gen_ev}
\mathbf{a}(u,Du,v,D^2u,Dv)\partial_t v=\mathbf{a}(u,Du,v,D^2u,Dv)\mathcal{L}_N v+N~\mathbf{b}(u,Du,v,D^2u,Dv)
\end{equation}
\end{subequations}

\noindent
where $u$ and $v$ are both column vectors of size $n$, corresponding to the dynamical variables; $\mathbf{a}$ and $\mathbf{b}$ are $n\times n$ matrices, depending on the fields $u$, $v$ (and their derivatives). We assume that the matrix $\mathbf{a}$ is invertible so that the hypersurfaces $\Sigma_t$ are non-characteristic and spacelike.

In Horndeski theories, for example, the variables $u$ include fields like the components of the induced metric $h_{ij}$ and the scalar field $\phi$, while the variables $v$ are auxiliary variables such as the extrinsic curvature $K_{ij}$ and the normal derivative of the scalar field $A$. Since the formulations to be discussed have different dynamical fields, this will be made explicit later, on a case-by-case basis.

The hyperbolicity of the system of equations (\ref{gen_def},\ref{gen_ev}) is found by analyzing its characteristic equation. This is obtained by linearizing the evolution equations, selecting the highest derivative (principal) terms and replacing all the derivatives with $\partial_\mu\to i\xi_\mu\equiv i(\xi_0,\xi_i)$. Note that in equation (\ref{gen_def}) the principal terms are the terms proportional to $v$ and first derivatives of $u$; whereas in equation (\ref{gen_ev}) the principal terms are first derivatives of $v$ and second derivatives of $u$.
 The characteristic equations have the general form

\begin{equation}\label{gen_char}
	i\xi_0\mathbb{A}~ U=\mathbb{L}(\xi_k)U
\end{equation}
	
	\noindent
	where $\mathbb{A}$ is a $2n\times 2n$ matrix depending on the background fields, $\mathbb{L}(\xi_k)$ is also a $2n\times 2n$ matrix depending on the background fields and the spatial Fourier variable $\xi_k$, $U$ is a size $2n$ column vector associated with the dynamical fields (i.e. $u$ and $v$). The condition that $t=$constant surfaces are non-characteristic implies that $\mathbb{A}$ is invertible.

Equation (\ref{gen_char}) can be regarded as the eigenvalue problem for the matrix $\mathbb{M}(\xi_k)\equiv\mathbb{A}^{-1}\mathbb{L}(\xi_k)$.
The system (\ref{gen_def},\ref{gen_ev}) is called \textit{weakly hyperbolic} if and only if the eigenvalues of $\mathbb{M}(\xi_k)$ are real for any $\xi_k$ of unit norm $\xi_k\xi_k=1$. When the matrix $\mathbb{M}(\xi_k)$ i) has real eigenvalues, ii) is diagonalizable and iii) has a complete set of eigenvectors that depend smoothly on $\xi_k$, for any $\xi_k$ of unit norm, then the system (\ref{gen_def},\ref{gen_ev}) is said to be \textit{strongly hyperbolic}. In order for the Cauchy problem of (\ref{gen_def},\ref{gen_ev}) to be well-posed, the system must be at least strongly hyperbolic. For more precise definitions and statements, see Appendix \ref{sec:back} and references cited therein.

In particular, for cubic Horndeski theories\footnote{For different formulations of the theory, the number of dynamical variables may be different so we continue the discussion with general $n$.}, we can write $U=(U_g,U_\phi,U_A)^T$ where $U_g$ is a size $2n-2$ vector that corresponds to the gravitational variables (e.g. $h_{ij}$ and $K_{ij}$); the components $U_\phi$ and $U_A$ correspond to the variables $\phi$ and $A$. Since the equations (\ref{def_A}) and (\ref{ev_sc_adm}) contain no principal terms associated with the spacetime metric $g$, the matrices $\mathbb{A}$ and $\mathbb{M}$ have the upper triangular form

\begin{equation}
\mathbb{A}(u,Du,v,D^2u,Dv)=\left(\begin{array}{cc}
\mathbb{A}_{gg}(u,Du,v) & \mathbb{A}_{g\phi}(u,Du,v) \\
0 & \mathbb{A}_{\phi\phi}(u,Du,v,D^2u,Dv)
\end{array}\right)
\end{equation}

\begin{equation}
\mathbb{L}(u,Du,v,D^2u,Dv)=\left(\begin{array}{cc}
\mathbb{L}_{gg}(u,Du,v) & \mathbb{L}_{g\phi}(u,Du,v) \\
0 & \mathbb{L}_{\phi\phi}(u,Du,v,D^2u,Dv).
\end{array}\right)
\end{equation}
	
\noindent
The matrix blocks labelled by subscripts $gg$, $g\phi$ and $\phi\phi$ have sizes $(2n-2)\times(2n-2)$, $2\times(2n-2)$ and $2\times 2$, respectively. It is also worth noting that the matrices $\mathbb{A}_{gg}$, $\mathbb{A}_{g\phi}$, $\mathbb{L}_{gg}$, $\mathbb{L}_{g\phi}$ depend only on the fields $u$, $Du$ and $v$, that is to say, the tensor (gravitational) evolution equations are quasilinear (see e.g. equation (\ref{g3_ev})).

	 It will be useful (especially in Section \ref{sec:ccz4}) to separate the Einstein-scalar-field theory and the Horndeski (i.e. $G_2$ and $G_3$-dependent) parts in $\mathbb{A}$ and $\mathbb{L}$:
	
\begin{align}	\label{char_split}
&\mathbb{A}=\mathbb{A}_0+\delta\mathbb{A}, \nonumber \\
&\mathbb{L}=\mathbb{L}_0+\delta\mathbb{L} 
\end{align}
	
	\noindent
	where $\mathbb{L}_0$ and $\mathbb{M}_0$ correspond to the Einstein-scalar-field theory, $\delta\mathbb{A}$ and $\delta\mathbb{L}$ are the Horndeski terms. The specific forms of these terms will also be given on a case-by-case basis.
	
	Let us consider in more detail the characteristic equation corresponding to (\ref{ev_sc}) (or (\ref{ev_sc_adm})). Selecting the second derivatives of $\phi$ in the linearized version of (\ref{ev_sc}) (recall that there are no second derivatives of $g$ in (\ref{ev_sc})) and substituting the derivatives $\partial_\mu\to i\xi_\mu\equiv i(\xi_0,\xi_i)$, the characterstic equation for the scalar mode is given by
	
	\begin{equation}\label{char}
	0=\left(P^\prime_{\phi\phi}\right)^{\mu\nu}\xi_\mu\xi_\nu\equiv P_{\phi\phi}(\xi)-\partial_XG_3\left( P_{g\phi}^{\mu\nu}(\xi)-\frac12g_{\rho\sigma}g^{\mu\nu}P_{g\phi}^{\rho\sigma}(\xi)\right)\nabla_\mu\phi\nabla_\nu\phi
	\end{equation}
	
	\noindent
	with
	
	\begin{equation}
	P_{g\phi}^{\mu\nu}(\xi)=\frac12\partial_X G_3\nabla^\mu\phi\nabla^\nu\phi |\xi|_g^2-\partial_X G_3\xi^\sigma\nabla_\sigma\phi\left(\xi^{(\mu}\nabla^{\nu)}\phi-\frac12g^{\mu\nu}\xi^\rho\nabla_\rho\phi\right)
	\end{equation}

\noindent
	and
			
	\begin{align}\nonumber
	P_{\phi\phi}(\xi)=&\left(1+\partial_XG_2+2X\partial_X^2G_2+2\partial_\phi G_3+2X\partial_{X\phi}^2G_3\right)|\xi|_g^2\\ \nonumber
	&+\left(\partial_X^2G_2+2\partial_{X\phi}^2G_3\right)\left((\partial\phi)^2|\xi|_g^2-\left(\xi_\mu\nabla^\mu\phi\right)^2\right)\\ \nonumber	
	&+2\partial_XG_3\left((\Box\phi)|\xi|_g^2-\xi^\mu\xi^\nu\nabla_\mu\nabla_\nu\phi\right)\\
	&-\partial_X^2G_3\nabla^\mu\phi\nabla^\nu\phi\left(\Box\phi~\xi_\mu\xi_\nu+|\xi|_g^2\nabla_\mu\nabla_\nu\phi-2\xi_{\rho}\xi_{(\mu}\nabla_{\nu)}\nabla^\rho\phi\right).
	\end{align}
	
\noindent
	The notations $P_{g\phi}$, $P_{\phi\phi}$ and $P^\prime_{\phi\phi}$ refer to the coefficients of the second derivatives of $\phi$ in the linearized versions of equations (\ref{eom_g3}), (\ref{original_sc}) and (\ref{ev_sc}), respectively.
	
	The same characteristic equation corresponding to the scalar degree of freedom has been previously found in \cite{gp_hsr} and \cite{bounce1}. More precisely, to make a comparison with \cite{gp_hsr}, we note that with the preferred gauge choice ${\mathcal{H}}_{ab}=-\partial_X G_3 \nabla_a\phi\nabla_b\phi$ (and $G_4=0$) made therein, equation (230) of \cite{gp_hsr} agrees with (\ref{char}).
	
	In a regime in which the fields are sufficiently weak, $P_{\phi\phi}^\prime$ is close to the spacetime metric $g$ and therefore, it is a Lorentzian metric. By sufficiently weak fields we mean field configurations such that the Horndeski terms are small compared to the Einstein-scalar-field terms. More precisely, let $E=\text{max}\left\{|\mathcal{R}_{\mu\nu\rho\sigma}|^{1/2},|\nabla_\mu\phi|,|\nabla_\mu\nabla_\nu\phi|^{1/2}\right\}$ in all orthonormal bases. Then the weak field condition is equivalent to
	
	\begin{align}\label{wfb}\nonumber
	&|\partial_X^k\partial_\phi^l G_2|E^{2k+2}\ll 1 & k=0,1,2;~ l=0,1;\\
	&|\partial_X^k\partial_\phi^l G_3|E^{2k}\ll 1 & k,l=0,1,2.
	\end{align}
	
	\noindent
	Regarding (\ref{char}) as an equation for the “characteristic speeds” $\xi_0$ for given $\xi_i\neq 0$, this equation has two distinct real solutions $\xi_0^{\phi,\pm}$. Furthermore, the weak field assumptions (\ref{wfb}) also ensure that the spacelike $t=$constant hypersurfaces are non-characteristic.

	\subsection{Constraint propagation}\label{sec:setup_prop}
	
	In addition to studying the hyperbolicity of the equations of motion in different formulations, we need to address the issue of constraint propagation. That is to say, we need to check whether solutions to the equations used in these formulations remain solutions of the original Horndeski equations of motion during the evolution. Here we present a fairly detailed derivation of the equations governing the propagation of gauge conditions and constraints, even though the individual steps are quite standard. The purpose of this is to demonstrate that the Bianchi identity (and its generalization) leads straightforwardly to a homogeneous system of PDE for the constraint variables, without making any reference to a specific form of the equations of motion. The only assumption we make is that the equations of motion are second order PDEs obtained by varying a diffeomorphism invariant action. This guarantees that the normal projection of the equations of motion is a constraint equation. Note that in this section we derive the equations without gauge fixing, the effect of the gauge fixing terms on the constraint propagation system will be discussed later.
	Let $E_{ab}=0$ be the equations of motion obtained by varying the action with respect to the spacetime metric $g_{ab}$. Let us decompose it as
	
	\begin{equation}
	E_{ab}=\mathbf{E}_{ab}-n_a\mathbf{M}_b-n_b\mathbf{M}_a+n_an_b \mathbf{H}
	\end{equation}
	
	\noindent
	with
	
	\begin{subequations}
		
		\begin{equation}
		\mathbf{E}_{ab}=E_{cd}h_a^ch_b^d,
		\end{equation}
		
		\begin{equation}
		\mathbf{H}=E_{ab}n^an^b
		\end{equation}

		\begin{equation}
		\mathbf{M}_a=E_{cb}h_a^cn^b.
		\end{equation}
		
	\end{subequations}
	
	\noindent
	These variables denote the spatial evolution equation, the Hamiltonian constraint and the momentum constraint, respectively.
	
	First, we consider
	
	\begin{equation}
	n^b\nabla^aE_{ab}=-\mathbf{E}_{ab}\nabla^an^b-n^bn_a\nabla^a\mathbf{M}_b+\nabla^a\mathbf{M}_a-n_a\nabla^a\mathbf{H}+K\mathbf {H},
	\end{equation}
	
	\noindent
	using $\mathbf{E}_{ab}n^b=0$, $\mathbf{M}_an^a=0$, $n_an^a=-1$ and $\nabla^an_a=-K$. Furthermore, the following identities hold:
	
	\begin{subequations}
		\begin{equation}
		\nabla^a\mathbf{M}_a=D^a\mathbf{M}_a+\mathbf{M}_bn^a\nabla_an^b
		\end{equation}
		\begin{equation}
		\mathbf{E}_{ab}\nabla^an^b=-\mathbf{E}_{ab}K^{ab}
		\end{equation}
		\begin{equation}
		n^bn_a\nabla^a\mathbf{M}_b=-\mathbf{M}_bn_a\nabla^an^b=-\mathbf{M}_b\frac{D^b N}{N}
		\end{equation}
		
	\end{subequations}
	
	We would like to use the spatial projection of the trace reversed version of $\mathbf{E}_{ab}$ as evolution equation\footnote{The reason why we prefer to use $\mathcal{E}_{ab}=0$ as evolution equation rather than $\mathbf{E}_{ab}=0$ is that in general relativity the latter approach yields only a weakly hyperbolic system of equations for the constraint variables \cite{sarbach}.}, i.e.
	
	\begin{equation*}
	\mathcal{E}_{ab}=h_a^ch_b^d\left(E_{cd}-\frac12g^{ef}E_{ef}g_{cd}\right)=\mathbf{E}_{ab}-\frac12 \mathbf{E}h_{ab}+\frac12 \mathbf{H}h_{ab}.
	\end{equation*}
	
	\noindent
	Hence, we set	
	
	\begin{equation}
	\mathbf{E}_{ab}=\mathcal{E}_{ab}+\mathbf{H}h_{ab}-\mathcal{E}h_{ab}.
	\end{equation}

	\noindent
	With these identities we have
	
	\begin{equation}\label{ham_ev}
	n^b\nabla^aE_{ab}=-\left(\partial_t-\mathcal{L}_N\right)\mathbf{H}+2NK\mathbf{H}+\frac{1}{N}D^i\left(N^2\mathbf{M}_i\right)+N\mathcal{E}_{ij}\left(K^{ij}-Kh^{ij}\right).
	\end{equation}
	
	\noindent
	Next, we consider the spatial projection
	
	\begin{equation}
	h^b_c\nabla^aE_{ab}=h^b_c\nabla^a\mathbf{E}_{ab}-\nabla^an_a \mathbf{M}_c-n_ah^b_c\nabla^a\mathbf{M}_b-\nabla_an_b h^b_c\mathbf{M}_a+\mathbf{H}h^b_cn_a\nabla^an_b.
	\end{equation}
	
	\noindent
	In this case we use
	
	\begin{subequations}
		\begin{equation}
		h^b_c\nabla^a\mathbf{E}_{ab}=D^a\mathbf{E}_{ac}+\mathbf{E}_{ac}n^b\nabla_b n^a
		\end{equation}
		\begin{equation}
		n_ah^b_c\nabla^a\mathbf{M}_b=h^b_c\left(\mathcal{L}_n\mathbf{M}_b-\mathbf{M}_a\nabla_bn^a\right)
		\end{equation}
		\begin{equation}
		h^b_c\mathbf{M}_a\nabla^an_b=h^b_c\mathbf{M}_a\nabla_bn^a=-\mathbf{M}_aK^a_b
		\end{equation}
		\begin{equation}
		\mathbf{H}h^b_cn_a\nabla^an_b=\mathbf{H}n_a\nabla^an_c
		\end{equation}
		
	\end{subequations}
	
	\noindent
	to obtain
	
	\begin{equation}\label{mom_ev}
	\nabla^\mu E_{\mu i}=-\left(\partial_t-\mathcal{L}_N\right)\mathbf{M}_i+NK\mathbf{M}_i+\frac{1}{N}D_i\left(N^2\mathbf{H}\right)+D^j\left[N\left(\mathcal{E}_{ij}-\mathcal{E}h_{ij}\right)\right]
	\end{equation}
	
	\noindent
	(cf. eqn. (103,104) in \cite{sarbach}).
	
	Now we consider the generalized version of the Bianchi identity, that is,
	
	\begin{equation}\label{bianchi}
	\nabla^aE_{ab}-E_\phi\nabla_b\phi=0
	\end{equation}
	
	\noindent
	which is a consequence of the diffeomorphism invariance of the Horndeski action. Recall that we wish to use $\tilde{ E}_\phi=0$ as the scalar evolution equation. For this reason we set 
	
	\begin{equation}\label{scalar_mod}
	E_\phi=\tilde E_\phi+\partial_X G_3\left(E_{cd}-\frac12Eg_{cd}\right)\nabla^c\phi\nabla^d\phi.
	\end{equation}

	Putting together equations (\ref{ham_ev}), (\ref{mom_ev}), (\ref{bianchi}) and (\ref{scalar_mod}) gives the equations governing the evolution of the momentum and Hamiltonian constraints
	
	\begin{align}\label{prop_ham}\nonumber
	\left(\partial_t-\mathcal{L}_N\right)\mathbf{H}=&~2NK\mathbf{H}+\frac{1}{N}D^i\left(N^2\mathbf{M}_i\right)+N\mathcal{E}_{ij}\left(K^{ij}-Kh^{ij}\right)\\
	&-\tilde E_\phi A-N\partial_X G_3 A\left(\mathcal{E}_{ij}A^iA^j+2\mathbf{H}A^2+\mathcal{E}A^2-2A\mathbf{M}_iA^i\right)
	\end{align}
	
	\begin{align}\label{prop_mom}\nonumber
	\left(\partial_t-\mathcal{L}_N\right)\mathbf{M}_i=&NK\mathbf{M}_i+\frac{1}{N}D^i\left(N^2\mathbf{H}\right)+D^j\left[N\left(\mathcal{E}_{ij}-\mathcal{E}h_{ij}\right)\right]\\
	&-\tilde E_\phi A_i-N\partial_X G_3 A_i\left(\mathcal{E}_{ij}A^iA^j+2\mathbf{H}A^2+\mathcal{E}A^2-2A\mathbf{M}_iA^i\right).
	\end{align}
	
	\noindent
	In both of these equations, the terms in the second line arise due to the fact that we use equation (\ref{ev_sc_adm}) instead of (\ref{original_sc}) as the scalar equation of motion. Note that in each of the formulations studied in this paper, the tensor evolution equation is modified with gauge fixing terms. In other words, the equation $\mathcal{E}_{ij}=0$ is replaced with a different equation which introduces additional terms into equations (\ref{prop_ham}-\ref{prop_mom}). This will be analysed on a case-by-case basis.

	\section{Elliptic-Hyperbolic formulation}\label{sec:ell_hyp}
	
	\subsection{Review of Andersson and Moncrief's results}
	
	In this subsection we briefly summarize the work done by Andersson and Moncrief in \cite{andersson} on the vacuum Einstein’s equations. Firstly, we describe how they derived a coupled elliptic-hyperbolic system equivalent to the vacuum Einstein's equations. Then we sketch their arguments establishing local well-posedness.

	We start from the ADM formulation in which the vacuum Einstein equations
	
	\begin{equation}
	\mathcal{R}_{ab}=0
	\end{equation}
	
	\noindent
	are rewritten as two sets of first order in time evolution equations
	
	\begin{subequations}\label{adm_einst}
		
		\begin{equation}\label{def_curv}
		\left(\partial_t-\mathcal{L}_N\right)h_{ij}=-2NK_{ij}
		\end{equation}
		\begin{equation}\label{einst_ev_00}
		\left(\partial_t-\mathcal{L}_N\right)K_{ij}=-D_iD_jN+N\left(R_{ij}+KK_{ij}-2K_{ik}K^{k}_j\right),
		\end{equation}
		
		\noindent
		complemented by the Hamiltonian constraint
		
		\begin{equation}\label{ham_einst}
		2\mathbf{H}\equiv 2E_{\mu\nu}n^\mu n^\nu=R+K^2-K_{ij}K^{ij}=0
		\end{equation}
		
		\noindent
		and the momentum constraint
		
		\begin{equation}\label{mom_einst}
		\mathbf{M}_i\equiv E_{\mu i}n^\mu=  D_iK-D^jK_{ij}=0.
		\end{equation}
		
	\end{subequations}

	Andersson and Moncrief consider a modified version of the system (\ref{adm_einst}) by imposing constant mean curvature (CMC) slicing\footnote{We see that the mean curvature $K$ is constant over the slices $\Sigma_t$, but not necessarily in time. As it is mentioned in \cite{andersson}, they could also have considered a prescribed mean curvature slicing, i.e. $K=s(t,x)$.}
	
	\begin{equation}
	K\equiv h^{ij}K_{ij}=t
	\end{equation}
	
	\noindent
	and a spatial harmonic (SH) gauge condition
	
	\begin{equation}\label{def:sh}
	V^i\equiv h^{kl}\left(\Gamma^i_{kl}-\tilde{\Gamma}^i_{kl}\right)=0.
	\end{equation}
	
	\noindent
	The evolution equations are the defining equation of the extrinsic curvature (\ref{def_curv}) and (\ref{einst_ev_00}) modified by adding $-D_{(i}V_{j)}$ to the RHS of (\ref{einst_ev_00}):
	
	\begin{subequations}
		
		\begin{equation}\label{einst_ev_0}
		\left(\partial_t-\mathcal{L}_N\right)h_{ij}=-2NK_{ij}
		\end{equation}
		\begin{equation}\label{einst_ev}
		\left(\partial_t-\mathcal{L}_N\right)K_{ij}=-D_iD_jN+N\left(R_{ij}-D_{(i}V_{j)}+KK_{ij}-2K_{ik}K^{k}_{~j}\right).
		\end{equation}
		
		\noindent
		Equations (\ref{ham_einst},\ref{mom_einst}) are replaced by the modified constraints which can be regarded as the equations which determine the lapse function and the shift vector:
		
		\begin{equation}\label{ell_lapse}
		-D^iD_i N+NK_{ij}K^{ij}=1
		\end{equation}
		
		\begin{equation}\label{ell_shift}
		D^kD_kN^i+R^i_{~j}N^j-\mathcal{L}_NV^i=\left(-2NK^{kl}+2D^kN^l\right)\left(\Gamma^i_{kl}-\mathring{\Gamma}^i_{kl}\right)+2K^{ij}D_jN-D^iNK
		\end{equation}
		
	\end{subequations}
	
	\noindent
	Equation (\ref{ell_lapse}) can  be obtained by taking the trace of (\ref{einst_ev}), using the Hamiltonian constraint to trade in the Ricci curvature $R$ for lower order terms and using the CMC condition to set $\left(\partial_t-\mathcal{L}_N\right)K=1$. 
	
	Equation (\ref{ell_shift}) can  be derived as follows. Taking the time derivative of $V^i$ and commuting $\partial_t$ with $h^{kl}$ and the spatial derivatives, one easily obtains
	
	\begin{equation}\label{eq:dtv}
	\partial_t V^i=D^kD_kV^i+R^i_{~j}N^j-\left(-2NK^{kl}+2D^kN^l\right)\left(\Gamma^i_{kl}-\mathring{\Gamma}^i_{kl}\right)-2D_j(NK^{ij}) +D^i\left(NK\right).
	\end{equation}
	
	Using the momentum constraint, the CMC slicing condition ($D_iK=0$) and the spatial harmonic condition $(\partial_t-\mathcal{L}_N)V^i=0$ to eliminate second derivatives of the spatial metric and first derivatives of the extrinsic curvature then yields (\ref{ell_shift}). It is worth emphasizing that the CMC slicing condition was used to arrive at both equation (\ref{ell_lapse}) and (\ref{ell_shift}).
	
	Now we move on to the question of well-posedness of the system (\ref{einst_ev_0}-\ref{ell_shift}) in Sobolev spaces and consider initial data $h_{ij}, N, N^i\in H^s$ and $K_{ij}\in H^{s-1}$ ($s>\tfrac52$ in 4-dimensional spacetime) that satisfies the Hamiltonian and momentum constraints.
	
	The modified constraints (\ref{ell_lapse},\ref{ell_shift}) are equations relating derivatives of $N,N^i$ up to second order to derivatives of $h_{ij}$ up to first order (including the extrinsic curvature), when written in coordinates. This statement is obvious for (\ref{ell_lapse}) but one can check that the second derivatives of $h_{ij}$ cancel each other out on the LHS of (\ref{ell_shift}). More precisely, the modified constraints have the form
	
	\begin{equation}
	\mathbf{A}(h,\partial h, K)u=\left(\begin{array}{c}
1\\
0
\end{array}\right)
	\end{equation}
	
	\noindent
	with $u=(N,N^i)^T$ and $\mathbf{A}$ being a second order, linear elliptic differential operator, with coefficients depending only on the spatial metric, its first spatial derivatives and the extrinsic curvature. Moreover, the elliptic operator $\mathbf{A}$ is lower triangular:

\begin{equation}
\mathbf{A}=\left(\begin{array}{cc}
-D^iD_i+K_{ij}K^{ij} & 0 \\
\mathbf{B}^i(h,\partial h, K) & \mathbf{C}^i_{~j}(h,\partial h, K)
\end{array}\right)
\end{equation}

\noindent
with

\begin{equation}\label{shift_op_gr}
\mathbf{C}^i_{~j}(h,\partial h, K)N^j\equiv -D^kD_k N^i-R^i_{~j}N^j+\mathcal{L}_NV^i-2D^kN^l\left(\Gamma^i_{kl}-\mathring{\Gamma}^i_{kl}\right).
\end{equation}

Standard results in the theory of elliptic PDEs (see Appendix \ref{sec:elliptic}) show that the scalar elliptic operator $-D^iD_i+K_{ij}K^{ij}$ is an isomorphism $H^{s}\to H^{s-2}$. Furthermore, it is proved in \cite{andersson} that the elliptic operator $\mathbf{C}^i_{~j}$ is an isomorphism $H^{s}\to H^{s-2}$, if $(\Sigma,h)$ is a compact manifold with negative Ricci curvature\footnote{Note that this formulation was ultimately used to prove a global existence theorem for non-linear perturbations of spatially compact versions of FRW spacetimes with $k=-1$ \cite{andersson2}.}. These results and the lower triangular structure of $\mathbf{A}$ then implies that $\mathbf{A}$ is also an isomorphism $H^{s}\to H^{s-2}$. Therefore, if we a priori assume that $h_{ij}\in H^s$ and $K_{ij}\in H^{s-1}$, then the unique solutions $N,N^i$ to the modified constraints are in $H^{s+1}$, i.e. they have an extra regularity compared to $h_{ij}$. This is necessary because this way the terms involving first derivatives of $N^i$ in (\ref{einst_ev_0}) and second derivatives of $N$ in (\ref{einst_ev}) are non-principal.

	It follows that by solving the modified constraints to determine $N$ and $N^i$, the evolution equations become a first order quasilinear system of pseudodifferential equations. It is easy to see now that this system of evolution equations is strongly hyperbolic. Linearizing the evolution equations and substituting the derivatives with the Fourier variables $(\partial_t,\partial_k)\to (i\xi_0,i\xi_k)$, the coefficients of the highest derivative (principal) terms define the principal symbol. Based on the definitions and general arguments presented in Appendix \ref{sec:hyp}, strong hyperbolicity means that the principal symbol has real eigenvalues with a complete set of eigenvectors with smooth dependence on $\xi_i$. Linearizing (\ref{einst_ev_0}) and (\ref{einst_ev}) around a generic solution $h_{ij}\to h_{ij}+\gamma_{ij}$, $K_{ij}\to K_{ij}+\kappa_{ij}$, the eigenvalue problem of the principal symbol reads as 
	
	\begin{equation}
	i\xi_0\left(\begin{array}{c}
	\hat \gamma_{ij}\\
	2\hat \kappa_{ij}
	\end{array}\right)=P_0(\xi)^{kl}_{ij}\left(\begin{array}{c}
	\hat{\gamma}_{kl}\\
	2\hat{\kappa}_{kl}
	\end{array}\right)
	\end{equation}
	
	\noindent
	where $\mathbb{M}_0$ is the $2\times 2$ block matrix (recall that the terms involving second derivatives of $N$ and first derivatives of $N^i$ are non-principal)
	
	\begin{equation}
	\mathbb{M}_0(\xi)^{kl}_{ij}=\left(\begin{array}{cc}
	i \left(N^m\xi_m\right)h^k_ih_j^l & -Nh^k_ih_j^l \\
	N\xi^2h^k_ih_j^l & i \left(N^m\xi_m\right)h^k_ih_j^l
	\end{array}\right).
	\end{equation}
	
	\noindent
	It is easy to see that the principal symbol has eigenvalues $\xi_0^\pm=N^k\xi_k\pm N|\xi|_h$ with a complete set of eigenvectors: for any symmetric matrix $u_{ij}$,
	
	\begin{equation}
	\left(\begin{array}{c}
	\hat \gamma_{ij}\\
	2\hat \kappa_{ij}
	\end{array}\right)=\left(\begin{array}{c}
	u_{ij}\\
	\mp i \left(n^\mu\xi_\mu\right) u_{ij}
	\end{array}\right)
	\end{equation}
	
	\noindent
	is an eigenvector with $\xi_\mu=(\xi_0,\xi_i)$. Note that this means that $\xi_\mu$ is a null vector. Since all eigenvalues are real and the eigenvectors can be chosen to be independent of $\xi_k$, the system of evolution equations is strongly hyperbolic when the modified constraints are solved. (In fact, it is symmetric hyperbolic so one can demonstrate well-posedness by standard energy methods in physical space, as was done in \cite{andersson}.) 
	
	\subsection{Equations of motion and gauge fixing in cubic Horndeski theories}

	We will now show that the above formalism can be extended to cubic Horndeski theories. For this, we first discuss the generalization of the SH-CMC gauge condition. Recall that the CMC condition in general relativity was used to set $(\partial_t-\mathcal{L}_N)K$ to $1$ in the trace of the evolution equation. When taking the trace of (\ref{g3_ev}), it is possible to get a constraint equation by a choice of an appropriate slicing condition which sets the terms involving $(\partial_t-\mathcal{L}_N)K$ and $(\partial_t-\mathcal{L}_N)A$ to an a priori fixed function. For this reason, we take an approach very similar to the one above: in the trace of (\ref{g3_ev}) we trade in $R$ using the Hamiltonian constraint to get
	
	\begin{align}\label{trace}\nonumber
	(\partial_t-\mathcal{L}_N)K-\frac14\partial_XG_3&(3A^2-A_kA^k)(\partial_t-\mathcal{L}_N)A=-D^i D_iN -\frac14\partial_X G_3\left(3A^2-A^kA_k\right)A^iD_i N \\ \nonumber
	&+N\left\{K_{ij}K^{ij}+\frac12 A^2+\frac12 G_2+\frac14\left(A^2+A^kA_k\right)\partial_XG_2+A^2\partial_\phi G_3\right. \\ 
	& \qquad \left. {}+\partial_XG_3\left(\frac14\left(A^2+A^kA_k\right)\left(D^iA_i-AK\right) -AA^iD_iA+AK_{ij}A^iA^j\right)\right\}
	\end{align}
	
	\noindent
	We are seeking a gauge condition of the form 
	
	\begin{equation}\label{cmc_g3}
	K+f(\phi,A,A_i,h_{kl})=s(x,t)
	\end{equation}

	\noindent
	to eliminate the time derivatives in equation (\ref{trace}). Taking the normal derivative of (\ref{cmc_g3}) gives

	\begin{equation}
	(\partial_t-\mathcal{L}_N)s(x,t)=(\partial_t-\mathcal{L}_N)K+N\partial_\phi f A+\partial_A f~(\partial_t-\mathcal{L}_N)A+\partial_{A_k}fD_k(NA)-2NK_{kl}\frac{\partial f}{\partial h_{kl}}.
	\end{equation}
	
	\noindent
	Therefore, the desired choice is an $f$ satisfying
	
	\begin{equation}\label{g3_slicing}
	\partial_A f=-\frac14\partial_XG_3(3A^2-A_kA^k).
	\end{equation}

\noindent
	Note that this slicing condition has an interesting relationship with the canonical momentum $\pi^{ij}$ conjugate to $h_{ij}$. If we switch to a Hamiltonian description, equation (\ref{trace}) (which is the trace of (\ref{g3_ev})) is equivalent to the trace of

	\begin{equation}
	\partial_t\pi^{ij}=-\frac{\delta \mathcal{H}}{\delta h_{ij}}
	\end{equation}

\noindent
where $\mathcal{H}$ is the Hamiltonian. Hence, it is clear that the time differentiated terms in (\ref{trace}) come from $\partial_t(h_{ij}\pi^{ij})$, that is, the preferred slicing condition is equivalent to $\pi=s(x,t)$.

	Rewriting $G_3(\phi,X)$ as a function depending on $\phi$, $A$, $A_i$ and $h^{ij}$, the condition (\ref{g3_slicing}) can be integrated in $A$ and so $f$ can be determined up to the addition of an arbitrary function of $\phi$, $h_{ij}$ and $A_i$. Hence, the elliptic equation for $N$ reads as
	
	\begin{align}\label{ell_N_g3}\nonumber
	(\partial_t-\mathcal{L}_N)s(x,t)=&-D^i D_iN -\left[\frac14\partial_X G_3\left(3A^2-A^kA_k\right)A^i-A\partial_{A_i}f\right]D_i N+N\biggl\{K_{ij}K^{ij}+\frac12 A^2 \\ \nonumber
	&+\partial_\phi f A-2K_{ij}\frac{\partial f}{\partial h_{ij}}+\partial_{A_i}f D_i A+\frac12 G_2+\frac14\left(A^2+A^kA_k\right)\partial_XG_2+A^2\partial_\phi G_3 \\ 
	&+\partial_XG_3\left(\frac14\left(A^2+A^kA_k\right)\left(D^iA_i-AK\right) -AA^iD_iA+AK_{ij}A^iA^j\right)\biggr\}.
	\end{align}
	
	\noindent
	Based on standard results in the theory of elliptic PDEs (see Appendix \ref{sec:elliptic} and the references cited therein), this equation has a unique solution for $N$ if the coefficient of the zeroth order term (i.e. terms enclosed by the curly brackets $\{\}$) is non-negative. This condition is satisfied if the Horndeski terms are small corrections to GR, i.e. in the weak field regime. Note that the weak field requirements on $f$ are

	\begin{align}
	&|\partial_\phi^l\partial_A^kf|E^{k+1}\ll 1 &k=0,1;~ l=0,1 \nonumber \\
&|\partial_{A^i}^kf|E^{k+1}\ll 1 &k=0,1 \nonumber \\
&|\partial_{h_{ij}}^kf|E^{k+1}\ll 1  &k=0,1;
	\end{align}

	\noindent
with $E=\text{max}\left\{|\mathcal{R}_{\mu\nu\rho\sigma}|^{1/2},|\nabla_\mu\phi|,|\nabla_\mu\nabla_\nu\phi|^{1/2}\right\}$ in an orthonormal basis.
	
	The generalization of the spatial harmonic gauge is more straightforward: we require
	
	\begin{equation}
	J^i\equiv ~V^i+H^i(h_{kl},\phi,A_k)=0
	\end{equation}
	
	\noindent
	for some $H^i$ ($V^i$ is as in equation (\ref{def:sh})). Once again, we can derive an elliptic equation for the shift vector by requiring $(\partial_t-\mathcal{L}_N)J^i=0$ and eliminating derivatives of the extrinsic curvature by using the momentum constraint and the generalized CMC condition. The result is
	
	\begin{align}\nonumber
	& D^kD_kN^i+R^i_jN^j-\mathcal{L}_NV^i-\left(-2NK^{kl}+2D^kN^l\right)\left(\Gamma^i_{kl}-\tilde{\Gamma}^i_{kl}\right)= 2K^{ij}D_jN -D^iNK\\ \nonumber
	& -(\partial_t-\mathcal{L}_N)H^i-ND^if-N\left\{ \left(1+\partial_XG_2+2\partial_\phi G_3\right)A^i A+\partial_XG_3\left(AA^i\left(D^kA_k-AK\right)+\right.\right.\\
	& \qquad \left. {} \qquad \left. {} K_{kl}A^kA^lA^i+A^2D^iA-A^iA^kD_kA-AA^kD_kA^i \right) \right\}.
	\end{align}
	
	\noindent
	Furthermore, using
	
	\begin{equation}
	(\partial_t-\mathcal{L}_N)H^i=N\partial_\phi H^i A+\partial_{A_k}H^i D_k(NA)-2NK_{kl}\frac{\partial H^i}{\partial h_{kl}}
	\end{equation}
	
	\noindent
	and
	
	\begin{equation}
	D^if=\partial_\phi f A^i+\partial_A f~D^iA+\partial_{A_k}fD^iA_k=\partial_\phi f A^i-\frac14\partial_XG_3(3A^2-A_kA^k)~D^iA+\partial_{A_k}fD^iA_k
	\end{equation}
	
	\noindent
	gives
	
	\begin{align}\label{shift_g3}\nonumber
	& D^kD_kN^i+R^i_jN^j-\mathcal{L}_NV^i-\left(-2NK^{kl}+2D^kN^l\right)\left(\Gamma^i_{kl}-\tilde{\Gamma}^i_{kl}\right)= 2K^{ij}D_jN -D^iNK-A\partial_{A_k}H^i D_kN\\ \nonumber
	& -N\Biggl\{ \left(1+\partial_XG_2+2\partial_\phi G_3\right)A^i A+A\partial_\phi H^i A+\partial_{A_k}H^i D_kA-2K_{kl}\frac{\partial H^i}{\partial h_{kl}}+\partial_\phi f A^i +\partial_{A_k}fD^iA_k\\
	& +\partial_XG_3\left(AA^i\left(D^kA_k-AK\right)+ K_{kl}A^kA^lA^i+\frac14\left(A^2+A^kA_k\right)D^iA-A^iA^kD_kA-AA^kD_kA^i \right)\Biggr\}.
	\end{align}
	
	\noindent
	The operator on the LHS acting on $N^i$ is exactly the same as in GR (c.f. (\ref{shift_op_gr})) which means that (\ref{shift_g3}) has a unique solution for $N^i$ on spatial slices with negative spatial Ricci curvature.

	Finally, we need to decide how to use the generalized spatial harmonic condition in the evolution equations. As mentioned before, we do not modify the scalar equation $\tilde E_\phi$, only the tensor equation in the most natural way, i.e. by replacing $\mathcal{E}_{ij}$ by $\tilde{\mathcal{E}}_{ij}\equiv\mathcal{E}_{ij}-D_{(i}J_{j)}$:

	\begin{align}\label{g3_ev_mod} \nonumber
	&\left(\partial_t-\mathcal{L}_N\right)K_{ij}-\frac12\partial_XG_3(Xh_{ij}+A_iA_j)\left(\partial_t-\mathcal{L}_N\right)A=-D_i D_jN +N\left\{R_{ij}-D_{(i}J_{j)}+KK_{ij}-2K_{ik}K^k_j \right. \\ \nonumber
	&\qquad \left. {}+\frac12h_{ij}\left(G_2-X\partial_XG_2-X\partial_XG_3\left(D^kA_k+AK+A^kD_k\ln N\right)\right)-\frac12\left(1+\partial_XG_2+2\partial_\phi G_3\right)A_iA_j\right. \\ 
	&\qquad \left. {}+\frac12\partial_XG_3\left(A^kD_k\left(A_iA_j\right)-A_iA_jD^kA_k-A_iA_jAK-A_iA_jA^kD_k\ln N-2AA_{(i}D_{j)}A\right)\right\}.
	\end{align}
	
	Therefore, the Cauchy problem for cubic Horndeski theories can be formulated as follows. Consider initial data $h_{ij}, N, N^i,\phi\in H^s$ and $K_{ij},A\in H^{s-1}$ ($s>\tfrac72$)\footnote{The lower bound on $s$ is stronger than in vacuum GR due to the fact that the scalar evolution equation (\ref{ev_sc_adm}) is not quasilinear (see Appendix \ref{sec:hyp}).} that satisfies the Hamiltonian and momentum constraints. Then the system of equations to be solved consists of the evolution equations (\ref{extr_curv_def}), (\ref{def_A}), (\ref{ev_sc_adm}) and (\ref{g3_ev_mod}), together with the elliptic equations (\ref{ell_N_g3}), (\ref{shift_g3}).



	

	\subsection{Constraint propagation}

	Before moving on to the question of well-posedness, in this section we explain how to get the equations describing the propgation of the gauge conditions and the original constraints using the gauge-fixed equations of motion. As described in the previous section, we use $\tilde{\mathcal{E}}_{ij}\equiv\mathcal{E}_{ij}-D_{(i}J_{j)}=0$ as tensor evolution equation. Following the argument started in section \ref{sec:setup_prop}, we set $\tilde{\mathcal{E}}_{ij}=\tilde{{E}}_{\phi}=0$ and switch to the new variables 
	
	\begin{subequations}
		\begin{equation}
		\mathbf{F}\equiv K+f-s(x,t),
		\end{equation}
		
		\begin{equation}
		\mathbf{J}_i \equiv V_i+H_i,
		\end{equation}
		
		\begin{equation}
		\mathbf{\tilde H}\equiv 2\mathbf{H}-D^k\mathbf{J}_k
		\end{equation}
		
		\noindent
		and 
		
		\begin{equation}
		\mathbf{ \tilde M}_i=2\mathbf{M}_i-D_i\mathbf{ F},
		\end{equation}
		
	\end{subequations}
	
	\noindent
	we have the following system of homogeneous linear evolution equations
	
	\begin{subequations}\label{constr_prop}
		
		\begin{equation}\label{propF}
		\left(\partial_t-\mathcal{L}_N\right)\mathbf{F}=N\mathbf{\tilde H}
		\end{equation}
		
		\begin{equation}\label{propJ}
		\left(\partial_t-\mathcal{L}_N\right)\mathbf{J}_i=N \mathbf{\tilde M}_i
		\end{equation}
		
		\begin{align}\label{propH}\nonumber
		\left(\partial_t-\mathcal{L}_N\right)\mathbf{\tilde H}=&~2NK\mathbf{\tilde H}+D^iN\mathbf{\tilde M}_i+2ND_{(i}\mathbf{J}_{j)}K^{ij}+\mathbf{J}^jD_j\left(NK\right)+ND^iD_i \mathbf{F}+2D_iND^i\mathbf{F}\\
		&-\partial_X G_3 A\left(D_i\mathbf{J}_jA^iA^j+\mathbf{\tilde H}A^2+2D^i\mathbf{J}_iA^2-A\mathbf{\tilde M}_iA^i-AA^iD_i\mathbf{F}\right)
		\end{align}
		
		\begin{align}\label{propM}\nonumber
		\left(\partial_t-\mathcal{L}_N\right)\mathbf{\tilde M}_i=&~NK\mathbf{\tilde M}_i+NKD_i \mathbf{F}+D_iN\mathbf{\tilde H}+2D^jN D_{(i}\mathbf{J}_{j)}+N\left(D^kD_k\mathbf{J}_i+R_{ij}\mathbf{J}^j\right)\\
		&-\partial_X G_3 A_i\left(D_k\mathbf{J}_jA^kA^j+\mathbf{\tilde H}A^2+2D^k\mathbf{J}_kA^2-A\mathbf{\tilde M}_kA^k-AA^kD_k\mathbf{F}\right).
		\end{align}
		
	\end{subequations}
	
	\noindent
	The first two equations follow easily by recalling the steps we used to get the elliptic equations (\ref{ell_lapse},\ref{ell_shift}) from the evolution equations and the constraints.
	To show that the quantities $(\mathbf{F},\mathbf{J}_i,\mathbf{\tilde H},\mathbf{\tilde M}_i)$ remain zero during the evolution, we first note that it follows from equations (\ref{propF},\ref{propJ}) that if $(\mathbf{F},\mathbf{J}_i,\mathbf{\tilde H},\mathbf{\tilde M}_i)$ vanish initially then $\partial_t \mathbf{F}=\partial_t \mathbf{J}_i=0$ on the initial surface. It turns out that one can obtain a simple energy estimate for the system (\ref{constr_prop}) without solving the eigenvalue problem of the principal symbol. Consider the energy	\footnote{The steps of the proof of constraint propagation, as well as the expression for the energy, are the same as in \cite{andersson} (Lemma 4.1.), since the extra terms entering to the equations due to the presence of a scalar field are non-principal. Nevertheless, we provide a sketch of the proof for completeness.}

	\begin{equation}\label{constr_energy}
	E_\text{constraint}[\Sigma_t]=\frac12\int\limits_{\Sigma_t}\mathrm{d}^3x\sqrt{h}\left(|\mathbf{F}|^2+|D\mathbf{F}|_h^2+|\mathbf{J}|_h^2+|D\mathbf{J}|_h^2+|\mathbf{\tilde H}|^2+|\mathbf{\tilde M}|_h^2\right).
	\end{equation}

	\noindent
	Specifically, we want to show that $|\partial_tE_\text{constraint}|\leq C E_\text{constraint}$ for some constant $C(h,K,N)$. Clearly, the action of $\partial_t$ on the volume form can be bounded by a constant. When the time derivative acts on the gauge and constraint quantities, we use (\ref{constr_prop}) to exchange the time derivatives. Since the energy (\ref{constr_energy}) is invariant under spatial diffeomorphisms, the terms involving Lie derivatives will vanish.
	
	The non-principal terms can be estimated by the energy itself. For example,
	
	\begin{equation}
	(\partial_t-\mathcal{L}_N)|\mathbf{F}|^2=2N\mathbf{F}\mathbf{\tilde H}\leq C \left(|\mathbf{F}|^2+|\mathbf{\tilde H}|^2\right)
	\end{equation}
	
	\begin{equation}
	(\partial_t-\mathcal{L}_N)|\mathbf{J}|_h^2=2Nh^{ij}\mathbf{J}_i\mathbf{\tilde M}_j+2NK^{ij}\mathbf{J}_i\mathbf{J}_j\leq C \left(|\mathbf{J}|_h^2+|\mathbf{\tilde M}|_h^2\right).
	\end{equation}
	
	\noindent
	The potentially problematic (principal) terms are
	
	\begin{equation}
	(\partial_t-\mathcal{L}_N)|\mathbf{\tilde H}|^2\simeq 2N\mathbf{\tilde H} D^iD_i \mathbf{F} \sim -2ND^i \mathbf{\tilde H} D_i \mathbf{F}
	\end{equation}
	
	\begin{equation}
	(\partial_t-\mathcal{L}_N)|D\mathbf{J}|_h^2\simeq 2N D_i \mathbf{J}_j D_k \mathbf{\tilde M}_l h^{ik} h^{jl}\sim -2ND^kD_k \mathbf{J}_j \mathbf{\tilde M}_l h^{jl}
	\end{equation}
	
	\begin{equation}
	(\partial_t-\mathcal{L}_N)|D\mathbf{F}|^2\simeq 2ND_i\mathbf{F}D^i \mathbf{\tilde H}
	\end{equation}
	
	\begin{equation}
	(\partial_t-\mathcal{L}_N)|\mathbf{\tilde M}|_h^2\simeq  2ND^kD_k \mathbf{V}_j \mathbf{\tilde M}_l h^{jl}
	\end{equation}
	
	\noindent
	where $\simeq$ denotes equivalence up to principal terms and $\sim$ denotes equivalence of the integrands up to integration by parts. We see that the terms containing higher derivatives cancel each other out, giving us the desired result. Therefore, if $E_\text{constraint}$ vanishes initially, then it remains zero during the evolution as well, implying $(\mathbf{F},\mathbf{J}_i,\mathbf{\tilde H},\mathbf{\tilde M}_i)=0$.

	\subsection{Proof of strong hyperbolicity}
	
	Now we linearize the equations of motion (\ref{extr_curv_def}), (\ref{def_A}), (\ref{ev_sc_adm}), (\ref{g3_ev_mod}), (\ref{ell_N_g3}), (\ref{shift_g3}) around a  generic weak field background with negative spatial Ricci curvature. The linearized quantities are as follows:
	
	\begin{align*}
	h_{ij}&\to h_{ij}+\gamma_{ij}\\
	K_{ij}&\to K_{ij}+\kappa_{ij}\\
	N&\to N+\alpha\\
	N^i&\to N^i+\beta^i\\
	\phi&\to\phi+\psi\\
	A&\to A+a
	\end{align*}

	Now we assume that the elliptic equations (\ref{ell_N_g3}) and (\ref{shift_g3}) have unique solution for any $(\gamma_{ij},\kappa_{ij},\psi,a)$ satisfying the gauge conditions and constraint equations. Recall that this is true for weak fields and when the background spacetime has negative spatial Ricci curvature. 
	
	The important difference compared to GR is that in this case the first derivatives of $\beta^i$ appearing in the defining equation of the extrinsic curvature (\ref{extr_curv_def}) and the second derivatives of $\alpha$ appearing in the tensor evolution equation (\ref{g3_ev_mod}) are principal terms. To see how these terms affect the hyperbolicity of the evolution equations, we have to look at the principal terms in the linearized versions of the elliptic equations (\ref{ell_N_g3}), (\ref{shift_g3}):
	
	\begin{equation}
	h^{ij}\partial_i\partial_j\alpha\simeq\frac14 N\partial_XG_3\left(A^2+A^kA_k\right)h^{ij}\partial_i\partial_j\psi-N\left(-\partial_{A_k}f+\partial_XG_3AA^k\right)\partial_k a
	\end{equation} 
	
	\begin{align}\nonumber
	h^{kl}\partial_k\partial_l \beta^i \simeq &~N\left[\left(-\partial_{A_j}H^i+\partial_XG_3 A^iA^j\right)-\frac14\partial_XG_3\left(A^2+A^kA_k\right)h^{ij}\right]\partial_ja\\
	&+N\left[\left(-\partial_{A_k}f+\partial_XG_3AA^k\right)h^{il}-\partial_XG_3AA^ih^{kl} \right]\partial_k\partial_l\psi
	\end{align}
	
	\noindent
	Solving the linearized elliptic equations for given $\psi$, $a$, $\gamma_{ij}$ and $\kappa_{ij}$, one obtains a (non-local) map $(\hat\alpha,\hat\beta): (\psi, a, \gamma,\kappa)\to (\alpha,\beta)$. This solution map is a pseudodifferential operator of class $\mathcal{OP}_\text{cl}^{-2}$ with principal symbol (see Appendix \ref{sec:back})
	
	\begin{equation}
	\hat \alpha^{(0)}[ \hat{\psi},\hat{a};\xi]= \frac14 N\partial_XG_3\left(A^2+A^kA_k\right)\hat\psi+iN\left(-\partial_{A_k}f+\partial_XG_3AA^k\right) \frac{\xi_k}{|\xi|_h^2} \hat a
	\end{equation}
	
	\begin{align}\nonumber
	\hat{\beta}^{(0)i}[ \hat{\psi},\hat{a};\xi] = &~-iN\left[\left(-\partial_{A_j}H^i+\partial_XG_3 A^iA^j\right)-\frac14\partial_XG_3\left(A^2+A^kA_k\right)h^{ij}\right]\frac{\xi_j}{|\xi|_h^2}\hat a\\
	&+N\left[\left(-\partial_{A_k}f+\partial_XG_3AA^k\right)\frac{\xi^i\xi_k}{|\xi|_h^2}-\partial_XG_3AA^i \right]\hat\psi.
	\end{align}
	
	\noindent
	where the $~\hat{ }~$ on the fields $\psi$ and $a$ denotes the Fourier transform. Note that the principal symbol of the solution map does not depend on $\hat\gamma$ and $\hat\kappa$.
	
	By solving the elliptic equations, the evolution equations take the form of a first order pseudodifferential system of evolution equations. To determine the hyperbolicity of this system, we consider the eigenvalue problem of the principal symbol, c.f. (\ref{gen_char}). The vector $U$ is now a column vector of size $14$

$$U=\left[\hat{\gamma}_{ij},\hat{\kappa}_{ij},\hat\psi,\hat a\right]^T. $$

The characteristic equation can be written as
	
	\begin{subequations}
		
		\begin{equation}\label{ev_gamma}
		\frac{i}{N}\left(\xi_0-N^k\xi_k\right)\hat{\gamma}_{ij}=-2\hat{\kappa}_{ij}+\frac{2i}{N}h_{l(i}\xi_{j)}\hat{\beta}^{(0)l}[ \hat{\psi},\hat{a};\xi],
		\end{equation}
		
		\begin{equation}\label{ev_psi}
		\frac{i}{N}\left(\xi_0-N^k\xi_k\right)\hat{\psi}=\hat{a},
		\end{equation}
		
		\begin{align}\label{ev_kappa}\nonumber
		&\frac{i}{N}\left(\xi_0-N^k\xi_k\right)~2\hat{\kappa}_{ij}-\frac{i}{N}\partial_X G_3\left(Xh_{ij}+A_iA_j\right)\left(\xi_0-N^k\xi_k\right)\hat{a}=|\xi|_h^2\hat{\gamma}_{ij}+\frac{2}{N}\xi_i\xi_j\hat{\alpha}^{(0)}[\hat\psi,\hat a;\xi]\\
		&-\partial_XG_3\left(-(Xh_{ij}+A_iA_j)|\xi|_h^2\hat{\psi}+A^k\xi_kA_{(i}\xi_{j)}\hat{\psi}-2iAA_{(i}\xi_{j)}\hat{a}\right)+2\partial_{A_k}H_{(i}\xi_{j)}\xi_k\hat{\psi},
		\end{align}
		
		\begin{equation}\label{ev_a}
		\mathcal{A}\frac{i}{N}\left(\xi_0-N^k\xi_k\right)\hat{a}=i\mathcal{B}(\xi)\hat{a}+i\mathcal{C}(\xi)\hat{\psi}
		\end{equation}
		
	\end{subequations}
	
	\noindent
	where the specific form of the linearized version of (\ref{ev_sc_adm}) is quite long and unessential for our purposes. Nevertheless, we note that substituting (\ref{ev_psi}) into (\ref{ev_a}) gives (recall equation (\ref{char}))
	
	\begin{equation}
	\hat\psi\left(P^\prime_{\phi\phi}\right)^{\mu\nu}\xi_\mu\xi_\nu=0.
	\end{equation}
	
	It follows from the upper triangular structure of the characteristic equation (see Section \ref{sec:setup}) that

\begin{equation}\label{gr_eig_vec}
\left[\hat{\gamma}_{ij},\hat{\kappa}_{ij},\hat\psi,\hat a\right]^T=\left[\hat{u}_{ij},-i\frac12\left(n^\mu\xi_\mu^{\pm}\right)\hat{u}_{ij},0,0\right]^T
\end{equation}

 is an eigenvector of (\ref{ev_gamma}-\ref{ev_a}) with eigenvalues $\xi_0^{\pm}=N^k\xi_k\pm N|\xi|_h$ for any symmetric $\hat{u}_{ij}$. (Recall that $\xi_\mu=(N^k\xi_k\pm N|\xi|_h,\xi_i)$ is a null vector with respect to the spacetime metric.) One can easily find $6$ linearly independent vectors $\hat{u}_{ij}$. Taking into account the two sign choices in $\xi_0^\pm$ and in (\ref{gr_eig_vec}), this gives $12$ eigenvectors: a $6$-dimensional eigenspace for both eigenvalues $\xi_0^{\pm}=N^k\xi_k\pm N|\xi|_h$. Note that these include the $2$ pairs of transverse-traceless modes corresponding to the gravitational degrees of freedom, i.e.

\begin{equation}
\left[\hat{\gamma}_{ij},\hat{\kappa}_{ij},\hat\psi,\hat a\right]^T=\left[\hat{u}_{ij}^{TT},-i\frac12\left(n^\mu\xi_\mu^{\pm}\right)\hat{u}_{ij}^{TT},0,0\right]^T
\end{equation}

with $\hat{u}_{ij}^{TT}$ satisfying $h^{ij}\hat{u}_{ij}^{TT}=0$ and $\xi^i\hat{u}_{ij}^{TT}=0$.
	
	The remaining eigenvalues $\xi_0^{\phi,\pm}$ are found by solving
	
	\begin{equation}\label{char_phi}
	\left(P^\prime_{\phi\phi}\right)^{\mu\nu}\xi_\mu\xi_\nu=0
	\end{equation}
	 
	\noindent
	(see Section \ref{sec:setup} for notation). Recall that for weak fields, $\xi_0^{\phi,\pm}$ are distinct and real.
	The corresponding eigenvectors have the form 

$$\left[\hat{\gamma}_{ij},\hat{\kappa}_{ij},\hat\psi,\hat a\right]^T=\left[\hat{\gamma}_{ij}^\phi,\hat\kappa_{ij}^\phi,1,i\left(n^\mu\xi_\mu\right)\right]^T$$ 

\noindent
with
	
	\begin{equation}\label{ell_hyp_gamma}
	\hat\gamma_{ij}^\phi=-\partial_X G_3\left(Xh_{ij}+A_iA_j\right)+\frac{2}{|\xi|_h^2}\left(\partial_X G_3 A^kA_{(i}\xi_{j)}-\partial_{A_k}H_{(i}\xi_{j)}\right)\xi_k-\frac{\xi_i\xi_j}{2|\xi|_h^2}\partial_XG_3\left(A^2+A^kA_k\right)
	\end{equation}
	
	\begin{equation}\label{ell_hyp_kappa}
	\hat\kappa_{ij}^\phi=i\left(n^\mu\xi_\mu^{\phi,\pm}\right)\partial_X G_3\left(Xh_{ij}+A_iA_j\right)+2i\left[\left(-\partial_{A_k}f+\partial_XG_3AA^k\right)\frac{\xi_i\xi_j\xi_k}{|\xi|_h^2}-\partial_XG_3AA_{(i}\xi_{j)}\right].
	\end{equation}
	
	These expressions are clearly smooth functions of $\xi_i$ for any choice of $H_i$, since $h_{ij}$ is a positive definite metric and $\xi_i\neq 0$ by assumption ($\xi_i$ has unit norm). 
	
	
	These two “physical” eigenvectors corresponding to the scalar degree of freedom satisfy the high frequency limit of the Hamiltonian and momentum constraints
	
	\begin{equation}\nonumber
	2\hat{\mathbf{H}}\equiv-2\xi^2h^{ij}\hat\gamma_{ij}+2\xi^i\xi^j\hat\gamma_{ij}+\partial_X G_3\left(A^2|\xi|_h^2-(A^k\xi_k)^2\right)\hat\psi=0,
	\end{equation}

	\begin{equation}\nonumber
	\hat{\mathbf{M}}_i\equiv i\xi_i\hat\kappa-i\xi^j\hat\kappa_{ij}+\frac12\partial_X G_3\left(A A_i|\xi|_h^2\hat\psi-AA^j\xi_j\xi_i\hat\psi+A_iA^k\xi_k i\hat a-A^2i\xi_i\hat a\right)=0.
	\end{equation}
	
	To summarize, we have found that the principal symbol (see equations (\ref{ev_gamma}-\ref{ev_a})) has $14$ real eigenvalues and the corresponding eigenvectors are linearly independent and have smooth dependence on $\xi\in\mathbb{S}^2$. This implies that the evolution equations are strongly hyperbolic when the modified constraint equations have a unique solution for arbitrary $(\phi,A,h,K)$. In particular, this is the case in a weak field regime and in spacetimes that can be foliated with generalized prescribed mean curvature slices with negative Ricci curvature. 
	

	\section{BSSN-type formulation with non-dynamical shift vector}\label{sec:bssn}
	
	\subsection{Equations of motion}
	
	The Baumgarte-Shapiro-Shibata-Nakamura (BSSN) formulation (together with its modifications) is a popular method used to numerically integrate Einstein’s equations. Several versions of this approach give rise to a strongly hyperbolic reformulation of the vacuum Einstein equations \cite{sarbach_bssn,beyer,gundlach}. Here, we extend this approach to cubic Horndeski theories.
	
	The equations of motion are obtained from the standard ADM equations given in Section \ref{sec:setup} as follows. First, we introduce the conformal metric $\tilde h_{ij}$ as a new variable, defined by
	
	\begin{equation}\label{defh}
	\tilde h_{ij}\equiv e^{-4\Omega} h_{ij}
	\end{equation}
	
	\noindent
	where the conformal factor $\Omega$ is
	
	\begin{equation}\label{defomega}
	\Omega\equiv\frac{1}{12}\ln \frac{h}{\mathring{h}}
	\end{equation}
	
	\noindent
	for an arbitrary smooth background metric $\mathring{h}_{ij}$. Note that this implies that $\det\mathring{h}=\det\tilde{h}$. The inverse conformal metric, denoted by $\tilde h^{ij}$ is then

	\begin{equation}\label{defh2}
	\tilde h^{ij}=e^{4\Omega} h^{ij}.
	\end{equation}
	
	\noindent
	Next, we define the quantity
	
	\begin{equation}\label{defv}
	\tilde V^i\equiv\tilde h^{kl}\left(\tilde \Gamma^i_{kl}-\mathring{\Gamma}^i_{kl}\right)= -\mathring{D}_j\tilde h^{ij}
	\end{equation}
	
	\noindent
	where $\mathring{\Gamma}$ and $\mathring{D}$ denote the Christoffel symbol and the covariant derivative corresponding to $\mathring{h}_{ij}$. Similarly, let $\tilde D$ be the covariant derivative corresponding to the metric $\tilde h$ and let 
	
	\begin{equation}
	\tilde A_i\equiv\tilde D_i\phi
	\end{equation}
	
	\noindent
	and
	
	\begin{equation}
	\tilde A^i\equiv \tilde h^{ij}\tilde A_j.
	\end{equation}
	
	\noindent
	Clearly, the definition of $\tilde{A}_i$ does not depend on $\tilde{h}$ but the index of the vector field $\tilde{A}$ is raised and lowered with $\tilde{h}$, whereas the index of $A$ is raised and lowered with $h$. We continue to use a similar convention in the further discussion: indices of tensor fields denoted by letters with a tilde are raised and lowered with $\tilde{h}$, whereas indices of tensor fields without a tilde are raised and lowered with $h$.

	The extrinsic curvature is decomposed to its trace and conformal traceless parts
	
	\begin{equation}\label{defk}
	K_{ij}\equiv e^{4\Omega}\left(\tilde Q_{ij}+\frac13\tilde h_{ij}K\right),
	\end{equation}
	
	\noindent
	or alternatively,
	
	\begin{equation}
	\tilde  Q_{ij}\equiv e^{-4\Omega}\left(K_{ij}-\frac13 h_{ij}K\right).
	\end{equation}

\noindent
	The evolution equations for the variables $\tilde h_{ij}$ and $\Omega$ are
	
	\begin{equation}\label{ev_hh}
	\partial_0\tilde h_{ij}=-2N\tilde Q_{ij}+2\tilde h_{k(i}\mathring{D}_{j)}N^k-\frac23\tilde h_{ij}\mathring{D}_kN^k
	\end{equation}
	
	\begin{equation}\label{ev_omega}
	\partial_0\Omega=-\frac{N}{6}K+\frac16\mathring{D}_kN^k
	\end{equation}
	
	\noindent
	where $\partial_0$ is given by
	
	\begin{equation}
	\partial_0\equiv \partial_t-N^k\mathring{D}_k.
	\end{equation}
	
	\noindent
	To write the remaining equations in a more compact way, we use the conformal versions of the auxiliary variables introduced in (\ref{Phi}):

	\begin{align}
	\tilde{\Phi}\equiv\Phi={}&\frac{1}{N}\left(\partial_0A-e^{-4\Omega}\tilde{h}^{ij}\tilde{A}_i\tilde{D}_jN\right), \\
	\tilde{\Phi}_i\equiv{\Phi}_i={}&\tilde{D}_iA+\tilde{Q}_{ik}\tilde{A}_j\tilde{h}^{jk}+\tfrac13 K\tilde{A}_i,\\
	\tilde{\Phi}_{ij}\equiv\Phi_{ij}={}&\tilde{D}_i\tilde{A}_j-2\bigl(\tilde{A}_j\tilde{D}_i\Omega+\tilde{A}_i\tilde{D}_j\Omega- \tilde{h}_{ij}\tilde{h}^{kl}\tilde{A}_k\tilde{D}_l\Omega\bigr)+Ae^{4\Omega}\bigl(\tilde{Q}_{ij}+\tfrac13 K\tilde{h}_{ij}\bigr).
	\end{align}
	
	The evolution equation for $K$ is the trace of the tensor evolution equation, i.e. the same as (\ref{trace}), except that the variables $h_{ij}$ and $K_{ij}$ (and the covariant derivatives) are now replaced by the corresponding expressions (\ref{defh}-\ref{defk}).
	
	\begin{align}\label{ev_K}\nonumber
	&\partial_0K+ N\Phi \partial_XG_3 \bigl(\tfrac{3}{4}  A^2
 -  \tfrac{1}{4} \tilde{A}_{i} \tilde{A}_{j} \
\tilde{h}^{ij}e^{-4\Omega}\bigr)= -\tilde{h}^{ij}\tilde{D}_{i}\tilde{D}_{j}N- 2 \tilde{h}^{ij} \tilde{D}_{i}N\tilde{D}_{j}\Omega
 \nonumber\\
& +N\Bigl\{\tilde{Q}_{ik} \tilde{Q}_{jl}\tilde{h}^{ij}\tilde{h}^{kl}
+  \tfrac{1}{3}   K^2+\tfrac12 A^2 + \partial_{\phi}G_3  A^2+\tfrac{1}{2}  G_2+ \tfrac14\partial_XG_2 \bigl(  A^2+ \tilde{A}_{i} \tilde{A}_{j} \tilde{h}^{ij}e^{-4\Omega}\bigr)
 \nonumber\\
& + \partial_XG_3e^{-4\Omega} \Bigl( \tfrac{1}{4} A^2 \tilde{\Phi}_{ij} \
\tilde{h}^{ij}
 -  \tilde{A}_{i} A \tilde{\Phi}_{j} \tilde{h}^{ij}+ \tfrac14 \tilde{A}_{i} \tilde{A}_{j} \tilde{\Phi}_{kl} \
\tilde{h}^{ij} \tilde{h}^{kl} e^{-4 \Omega}\Bigr)\Bigr\}
\end{align}

	\noindent
	The equation describing the evolution of $\tilde Q_{ij}$ is obtained by taking the trace free part of (\ref{g3_ev})

	\begin{align}\label{ev_Q}\nonumber
		&\partial_0\tilde Q_{ij}-\frac{N}{2}e^{-4\Omega}\partial_X G_3\left(\tilde A_i\tilde A_j-\frac13\tilde h_{ij}\tilde A^k\tilde A_k\right)\Phi=Ne^{-4\Omega}\Bigl[ R_{ij}-\tfrac{1}{N}\tilde D_i\tilde D_j N+4\tilde D_{(i}\Omega\tilde D_{j)}\ln N \nonumber \\ 
		& -\frac12\bigl(1+ \partial_XG_2+2\partial_{\phi}G_3\bigr) \tilde{A}_{i} \tilde{A}_{j}  
 + \partial_XG_3 \Bigr(- A\tilde{A}_{(i}\tilde{\Phi}_{j)}  -  \tfrac12 \tilde{A}_{i} \tilde{A}_{j} \tilde{\Phi}_{kl}\tilde{h}^{kl}e^{-4 \Omega}  +  \tilde{A}_{k} \tilde{A}_{(i} \tilde{\Phi}_{j)l} \
\tilde{h}^{kl}e^{-4 \Omega}\Bigr)\Bigr]^\text{TF}\nonumber \\
		&+NK\tilde Q_{ij}-2N\tilde Q_{ik}\tilde Q^k_{~j}+2\tilde Q_{k(i}\mathring{D}_{j)}N^k-\frac23\tilde Q_{ij}\mathring{D}_kN^k.
		\end{align}
	
	\noindent
	where $\tilde T_{ij}^{TF}$ denotes the trace free part of a symmetric tensor $T_{ij}$,
	
	\begin{equation}
	\tilde T_{ij}^{TF}\equiv \tilde T_{ij}-\frac13 \tilde T_{kl}\tilde h^{kl} \tilde h_{ij},
	\end{equation}
	
	\noindent
	and the conformal decomposition of the spatial Ricci tensor is
	
	\begin{align}\nonumber
	R_{ij}=&~-\frac12 \tilde h^{kl}\mathring{D}_k\mathring{D}_l\tilde h_{ij}+\tilde h_{k(i}\mathring{D}_{j)}\tilde V^k-\frac12\tilde V^k \mathring{D}_k\tilde h_{ij}+\left(\tilde \Gamma^k_{il}-\mathring{ \Gamma}^k_{il}\right)\left(\tilde \Gamma^l_{kj}-\mathring{ \Gamma}^l_{kj}\right)\\ 
	&-\mathring{D}_k\tilde h_{l(i}\mathring{D}_{j)}\tilde h^{kl}-2\tilde D_i\tilde D_j\Omega-2\tilde h_{ij}\tilde D^k\tilde D_k\Omega+4\tilde D_i\Omega\tilde D_j\Omega-4\tilde h_{ij}\tilde D^k\Omega\tilde D_k\Omega.
	\end{align}

	Finally, the equation for $\tilde V^i$ is obtained by commuting $\partial_0$ with $\mathring{D}$ to get 
	
	\begin{equation}\label{ev_v0}
	\partial_0 \tilde V^i=-2N\mathring{D_j}\tilde Q^{ij}-2\tilde Q^{ij}\mathring{D}_jN-\tilde V^k\mathring{D}_kN^i+\frac23\tilde V^i\mathring{D}_kN^k+\tilde h^{kl}\mathring{D}_k\mathring{D}_lN^i+\frac13\tilde h^{ij}\mathring{D}_j\mathring{D}_kN^k
	\end{equation}
	
	\noindent
	and then adding the momentum constraint times $2m N$ to (\ref{ev_v0}) gives
	
	\begin{align}\label{ev_v}\nonumber
	\partial_0{\tilde V}^i=&~ -2\tilde Q^{ij}\mathring{D}_jN-\tilde V^k\mathring{D}_kN^i+\frac23\tilde V^i\mathring{D}_kN^k+\tilde h^{kl}\mathring{D}_k\mathring{D}_lN^i+\frac13\tilde h^{ij}\mathring{D}_j\mathring{D}_kN^k+N\Bigl\{2(m-1)\tilde D_k{\tilde Q}^{ki} \\ \nonumber
	&-\frac{4m}{3}\tilde D^iK+2m\tilde{h}^{ij}\Bigl[\frac12\bigl(1+   \partial_XG_2+2\partial_{\phi}G_3\bigr)A\tilde{A}_j
 + \frac12\partial_XG_3 \Bigl(-  A^2 \tilde{\Phi}_{j}
 +\tilde{A}_{l} A \tilde{\Phi}_{jk} \tilde{h}^{kl}e^{-4 \Omega}\nonumber\\
& -  \tilde{A}_{j} A \tilde{\Phi}_{kl}\tilde{h}^{kl}e^{-4 \Omega}
 +\tilde{A}_{k} \tilde{A}_{j} \tilde{\Phi}_{l} \
\tilde{h}^{kl} e^{-4 \Omega}\Bigr)\Bigr]\Bigr\}
	\end{align}
	
	\noindent
	where $m$ is arbitrary. We will show that for a range of values for the parameter $m$, the system is strongly hyperbolic. 
	Finally, once again we use $\tilde E_\phi=0$ as the scalar evolution equation (defined in (\ref{ev_sc})) which can also be rewritten in terms of the variables introduced here:
	
	\begin{equation}\label{ev_def_a}
	\partial_0\phi=NA
	\end{equation}
	
	\begin{align}\label{ev_sc_bssn}\nonumber
	\tilde E_\phi\equiv{} &\Phi \biggl(1
	+ \partial_XG_2 + A^2 \partial_{{X}}^2G_2
	+ 2 \partial_{\phi}G_3
	+  \partial_{X\phi}^2G_3 ( A^2
	+  \tilde{A}_{i} \tilde{A}_{j} \tilde{h}^{ij}e^{-4 \Omega})
	+ 2 \tilde{\Phi}_{ij} \partial_XG_3 \tilde{h}^{ij}e^{-4 \Omega}\nonumber\\
	& + \tfrac{1}{4} \bigl(\partial_XG_3\bigr)^2 \Bigl(3 A^4
	-  2 \tilde{A}_{i} \tilde{A}_{j} A^2 \tilde{h}^{ij}e^{-4 \Omega}
	-  \tilde{A}_{i} \tilde{A}_{j} \tilde{A}_{k} \tilde{A}_{l} \tilde{h}^{ik} \tilde{h}^{jl}e^{-8 \Omega}\Bigr)\nonumber\\
	& +  \partial_{{X}}^2G_3 \Bigl( A^2 \tilde{\Phi}_{ij} \tilde{h}^{ij}e^{-4 \Omega}
	-   \tilde{A}_{k} \tilde{A}_{l} \tilde{\Phi}_{ij} \tilde{h}^{ik} \tilde{h}^{jl}e^{-8 \Omega}\Bigr)\biggr)\nonumber\\
	&- \partial_{\phi}G_2
	-  \tilde{\Phi}_{ij} \tilde{h}^{ij}e^{-4 \Omega}
	- 2 \tilde{\Phi}_{ij} \partial_{\phi}G_3 \tilde{h}^{ij}e^{-4 \Omega}
	+  \partial_{X\phi}^2G_2 (A^2 -   \tilde{A}_{i} \tilde{A}_{j} \tilde{h}^{ij}e^{-4\Omega})\nonumber\\
	& +  \partial_{\phi}^2G_3 ( A^2 -   \tilde{A}_{i} \tilde{A}_{j} \tilde{h}^{ij}e^{-4\Omega})
	+ \tfrac{1}{2} G_2 \partial_XG_3 (A^2 -   \tilde{A}_{i} \tilde{A}_{j} \tilde{h}^{ij}e^{-4\Omega})\nonumber\\
	& +  \partial_{{X}}^2G_2 (- 2\tilde{A}_{i} A \tilde{\Phi}_{j} \tilde{h}^{ij}e^{-4 \Omega}
	+  \tilde{A}_{k} \tilde{A}_{l} \tilde{\Phi}_{ij} \tilde{h}^{ik} \tilde{h}^{jl}e^{-8 \Omega})
	\nonumber\\
	&  + \partial_XG_2 \Bigl(- \tilde{\Phi}_{ij} \tilde{h}^{ij}e^{-4 \Omega}+ \tfrac{1}{4} \partial_XG_3 \bigl(A^4
	-  2 \tilde{A}_{i} \tilde{A}_{j} A^2 \tilde{h}^{ij}e^{-4 \Omega}
	+ \tilde{A}_{i} \tilde{A}_{j} \tilde{A}_{k} \tilde{A}_{l}\tilde{h}^{ik} \tilde{h}^{jl}e^{-8 \Omega}\bigr)\Bigr)
	\nonumber\\
	& +  \partial_{X\phi}^2G_3 \Bigl(- 4 \tilde{A}_{i} A \tilde{\Phi}_{j} \tilde{h}^{ij}e^{-4 \Omega}
	+ A^2 \tilde{\Phi}_{ij} \tilde{h}^{ij}e^{-4 \Omega}
	+2 \tilde{A}_{k} \tilde{A}_{l} \tilde{\Phi}_{ij} \tilde{h}^{ik} \tilde{h}^{jl}e^{-8 \Omega}-  \tilde{A}_{i} \tilde{A}_{j} \tilde{\Phi}_{kl} \tilde{h}^{ij} \tilde{h}^{kl}e^{-8 \Omega}\Bigr)\nonumber\\
	&  + \partial_XG_3 \Bigl( \partial_{\phi}G_3 \bigl( A^4
	-  2 \tilde{A}_{i} \tilde{A}_{j} A^2 \tilde{h}^{ij}e^{-4\Omega}+  \tilde{A}_{i} \tilde{A}_{j} \tilde{A}_{k}\tilde{A}_{l} \tilde{h}^{ik} \tilde{h}^{jl}e^{-8 \Omega}\bigr)+ A^4
	-  2 \tilde{\Phi}_{i} \tilde{\Phi}_{j} \
	\tilde{h}^{ij}e^{-4 \Omega}
	\nonumber\\
	&  -  2 \tilde{A}_{i} \tilde{A}_{j} A^2 \tilde{h}^{ij}e^{-4 \Omega}
	+  \tilde{A}_{i} \tilde{A}_{j} \tilde{A}_{k} \
	\tilde{A}_{l} \tilde{h}^{ik} \tilde{h}^{jl}e^{-8 \Omega}
	-   \tilde{\Phi}_{ij} \tilde{\Phi}_{kl} \tilde{h}^{ij} \
	\tilde{h}^{kl}e^{-8 \Omega}
	+  \tilde{\Phi}_{ij} \tilde{\Phi}_{kl} \tilde{h}^{ik} \
	\tilde{h}^{jl}e^{-8 \Omega}\Bigr)\nonumber\\
	& +  \partial_{{X}}^2G_3 \Bigl(-  A^2 \tilde{\Phi}_{i} \tilde{\Phi}_{j} \tilde{h}^{ij}e^{-4 \Omega}
	-  2 \tilde{A}_{i} A \tilde{\Phi}_{jl} \tilde{\Phi}_{k} \
	\tilde{h}^{ik} \tilde{h}^{jl}e^{-8 \Omega}+ \tilde{A}_{i} \tilde{A}_{j} \tilde{\Phi}_{k} \
	\tilde{\Phi}_{l} \tilde{h}^{ik} \tilde{h}^{jl}e^{-8 \Omega}\nonumber\\
	& 
	+ 2 \tilde{A}_{k} A \tilde{\Phi}_{ij} \
	\tilde{\Phi}_{l} \tilde{h}^{ik} \tilde{h}^{jl}e^{-8 \Omega}-   \tilde{A}_{l} \tilde{A}_{m} \tilde{\Phi}_{in} \
	\tilde{\Phi}_{jk} \tilde{h}^{il} \tilde{h}^{jm} \tilde{h}^{kn}e^{-12 \Omega}
	+  \tilde{A}_{m} \tilde{A}_{n} \tilde{\Phi}_{ij} \
	\tilde{\Phi}_{kl} \tilde{h}^{im} \tilde{h}^{jn} \
	\tilde{h}^{kl}e^{-12 \Omega}\Bigr)\nonumber\\
	&+\tfrac14  \bigl(\partial_XG_3\bigr)^2 \Bigl(- 8 \tilde{A}_{i} A^3 \tilde{\Phi}_{j} \tilde{h}^{ij}e^{-4 \Omega}
	+ A^4 \tilde{\Phi}_{ij} \tilde{h}^{ij}e^{-4 \Omega}
	+ 4 \tilde{A}_{k} \tilde{A}_{l} A^2 \tilde{\Phi}_{ij} \
	\tilde{h}^{ik} \tilde{h}^{jl}e^{-8 \Omega}\nonumber\\
	& 
	{}-  2 \tilde{A}_{i} \tilde{A}_{k} A^2 \tilde{\Phi}_{jl} \tilde{h}^{ik} \tilde{h}^{jl}e^{-8 \Omega} +
	\tilde{A}_{i} \tilde{A}_{j} \tilde{A}_{k} \tilde{A}_{l} \
	\tilde{\Phi}_{mn} \tilde{h}^{ik} \tilde{h}^{jl} \
	\tilde{h}^{mn}e^{-12 \Omega}\nonumber\\
	& {}-  4 \tilde{A}_{i} \tilde{A}_{k} \tilde{A}_{l} \
	\tilde{h}^{ik} \tilde{h}^{jl} \bigl(-2 A \tilde{\Phi}_{j} + \
	\tilde{A}_{m} \tilde{\Phi}_{jn} \tilde{h}^{mn}e^{-4 \Omega}\bigr)e^{-8 \Omega}\Bigr)=0.
	\end{align}

	Equations (\ref{ev_hh}), (\ref{ev_omega}), (\ref{ev_K}), (\ref{ev_Q}), (\ref{ev_v}) and (\ref{ev_sc_bssn}) must be complemented with the evolution equations for the lapse function and the shift vector. This can be done by choosing an appropriate slicing condition and a spatial coordinate condition. A popular choice for the slicing condition is harmonic slicing and its generalizations. Harmonic slicing means that the harmonic coordinate condition is imposed only on the time coordinate
	
	$$\Box_g t=0.$$
	
	\noindent
	Writing this out in terms of the ADM variables gives an evolution equation for the lapse function:
	
	\begin{equation}
	\partial_0N=-N^2 K.
	\end{equation}
	
	\noindent
	Sometimes it is more convenient to consider a generalization of this condition, called the Bona-Mass\'o slicing condition
	
	\begin{equation}
	\partial_0N=-N^2F(N)K
	\end{equation}
	
	\noindent
	for a suitable function $F$. The choice $F(N)=\frac{2}{N}$ called the 1+log slicing is the most widely used in numerical applications. We will generalize this condition even further for cubic Horndeski theories
	
	\begin{equation}\label{ev_N}
	\partial_0N=-2N^2 F(t,x,N,\phi, A,\tilde{A}_k,K).
	\end{equation}

	To simplify the discussion of the linearized equations, we introduce
	
	\begin{equation}
	\sigma\equiv \frac{\partial F}{\partial K}; ~~~\tilde \rho\equiv \frac{1}{\partial_XG_3}\frac{\partial F}{\partial A}; ~~~\tilde{\rho}^k\equiv \frac{1}{\partial_XG_3} \frac{\partial F}{\partial \tilde{A}_k}.
	\end{equation}
	
	There are a number of ways to impose a dynamical gauge condition on the shift vector in general relativity. However, it has also been demonstrated that it is possible to write Einstein’s equations in a strongly (even symmetric) hyperbolic form by choosing an arbitrary (but a priori fixed) shift vector \cite{beyer}. In these latter formulations, the shift vector is only a source term. In this section we take this latter approach and show that the equations of motion of the subclass of Horndeski theories under consideration can be written in a strongly hyperbolic form with arbitrarily fixed shift vector.

	\subsection{Proof of strong hyperbolicity}\label{sec:bssn_proof}
	
	In this section, we are going to show that the BSSN equations of motion (consisting of equations (\ref{ev_hh}), (\ref{ev_omega}), (\ref{ev_K}), (\ref{ev_Q}), (\ref{ev_v}), (\ref{ev_N}), (\ref{ev_def_a}), (\ref{ev_sc_bssn})) is a strongly hyperbolic system for the dynamical variables. Our strategy will be as follows. First, we linearize the equations and select the highest derivative (principal) terms in the equations. (The list of the variables and their linearized versions is summarized in Table \ref{tab:not}.) These are the terms that are at the first derivative level in equations (\ref{ev_hh}), (\ref{ev_omega}), (\ref{ev_N}), (\ref{ev_def_a}) and the ones at the second derivative level in equations (\ref{ev_K}), (\ref{ev_Q}), (\ref{ev_v}), (\ref{ev_sc_bssn})). The second step is to convert the principal terms in the equations to the eigenvalue problem of the principal symbol. This involves switching to Fourier variables: 
	
	$$(\partial_0,\partial_k)\to(i\xi_0,i\xi_k)$$
	
	$$(\alpha,\omega,\kappa,{\tilde\gamma}_{ij},{q}_{ij},{\tilde v}_i,\psi,a)\to(\hat\alpha,\hat\omega,\hat\kappa,\hat{\tilde\gamma}_{ij},\hat{q}_{ij},\hat{\tilde v}_i,\hat\psi,\hat a).$$
	
	Note that in this section our choice of basis is slightly different from the one used in Section \ref{sec:ell_hyp}: here we use $\partial_0\equiv \partial_t-N^k\partial_k$, rather than $\partial_0\equiv\partial_t$.  Clearly, this only amounts to a shift in the variable $\xi_0\to\xi_0-N^k\xi_k$. Nevertheless, from now on, we will denote the eigenvalues of the principal symbol by $\xi_0$ in this basis. This includes the solutions $\xi_0^{\pm,\phi}$ to the scalar characteristic equation (\ref{char}).
	
	Next, we solve the eigenvalue problem, by determining the eigenvalues $\xi_0$ and the corresponding eigenvectors explicitly. Finally, we show that the conditions of strong hyperbolicity are met for an appropriate choice of the parameters $m$ and $\sigma$: the principal symbol has real eigenvalues and a complete set of eigenvectors with smooth dependence on $\xi_k$.

	 \begin{table}[htbp]
		\centering
		\begin{tabular}{ccc}
			\hline
			Quantity/Definition & Notation  & Linearized version \\
			\hline
			Conformal factor&$\Omega$& $\omega$ \\
			\\
			Conformal metric & $\tilde h_{ij}$ & $\tilde\gamma_{ij}$ \\
			\\
			Lapse function & $N$ & $\alpha$ \\
			\\
			Scalar field & $\phi$ & $\psi$  \\
			\\
			$A\equiv\mathcal{L}_n\phi$ & $A$ & $a$   \\
			\\
			Trace of the extrinsic curvature & $K$ & $\kappa$  \\
			\\
			Conformal traceless extrinsic curvature & $\tilde Q_{ij}$ & $\tilde q_{ij}$ \\
			\\
			$\tilde V^i\equiv-\mathring{ D} _j\tilde h^{ij}$ & $\tilde V^i$ & $\tilde v^i$\\
			\hline
		\end{tabular}
		\caption{The list of variables used in the BSSN formulation}
		\label{tab:not}
	\end{table}

	  Since the linearization is straightforward, we simply just state the eigenvalue problem for the $20$ variables $(\hat\alpha,\hat\omega,\hat\kappa,\hat{\tilde\gamma}_{ij},\hat{q}_{ij},\hat{\tilde v}_i,\hat\psi,\hat a)$:
	
	\begin{subequations}\label{eig_bssn}
		
		\begin{equation}
		i\xi_0 \hat\alpha=-2N^2\left(\sigma \hat\kappa +\partial_XG_3\tilde \rho\hat a+\partial_XG_3\tilde \rho^k~ i\xi_k\hat \psi\right)
		\end{equation}
		
		\begin{equation}
		i\xi_0 \hat \kappa-\frac14\partial_XG_3(3A^2-A_kA^k)i\xi_0\hat a=Ne^{-4\Omega} \left\{\frac{1}{N}|\xi|_{\tilde h}^2\hat \alpha-\frac14\partial_XG_3(A^2+A^kA_k)|\xi|_{\tilde h}^2\hat\psi-\partial_XG_3A\tilde A^k ~i\xi_k\hat a\right\}
		\end{equation}
		
		\begin{equation}
		i\xi_0 \hat\omega=-\frac{N}{6}\hat\kappa
		\end{equation}
		
		\begin{equation}\label{eig_def}
		i\xi_0\hat\psi=N\hat a
		\end{equation}
		
		\begin{equation}
		i\xi_0\hat{\tilde \gamma}_{ij}=-2N\hat{\tilde q}_{ij}
		\end{equation}

		\begin{align}\nonumber
		&i\xi_0\hat{\tilde q}_{ij}-\frac12e^{-4\Omega}\partial_X G_3\left(\tilde A_i\tilde A_j-\frac13\tilde h_{ij}\tilde A^k\tilde A_k\right)i\xi_0\hat a=Ne^{-4\Omega}\left\{\frac12|\xi|_{\tilde h}^2\hat{\tilde \gamma}_{ij}+i\xi_{(i}\hat{\tilde v}_{j)}-\frac13i\xi_k\hat{\tilde v}^k\tilde h_{ij}\right. \\ \nonumber
		&\qquad \left. {} +2\xi_i\xi_j\hat\omega-\frac23 |\xi|_{\tilde h}^2\hat\omega \tilde h_{ij}+\frac{1}{N}\xi_i\xi_j\hat \alpha-\frac13\frac{1}{N} |\xi|^2_{\tilde h}\hat\alpha \tilde h_{ij}-A\partial_XG_3\hat a\left(\tilde A_{(i}~i\xi_{j)}-\frac13\tilde A^k~i\xi_k\tilde h_{ij}\right)\right. \\ 
		&\qquad \left. {} +e^{-4\Omega}\partial_X G_3\hat\psi\left(\frac12(\tilde A_i\tilde A_j-\frac13\tilde h_{ij}\tilde A^k\tilde A_k)|\xi|^2_{\tilde h}-\tilde A^k \tilde A_{(i} \xi_{j)}\xi_k+\frac13\tilde A^k\tilde A^l \xi_k\xi_l\tilde h_{ij}\right)\right\}
		\end{align}
		
		\begin{align}\nonumber
		i\xi_0\hat{\tilde v}^i= &~N\left(2(m-1)i\xi_k\hat{\tilde q}^{ki}-\frac{4m}{3}i\xi^i\hat\kappa+m\partial_XG_3A^2~i\xi^i\hat a \right.\\
		&\qquad \left. {} +e^{-4\Omega}m\partial_X G_3\left(A\tilde A^k\xi_k\xi^i\hat\psi-A\tilde A^i|\xi|_{\tilde h}^2\hat\psi-\tilde A^i\tilde A^k~i\xi_k\hat a\right) \right).
		\end{align}
		
		\begin{equation}\label{eig_sc}
		i\xi_0\mathcal{A}\hat a=i\mathcal{B}\hat a+\mathcal{C}\hat\psi
		\end{equation}
		
	\end{subequations}
	
	\noindent
	where the coefficients $\mathcal{A}$, $\mathcal{B}$ and $\mathcal{C}$ depend on the background fields and $\xi_i$ and their explicit form is not important for our purposes (although it is straightforward to obtain them by linearizing (\ref{ev_sc_bssn})). Substituting (\ref{ev_psi}) into (\ref{ev_a}) gives 
	
	\begin{equation}
	\hat\psi\left(P^\prime_{\phi\phi}\right)^{\mu\nu}\xi_\mu\xi_\nu=0,
	\end{equation}
	
	\noindent
	that is to say, equation (\ref{char}).
	In the discussion of the eigenvalues and eigenvectors it is simpler to use the variable 
	
	\begin{equation}\label{xi_0_shift}
	\bar\xi_0\equiv \frac{\xi_0}{Ne^{-2\Omega}|\xi|_{\tilde h}}=\frac{\xi_0}{N|\xi|_h}
	\end{equation}
	
	\noindent
	instead of $\xi_0$.
	
	We can identify a subset of eigenvalues and eigenvectors of (\ref{eig_bssn}) by setting $\hat\psi=\hat a=0$. These $18$ eigenvalues and eigenvectors are the same as in the formulation described in section IV of \cite{beyer} for vacuum GR:
	
	\begin{itemize}
		\item[I.] Transverse-traceless (physical) modes with null characteristics $\bar\xi_0=\pm 1$. Since there are two linearly independent transverse-traceless symmetric tensors $\hat{\tilde\gamma}_{ij}^{TT}$, the eigenvectors span a $4$-dimensional space.
		\item[II.] Transverse modes with null characteristics, spanning a $2$-dimensional space.
		\item[III.] A $4$-dimensional space of modes with $\hat{\tilde \gamma}_{ij}=\xi_{(i}e_{j)}$ for any $e_i$ orthogonal to $\xi_i$ w.r.t. $\tilde h_{ij}$ and $\bar\xi_0=\pm\sqrt{m}$.
		\item[IV.] Zero speed modes ($\bar\xi_0=0$) with $\hat{\tilde \gamma}_{ij}=-\frac{2}{|\xi|_{\tilde h}^2}\left(i\xi_{(i}\hat{\tilde v}_{j)}+2\xi_i\xi_j\hat\omega\right)^\text{TF}$ where $\hat\omega$ and $\hat{\tilde v}_i$ are arbitrary, spanning a space with dimension $4$.
		\item[V.] A two-dimensional space of modes with $\bar\xi_0=\pm\sqrt{2\sigma}$.
		\item[VI.] A two-dimensional space of modes with $\bar\xi_0=\pm\sqrt{\frac{4m-1}{3}}$.
	\end{itemize}
	
	The expressions for the eigenvectors are listed in Table \ref{tab:eig}. These eigenvalues are real and the eigenvectors are smooth functions of their arguments when $\sigma>0$ and $m>\frac14$. 
	
	\begin{table}[htbp]
		\centering
		\begin{tabular}{ccccccc}
			\vspace{6pt}
			&$\hat\alpha$ &$\hat\omega$& $\hat\kappa$  & $\hat{\tilde \gamma}_{ij}$ &  $\hat{\tilde v}_i$ & $\bar \xi_0$ \\
			\hline
			\hline
			\\
			I. &$0$ & $0$ & $0$ & $\tilde\gamma_{ij}^{TT}$ &  $0$ & $\pm 1$    \\
			\\
			II. &$0$ & $0$ & $0$ & $\tilde h_{ij}-\frac{\xi_i\xi_j}{|\xi|^2_{\tilde h}}$ &  $0$ &  $\pm 1$  \\
			\\
			III. & $0$ & $0$ & $0$ & $\xi_{(i}e_{j)}$ &  $-i\frac{m-1}{2}|\xi|_{\tilde h}^2e_i$ &  $\pm \sqrt{m} $  \\
			\\
			IV. & $0$ & $\hat\omega$ & $0$ & $-\frac{2}{|\xi|_{\tilde h}^2}\left(i\xi_{(i}\hat{\tilde v}_{j)}+2\xi_i\xi_j\hat\omega\right)^\text{TF}$ &  $\hat{\tilde v}_i$ &  $0$   \\
			\\
			V. & $N\sigma$ & $\frac{1}{12}$ & $-\frac{i\xi_0}{2N}$ & $\left(\frac{\xi_i\xi_j}{|\xi|_{\tilde h}^2}\right)^\text{TF}$ & $i\frac23\xi_i$ &  $\pm\sqrt{2\sigma}$   \\
			\\
			VI. & $0$ & $0$ & $0$ & $\left(\frac{\xi_i\xi_j}{|\xi|_{\tilde h}^2}\right)^\text{TF}$ &  $2i(m-1)\xi_i$ &  $\pm\sqrt{\frac{4m-1}{3}}$  \\
			\\
			\hline
		\end{tabular}
		\caption{The list of eigenvalues and eigenvectors of the principal symbol with $\hat\psi=\hat a=0$.}
		\label{tab:eig}
	\end{table}
	
	For strong hyperbolicity, we need $20$ linearly independent eigenvectors with real eigenvalues and smooth dependence on $\xi_i$. Since we have already found $18$, this amounts to finding two additional eigenvalues and eigenvectors with non-trivial $\hat\psi$ and $\hat a$. The eigenvalues corresponding to these eigenvectors are found by solving the system consisting of equations (\ref{eig_def}) and (\ref{eig_sc}), or equivalently, (\ref{char}). As mentioned before, this equation has two distinct real solutions $\xi_0^{\phi,\pm}$ in a weakly curved background (see Section \ref{sec:setup}). Therefore, for a weakly curved background, there must indeed be 2 additional eigenvectors. The only thing that needs to be shown is the condition on smooth dependence.
	
	The easiest way to obtain the corresponding eigenvectors is to derive a closed equation containing $\hat \psi$, $\hat a^\phi\equiv\tfrac{1}{N}i\xi_0^{\phi,\pm}\hat\psi$ and the Fourier transform of the linearized extrinsic curvature
	
	\begin{equation}\label{decomp}
	\hat\kappa_{ij}=e^{4\Omega}\hat{\tilde q}_{ij}+\frac13\hat\kappa h_{ij}.
	\end{equation}
	
	\noindent
	Equations (\ref{eig_bssn}) imply
	
	\begin{align}\label{closed}\nonumber
	&|\xi^{\phi,\pm}|_g^2\left(\hat\kappa_{ij}-\frac12\partial_X G_3(Xh_{ij}+A_i A_j)\hat a^{\phi}+i\partial_XG_3AA_{(i}\xi_{j)}\hat\psi\right)=(2m-2\sigma-1)\xi_i\xi_j\hat\kappa\\ \nonumber
	&-2(m-1)\xi_{(i}\hat\kappa_{j)\xi}-\frac23(m-1)(|\xi|_h^2\hat\kappa-\hat\kappa_{\xi\xi})h_{ij}+\partial_XG_3\Bigl\{(m-1)A_{(i}\xi_{j)}\left(A^k\xi_k\hat a^{\phi}-iA|\xi|_h^2\hat\psi\right) \\
	& -\xi_i\xi_j\left((mA^2+2\rho)\hat a^\phi-(mAA^k-2\rho^k)i\xi_k\hat\psi\right)+\frac13(m-1)\hat a^\phi \left(A^2|\xi|_h^2-(A^k\xi_k)^2\right)h_{ij} \Bigr\}
	\end{align}
	
	\noindent
	with $\hat\kappa_{i\xi}\equiv \hat\kappa_{ij}\xi^j$ and 

$$|\xi^{\phi,\pm}|_g^2=-(\xi_0^{\phi,\pm})^2+N^2|\xi|_h^2.$$

\noindent
Note that in (\ref{closed}) we raise and lower indices with $h$.
	
	It follows from (\ref{decomp}) and (\ref{eig_bssn}) that if the solutions $\hat\kappa_{ij}^{\phi,\pm}$ of (\ref{closed}) are smooth functions of $\xi_i$, then the same is true for the auxiliary variables $\hat{\tilde q}_{ij}$, $\hat \kappa$, $\hat{\tilde \gamma}_{ij}$, $\hat{\tilde v}_i$, $\hat\omega$ and $\hat\alpha$. Based on the tensorial structure of (\ref{closed}), we look for an eigenvector of the form
	
	\begin{equation}\label{kappa_ansatz}
	\hat\kappa_{ij}^\phi=c_1\xi_i\xi_j+c_2|\xi|_h^2h_{ij}+c_3A_iA_j+2c_4A_{(i}\xi_{j)}.
	\end{equation}
	
	\noindent
	Note that
	
	\begin{equation}
	\hat\kappa_{i\xi}^\phi=\left((c_1+c_2)|\xi|_h^2+c_4(A^k\xi_k)\right)\xi_i+\left(c_3(A^k\xi_k)+c_4|\xi|_h^2\right)A_i
	\end{equation}
	
	\noindent
	and
	
	\begin{equation}
	\hat\kappa^\phi=(c_1+3c_2)|\xi|_h^2+c_3A^kA_k+2c_4A^k\xi_k.
	\end{equation}
	
	\noindent
	Plugging this into (\ref{closed}) we get the following system of linear equations for the coefficients
	
	\begin{align}\nonumber
	c_1|\xi|_g^2=&~(2m-2\sigma-1)\left((c_1+3c_2)|\xi|_h^2+c_3A^kA_k+2c_4A^k\xi_k\right)-2(m-1)\left((c_1+c_2)|\xi|_h^2+c_4(A^k\xi_k)\right)\\
	&-\partial_XG_3\left((mA^2+2\rho)\hat a^\phi-(mAA^k-2\rho^k)i\xi_k\hat\psi\right)
	\end{align}
	
	\begin{align}\nonumber
	|\xi|_g^2\left(c_2|\xi|_h^2-\frac12\partial_XG_3X\hat a\right)=&~-\frac23(m-1)\left(2c_2|\xi|_h^4+c_3(A^kA_k|\xi|_h^2-(A^k\xi_k)^2)\right)\\
	&+\frac13(m-1)\hat a^\phi \partial_XG_3\left(A^2|\xi|_h^2-(A^k\xi_k)^2\right)
	\end{align}
	
	\begin{equation}
	|\xi|_g^2\left(c_3-\frac12\partial_XG_3\hat a^\phi\right)=0
	\end{equation}
	
	\begin{equation}
	|\xi|_g^2\left(2c_4+i\partial_XG_3A\hat \psi\right)=-2(m-1)\left(c_3(A^k\xi_k)+c_4|\xi|_h^2\right)+\partial_XG_3(m-1)\left(A^k\xi_k\hat a^\phi-iA|\xi|_h^2\hat\psi\right)
	\end{equation}
	
	\noindent
	The solution is
	
	\begin{equation}\label{eq_c3}
	c_3=\frac12\partial_XG_3\hat a^\phi
	\end{equation}
	
	\begin{equation}\label{eq_c4}
	c_4=-i\frac12\partial_XG_3A\hat\psi
	\end{equation}
	
	\begin{equation}\label{eq_c2}
	c_2=\frac12\frac{1}{|\xi|_h^2}\partial_XG_3 X\hat a^\phi
	\end{equation}
	
	\begin{equation}\label{eq_c1}
	c_1=\frac{1}{-(\xi_0^{\phi,\pm})^2+2\sigma N^2|\xi|_h^2}\partial_XG_3\left[\left(-\frac{6\sigma+1}{4}A^2+\frac{2\sigma-1}{4}A^kA_k-2\tilde \rho\right)\hat a^\phi+\left(2\sigma AA^k-2\tilde\rho^k\right)i\xi_k\hat\psi\right]
	\end{equation}
	
	\noindent
	In order for $c_1$ to be smooth for any $\xi_i\in\mathbb{S}^2$, $-(\xi_0^{\phi,\pm})^2+2\sigma N^2 |\xi|_h^2$ must be non-zero for any $\xi_i$. This can be achieved by choosing $\sigma$ to be large enough. To see why this is true, we first note that the zeros of the function
	
	\begin{equation}
	\mathcal{F}_\sigma(\xi_0;\xi_i)=-(\xi_0)^2+2\sigma N^2 |\xi|_h^2
	\end{equation}
	
	\noindent
	define a cone for any $\sigma$, $\xi_0$. In the weak field regime, the null cone of $P^\prime_{\phi\phi}$ is close to the null cone of the spacetime metric $g$ for any $\xi_i\in\mathbb{S}^2$, they might even intersect for special values of $\xi_i$. Recall that
	
	\begin{equation}
	g^{\mu\nu}\xi_\mu\xi_\nu=-(\xi_0)^2+N^2|\xi|_h^2=\mathcal{F}_{\sigma=1/2}(\xi_0;\xi_i).
	\end{equation}

	 \noindent
	 Since $P^\prime_{\phi\phi}(\xi_0;\xi_i)$ and $\mathcal{F}_{\sigma=1/2}(\xi_0;\xi_i)$ may intersect for some $\xi_i$, and $\xi^{\phi,\pm}_0$ are the solutions of $P^\prime_{\phi\phi}(\xi_0;\xi_i)=0$,
	 this means that the expression on the RHS of (\ref{eq_c1}) could blow up for some $\xi_i$ if $\sigma$ is close to (or equal to) $\frac12$.  To avoid this, we can just choose $\sigma$ to be large enough (i.e. larger than $\frac12$), so that the cones given by $\mathcal{F}_{\sigma}(\xi_0;\xi_i)$ lie entirely “inside” the null cones of $g$ and $P’_{\phi\phi}$. In other words, for an appropriate $\sigma>\tfrac12$, $\mathcal{F}_{\sigma}(\xi_0^{\phi,\pm};\xi_i)$ vanishes for no choice of $\xi_i$. Choosing larger values of $\sigma$ makes it possible to deal with stronger background fields.
	
	To summarize, we have shown that the equations of motion for cubic Horndeski theories form a strongly hyperbolic system in a version of the BSSN formulation. The system was obtained using a generalization of the harmonic slicing condition and an arbitrary but (non-dynamically) fixed shift vector. The system is strongly hyperbolic for any $m>\frac14$ and for suitable $\sigma\equiv \partial_KF>\frac12$, regardless of how the source function $F$ depends on the scalar field and its derivatives. (For weakly curved backgrounds, choosing a large enough constant, e.g. $\sigma=1$ is enough.) This means that the original harmonic slicing $\sigma=\frac12$ does not work for cubic Horndeski theories.\footnote{One might wonder if $c_1$ could be set to $0$ in (\ref{eq_c1}) by an appropriate choice of $\rho$ and $\rho_k$, even for $\sigma=\frac12$. However, this particular choice of $\rho$ and $\rho_k$ fails to satisfy the integrability condition $\partial_A\rho^k=\partial_{A_k}\rho$, a similar issue to the one encountered in \cite{gp_hsr} (see equations (237-241) and the corresponding discussion).} On the other hand, the so-called 1+log slicing often used in numerical general relativity, corresponds to the choice $\sigma=\tfrac{1}{N}$ and hence remains a good slicing condition as long as $N<2$.
	Note that for GR the condition for strong hyperbolicity is $m>\frac14$ and $\sigma>0$, whereas the condition for symmetric hyperbolicity is $6\sigma=4m-1>0$. 
	
	\subsection{Propagation of constraints}
	
	To show that the solutions of the BSSN system are also solutions of the original Horndeski equations of motion, we derive a system of evolution equations for the Hamiltonian constraint, the momentum constraint and the variable
	
	\begin{equation}
	\mathbf{\tilde W}^k\equiv{\tilde V}^k+\mathring{D}_l\tilde h^{kl};
	\end{equation}
	
	\noindent
	and show that the system of equations is strongly hyperbolic. By uniqueness of the solutions to strongly hyperbolic systems, it follows that if the constraints are satisfied initially then they continue to hold throughout the evolution.
	
	Starting from equations (\ref{prop_ham}) and (\ref{prop_mom}) and setting
	
	\begin{equation}
	\mathcal{E}_{ij}\to\mathcal{E}_{ij}-N\left(\tilde h_{k(i}\partial_{j)}\mathbf{\tilde {W}}^k\right)^{TF}+\frac23 N \mathbf{H}~{h}_{ij},
	\end{equation}
	
	\noindent
	the constraint evolution equations become
	
	\begin{subequations}\label{bssn_constr}
	\begin{align}\nonumber
	\left(\partial_t-\mathcal{L}_N\right)\mathbf{H}=&~\frac23NK\mathbf{H}+\frac{1}{N}D^i\left(N^2\mathbf{M}_i\right)-Ne^{-4\Omega}\tilde Q^{ij}\tilde h_{k(i}\partial_{j)}{ \mathbf{\tilde W}}^k\\
	&-N\partial_X G_3 A\left[\left(-\left(\tilde h_{k(i}\partial_{j)}{ \mathbf{\tilde W}}^k\right)^{TF}+\frac23\mathbf{H}h_{ij}\right)A^iA^j+4\mathbf{H}A^2-2A\mathbf{M}_iA^i\right]
	\end{align}
	
	\begin{align}\nonumber
	\left(\partial_t-\mathcal{L}_N\right)\mathbf{M}_i=&NK\mathbf{M}_i-\frac{N^3}{3}D^i\left(N^{-2}\mathbf{H}\right)-D^j\left[N\left(\tilde h_{k(i}\partial_{j)}{ \mathbf{\tilde W}}^k\right)^{TF}\right]\\
	&-N\partial_X G_3 A_i\left[\left(-\left(\tilde h_{k(i}\partial_{j)}{ \mathbf{\tilde W}}^k\right)^{TF}+\frac23\mathbf{H}h_{ij}\right)A^iA^j+4\mathbf{H}A^2-2A\mathbf{M}_iA^i\right]
	\end{align}
	
	\begin{equation}
	\left(\partial_t-\mathcal{L}_N\right)\mathbf{\tilde W}^i=-2Nm\tilde h^{ij}\mathbf{M}_j.
	\end{equation}
	
	\end{subequations}

\noindent
We remind the reader that indices of tensors denoted with letters with tilde are raised and lowered with the conformal metric $\tilde{h}$, indices of tensors without a tilde are raised and lowered with the original induced metric $h$.

	Note that we have two additional constraints: the tracelessness of $\tilde{Q}_{ij}$ and $\det\mathring{h}=\det\tilde{h}$. Introducing the constraint variables

\begin{equation}
\mathbf{T}\equiv \tilde{h}^{ij}\tilde{Q}_{ij}
\end{equation}

and

\begin{equation}
\mathbf{D}\equiv \ln \frac{\det\tilde{h}}{\det\mathring{h}},
\end{equation}

it follows easily that

\begin{equation}\label{trace_constr}
\partial_0\mathbf{T}=-\frac12 e^{-4\Omega}\tilde{h}^{ij}\mathring{D}_i\mathring{D}_j\mathbf{D}
\end{equation}

\begin{equation}\label{det_constr}
\partial_0\mathbf{D}=-2\mathbf{T}.
\end{equation}

Substituting (\ref{trace_constr}) into the time derivative of (\ref{det_constr}) implies a wave equation for $\mathbf{D}$, decoupled from the rest of the constraint propagation system (\ref{bssn_constr}):

\begin{equation}
\partial_0^2\mathbf{D}=e^{-4\Omega}\tilde{h}^{ij}\mathring{D}_i\mathring{D}_j\mathbf{D}.
\end{equation}

Therefore, it is clear that starting from initial data that satisfies $\mathbf{D}=0$ and $\mathbf{T}=0$, these conditions will continue to hold throughout the evolution. For this reason, now we only need to deal with the system (\ref{bssn_constr}).
	
	The eigenvalue problem of the principal symbol of (\ref{bssn_constr}) can be written as
	
	\begin{subequations}
	\begin{equation}
	i\xi_0\mathbf{\hat{H}}= iN\xi^i\mathbf{\hat{M}}_i
	\end{equation}
	
	\begin{equation}
	i\xi_0\mathbf{\hat{M}}_i= -\frac13Ni\xi_i\mathbf{H}+\frac12N|\xi|_{\tilde h}^2\mathbf{\hat{\tilde W}}_i+\frac16 N\xi_i\xi_j\mathbf{\hat{\tilde W}}^j
	\end{equation}
	
	\begin{equation}
	i\xi_0\mathbf{\hat{\tilde W}}_i= -2Nm\mathbf{\hat{M}}_i.
	\end{equation}
	\end{subequations}
	
	\noindent
	The eigenvalues and eigenvectors of the principal symbol are as follows

	\begin{equation}
	\xi_0=0;~~\left(\begin{tabular}{c}
	\vspace{6pt}
	$\mathbf{\hat{H}}$\\
	\vspace{6pt}
	$\mathbf{\hat{\tilde W}}_i$\\
	
	$\mathbf{\hat{M}}_i$
	\end{tabular}\right)=\left(\begin{tabular}{c}
	\vspace{6pt}
	$-2iN|\xi|_h^2$\\
	\vspace{6pt}
	$N\xi_i$\\
	
	$0$
	\end{tabular}\right),
	\end{equation}

	\begin{equation}
	\xi_0=\pm\sqrt{\frac{4m-1}{3}}N|\xi|_h;~~\left(\begin{tabular}{c}
	\vspace{6pt}
	$\mathbf{\hat{H}}$\\
	\vspace{6pt}
	$\mathbf{\hat{\tilde W}}_i$\\
	
	$\mathbf{\hat{M}}_i$
	\end{tabular}\right)=\left(\begin{tabular}{c}
	\vspace{6pt}
	$-iN|\xi|_h^2$\\
	\vspace{6pt}
	$2Nm\xi_i$\\
	
	$-i\xi_0\xi_i$
	\end{tabular}\right)
	\end{equation}
	
	\noindent
	and finally,

	\begin{equation}
	\xi_0=\pm\sqrt{m}N|\xi|_h;~~\left(\begin{tabular}{c}
	\vspace{6pt}
	$\mathbf{\hat{H}}$\\
	\vspace{6pt}
	$\mathbf{\hat{\tilde W}}_i$\\
	
	$\mathbf{\hat{M}}_i$
	\end{tabular}\right)=\left(\begin{tabular}{c}
	\vspace{6pt}
	$0$\\
	\vspace{6pt}
	$2Nme_i$\\
	
	$-i\xi_0 e_i$
	\end{tabular}\right)
	\end{equation}
	
	\noindent
	for any vector $e_i$ orthogonal to $\xi_i$ with respect to $h_{ij}$. Therefore, the principal symbol of the system describing the evolution of the constraints possesses a complete set of smooth eigenvectors with real eigenvalues, provided that $m>\frac14$. Hence, the system (\ref{bssn_constr}) is strongly hyperbolic.

	\section{CCZ4-type formulation}\label{sec:ccz4}

	\subsection{Constraints}

	In this section we discuss how the so-called covariant conformal Z4 (CCZ4) \cite{ccz4,ccz4_2} formulation extends to the class of theories under consideration. 
	
	The idea behind the CCZ4 formulation is to introduce a 4-vector field $\mathcal{Z}^a$ that measures the deviations from the actual tensor equations of motion $E_{ab}$ (see (\ref{eom_g3})) and add terms containing $\mathcal{Z}^a$ and its first derivatives to the equations so that $\mathcal{Z}_a=0$ is an attractor of the modified equations. The modification is carried out in the following way:
	
	\begin{equation}\label{mod_ccz4}
	E_{ab}=0\to E_{ab}+\nabla_a \mathcal{Z}_b+\nabla_b \mathcal{Z}_a-g_{ab}\nabla^c\mathcal{Z}_c-k_1\left(n_a\mathcal{Z}_b+n_b\mathcal{Z}_a+k_2 n^c\mathcal{Z}_c g_{ab}\right)=0,
	\end{equation}
	
	\noindent
	or in the trace reversed version
	
	\begin{equation}\label{mod_ccz4_tr}
	E_{ab}-\frac12 E g_{ab}=0\to E_{ab}-\frac12 E g_{ab}+\nabla_a \mathcal{Z}_b+\nabla_b \mathcal{Z}_a-k_1\left(n_a\mathcal{Z}_b+n_b\mathcal{Z}_a-(1+k_2) n^c\mathcal{Z}_c g_{ab}\right)=0.
	\end{equation}
	
	The parameters $k_1$ and $k_2$ here are real constants. Splitting the four vector $\mathcal{Z}^a$ as $\mathcal{Z}^a\equiv Z^a+n^a\Theta$ with $Z^a\equiv h^a_b \mathcal{Z}^b$ and $\Theta\equiv -n^a\mathcal{Z}_a$, one can write the normal-normal and normal-spatial projections of (\ref{mod_ccz4}) as
	
	\begin{equation}\label{prop_theta}
	\left(\partial_t-\mathcal{L}_N\right)\Theta= N\mathbf{H}-N \Theta K + ND_kZ^k- D_kNZ^k -(2 + k_2)k_1 N\Theta,
	\end{equation}
	
	\begin{equation}\label{prop_z}
	\left(\partial_t-\mathcal{L}_N\right)Z_i = -N\mathbf{M}_i+ ND_i\Theta- D_iN\Theta- 2K_{ik}Z^k- k_1Z_i
	\end{equation}
	
	\noindent
	where the expressions for the Hamiltonian and momentum constraints ($\mathbf{H}$ and $\mathbf{M}_i$) are given by equations (\ref{ham_g3}) and (\ref{mom_g3}).

	When the generalized Bianchi-identity (\ref{bianchi}) holds, the evolution equations for the Hamiltonian and momentum constraints are
	
	\begin{align}\label{prop_ham_ccz4}\nonumber
	\left(\partial_t-\mathcal{L}_N\right)\mathbf{H}=&~2NK\mathbf{H}+\frac{1}{N}D^i\left(N^2\mathbf{M}_i\right)-2Nk_1(1 + k_2)K\Theta
	+ 2N \left(K h^{kl} - K^{kl}\right)\left(D_lZ_k - \Theta K_{kl}\right)
	\\ \nonumber
	&-N\partial_X G_3 A\left[-2A^iA^jD_iZ_j-D^kZ_k A^2+2A^iA^jK_{ij}\Theta+2K\Theta A^2\right.\\
	&\qquad \left. {}+k_1(1+k_2)(3A^2+A^kA_k)\Theta+2\mathbf{H}A^2-2A\mathbf{M}_iA^i\right]
	\end{align}
	
	\begin{align}\label{prop_mom_ccz4}\nonumber
	\left(\partial_t-\mathcal{L}_N\right)\mathbf{M}_i=&~NK\mathbf{M}_i+2D^j\left[N\left(D^kZ_kh_{ij}-D_{(i}Z_{j)}+\Theta (K_{ij}-Kh_{ij})-k_1(1+k_2)\Theta h_{ij}\right)\right]\\ \nonumber
	&+\frac{1}{N}D^i\left(N^2\mathbf{H}\right)-N\partial_X G_3 A_i\left[-2A^iA^jD_iZ_j-D^kZ_k A^2+2A^iA^jK_{ij}\Theta+2K\Theta A^2\right.\\
	&\qquad \left. {}+k_1(1+k_2)(3A^2+A^kA_k)\Theta+2\mathbf{H}A^2-2A\mathbf{M}_iA^i\right].
	\end{align}
	
	Once again, the system (\ref{prop_theta}-\ref{prop_mom_ccz4}) describing the propagation of constraint violations has the same principal symbol as in general relativity, cf. equations (7), (8), (11), (12) of \cite{constr_damp}. Therefore, the hyperbolicity and the high frequency behaviour of that system is not altered by the Horndeski terms. This has the following implications for constraint damping. Similarly to \cite{constr_damp}, one can carry out a preliminary mode analysis by linearizing around a generic weak field configuration and studying the high frequency limit of (\ref{prop_theta}-\ref{prop_mom_ccz4}). For large frequencies, the Horndeski terms become insignificant. Making a plane wave ansatz for the constraint variables then reduces the high frequency limit of (\ref{prop_theta}-\ref{prop_mom_ccz4}) to the same eigenvalue problem as in \cite{constr_damp} (see equation (19)). Hence, we come to the same conclusion as in vacuum GR: the real parts of all eigenfrequencies are negative if $k_1>0$ and $k_2>-1$. This suggests that with such choice of the parameters $k_1$ and $k_2$, large frequency constraint violating modes will be damped away in cubic Horndeski theories.
	
	\subsection{Equations of motion}
	
	Next, we provide the full system of evolution equations, in the conformal decomposition. We introduce $\tilde Z_i\equiv Z_i$, $\tilde Z^i\equiv \tilde{h}^{ij}\tilde{Z}_j$ and
	
	\begin{equation}
	\tilde U^i\equiv h^{kl}\left(\tilde \Gamma^i_{kl}-\mathring{\Gamma}^i_{kl}\right)+2\tilde Z^i=\tilde V^i+2\tilde Z^i.
	\end{equation}

	Similarly to the BSSN case, we use auxiliary variables

	\begin{align}
	\tilde{\Phi}\equiv\Phi={}&\frac{1}{N}\left(\partial_0A-e^{-4\Omega}\tilde{h}^{ij}\tilde{A}_i\tilde{D}_jN\right), \\
	\tilde{\Phi}_i\equiv{\Phi}_i={}&\tilde{D}_iA+\tilde{Q}_{ik}\tilde{A}_j\tilde{h}^{jk}+\tfrac13 K\tilde{A}_i,\\
	\tilde{\Phi}_{ij}\equiv\Phi_{ij}={}&\tilde{D}_i\tilde{A}_j-2\bigl(\tilde{A}_j\tilde{D}_i\Omega+\tilde{A}_i\tilde{D}_j\Omega- \tilde{h}_{ij}\tilde{h}^{kl}\tilde{A}_k\tilde{D}_l\Omega\bigr)+Ae^{4\Omega}\bigl(\tilde{Q}_{ij}+\tfrac13 K\tilde{h}_{ij}\bigr)\\
	\tilde{\rho}={}&e^{4\Omega}\Bigl(R+\tfrac23K^2-\tilde{Q}_{ac}\tilde{Q}_{bd} \tilde{h}^{ab}\tilde{h}^{cd}\Bigr)
	\end{align}

	to write the equations more compactly. (For a more complete list of formulas for the ADM and conformal decompositions, see Appendix \ref{app:adm}.)
	
	\noindent
	We use a natural generalization of the harmonic slicing condition (recall $\partial_0\equiv \partial_t-N^k \mathring{D}_k$)
	
	\begin{subequations}\label{sys_ccz4}
		\begin{equation}
		\partial_0 N=-2N^2\sigma (K-2\Theta),
		\end{equation}
		
		\noindent
		the shift vector is evolved using the standard "Gamma driver" condition \cite{num}
		
		\begin{equation}
		\partial_0 N^i=f N^2 e^{-4\Omega} B^i
		\end{equation}
		
		\begin{equation}
		\partial_0 B^i=\partial_0 \tilde U^i-\eta B^i.
		\end{equation}
		
		\noindent
		The parameters $f$ and $\eta$ are to be chosen and $B^i$ is an auxiliary variable. The evolution equations for the variables $\tilde h_{ij}$ and $\Omega$ are just defining equations and hence left unaltered compared to (\ref{ev_hh}) and (\ref{ev_omega}):

		\begin{equation}
		\partial_0\tilde h_{ij}=-2N\tilde Q_{ij}+2\tilde h_{k(i}\mathring{D}_{j)}N^k-\frac23\tilde h_{ij}\mathring{D}_kN^k
		\end{equation}
		
		\begin{equation}
		\partial_0\Omega=-\frac{N}{6}K+\frac16\mathring{D}_kN^k.
		\end{equation}
		
		\noindent
		The evolution equation for $\Theta$ is the same as (\ref{prop_theta}) (using the expression of the Hamiltonian constraint)  
		
		\begin{align}
		\partial_0\Theta =& \tfrac{1}{2}N\Bigl( \tilde{\rho}e^{-4\Omega}-  \tfrac12 A^2 - \tfrac12 e^{-4\Omega} \tilde{A}_{i} 			\tilde{A}_{j} \tilde{h}^{ij} +  G_2 -  A^2 \partial_XG_2 - \partial_{\phi}G_3 \bigr(A^2+ e^{-4\Omega} \tilde{A}_{i} \tilde{A}_{j} \tilde{h}^{ij}\bigl)\nonumber\\
&  + \partial_XG_3 \bigl(-  A^2e^{-4\Omega} \tilde{\Phi}_{ij} \
\tilde{h}^{ij}
 +e^{-8\Omega}{\tilde{A}_{k} \tilde{A}_{l} \tilde{\Phi}_{ij} \
\tilde{h}^{ik} \tilde{h}^{jl}}\bigr)\Bigr)\nonumber\\
		&+ Ne^{-4\Omega}\tilde D_i \tilde Z^i+2Ne^{-4\Omega}\tilde Z^i\tilde D_i\Omega -N\Theta K-e^{-4\Omega}\tilde Z^i\mathring{D}_iN-Nk_1(2+k_2)\Theta.
		\end{align}
		
		\noindent
		Equation (\ref{prop_mom_ccz4}) is no longer kept as a separate equation, instead, it is added to the evolution equation for $\tilde V^i$:
		
		\begin{align}
		\partial_0{\tilde U}^i ={}&2N\left(-\tfrac{2}{3}\tilde D^iK+\tilde \Gamma^i_{jk}\tilde Q^{jk}+6\tilde Q^{ij}\tilde D_j\Omega+\tilde D^i\Theta-\Theta\tilde D^i\ln N-\tfrac23K \tilde Z^i\right)-2\tilde Q^{ij}\tilde{D}_jN-\tilde U^k\mathring{D}_kN^i \nonumber \\ 
		&+\tfrac23\tilde U^i\mathring{D}_kN^k+\tilde h^{kl}\mathring{D}_k\mathring{D}_lN^i+\tfrac13\tilde h^{ij}\mathring{D}_j\mathring{D}_kN^k-2Nk_1 \tilde Z^i+2k_3\Bigl(\tfrac23\tilde{Z}^i\mathring{D}_kN^k-\tilde{Z}^k\mathring{D}_kN^i\Bigr)\nonumber \\
		&+2N\Bigl\{- \tfrac12 \tilde{A}^i A
 -  \tfrac{1}{2} \tilde{A}^i A \partial_XG_2
 -  \tilde{A}^i A \partial_{\phi}G_3
 + \tfrac12\partial_XG_3 \Bigr(-  A^2 \tilde{\Phi}^i
 +\tilde{A}_{k} A \tilde{\Phi}_{jl} \tilde{h}^{kl}\tilde{h}^{ij}e^{-4 \Omega}\nonumber\\
& -  \tilde{A}^i A \tilde{\Phi}_{kl}\tilde{h}^{kl}e^{-4 \Omega}
 +\tilde{A}_{k} \tilde{A}^{i} \tilde{\Phi}_{l} \
\tilde{h}^{kl} e^{-4 \Omega}\Bigl)\Bigr\}
		\end{align}
		
		\noindent
		This is equivalent to adding the momentum constraint times $2N$ to (\ref{ev_v0}).
		
		The evolution equations for $K$ and $\tilde Q_{ij}$ are the same as before, except for the constraint damping terms.
		
		\begin{align} 
		&\partial_0K-N\frac14\partial_XG_3(3A^2-e^{-4\Omega}\tilde A^k\tilde A_k)\Phi=N\Bigl\{R+2e^{-4\Omega}\tilde D_k\tilde Z^k+4e^{-4\Omega}\tilde Z^k\tilde D_k\Omega+K^2-2\Theta K \nonumber \\ 
		&+\tfrac{1}{N}e^{-4 \Omega}\Bigl(\tilde{h}^{ij}\tilde{D}_{i}\tilde{D}_{j}N+ 2 \tilde{h}^{ij} \tilde{D}_{i}N\tilde{D}_{j}\Omega\Bigr)-3 k_1(1+k_2)\Theta  -  \tfrac12 \tilde{A}_{i} \tilde{A}_{j}\tilde{h}^{ij}e^{-4\Omega}
 -  \partial_{\phi}G_3  \tilde{A}_{i} \tilde{A}_{j}\tilde{h}^{ij}e^{-4\Omega} \nonumber\\
& +\tfrac{3}{2}  G_2+ \partial_XG_2 \bigl(- \tfrac{3}{4}  A^2
 + \tfrac{1}{4} \tilde{A}_{i} \tilde{A}_{j} \tilde{h}^{ij}e^{-4\Omega}\bigr)
 + \partial_XG_3e^{-4\Omega} \Bigl(- \tfrac{3}{4} A^2 \tilde{\Phi}_{ij} \
\tilde{h}^{ij}
 -  \tilde{A}_{i} A \tilde{\Phi}_{j} \tilde{h}^{ij}\nonumber\\
& + \tfrac14 \tilde{A}_{i} \tilde{A}_{j} \tilde{\Phi}_{kl} \
\tilde{h}^{ij} \tilde{h}^{kl} e^{-4 \Omega}
 + \tilde{A}_{i} \tilde{A}_{j} \tilde{\Phi}_{kl} \
\tilde{h}^{ik} \tilde{h}^{jl}e^{-4 \Omega}\Bigr)\Bigr\}
		\end{align}

		\begin{align}
		&\partial_0\tilde Q_{ij}-N\tfrac12e^{-4\Omega}\partial_X G_3\left(\tilde A_i\tilde A_j- \frac13\tilde h_{ij}\tilde A^k\tilde A_k\right)\Phi=Ne^{-4\Omega}\Bigl[ R_{ij}+2\tilde D_{(i}\tilde{Z}_{j)}-8\tilde{Z}_{(i}\tilde{D}_{j)}\Omega \nonumber \\ 
		&-\tfrac{1}{N}\tilde D_i\tilde D_j N+4\tilde D_{(i}\Omega\tilde D_{j)}\ln N -\frac12\bigl(1+ \partial_XG_2+2\partial_{\phi}G_3\bigr) \tilde{A}_{i} \tilde{A}_{j}  
		\nonumber \\
		& + \partial_XG_3 \Bigr(- A\tilde{A}_{(i}\tilde{\Phi}_{j)}  -  \tfrac12 \tilde{A}_{i} \tilde{A}_{j} \tilde{\Phi}_{kl}\tilde{h}^{kl}e^{-4 \Omega}  +  \tilde{A}_{k} \tilde{A}_{(i} \tilde{\Phi}_{j)l} \
		\tilde{h}^{kl}e^{-4 \Omega}\Bigr)\Bigr]^\text{TF}\nonumber\\
		&+N(K-2\Theta)\tilde Q_{ij}-2N\tilde Q_{ik}\tilde Q^k_{~j}+2\tilde Q_{k(i}\mathring{D}_{j)}N^k-\frac23\tilde Q_{ij}\mathring{D}_kN^k.
		\end{align}
		
\noindent		
Finally, we also have a pair of scalar evolution equations: (\ref{ev_def_a}), (\ref{ev_sc_bssn}). 

We conclude this section with a technical remark. For general relativity, it has been noted that the CCZ4 equations of motion can be derived from an action principle \cite{ccz4} (at least if we ignore the lower order terms with $k_1$ and $k_2$). If we insisted on a similar action principle for cubic Horndeski theories, then upon taking the linear combination (\ref{ev_sc}) of the gravitational and scalar equations of motion, the resulting equation would contain principal terms from the ADM decomposition of $\nabla Z$. However, for cubic Horndeski theories, it is more beneficial to keep equation (\ref{ev_sc_bssn}) as the scalar evolution equation. It appears that for these theories, it is more useful to introduce the $Z$-terms at the level of the equations, rather than at the level of the action.

	\end{subequations}

	\subsection{Strong hyperbolicity}
	
	Now we linearize the system of equations (\ref{sys_ccz4}),(\ref{ev_def_a}), (\ref{ev_sc_bssn}) and study its hyperbolicity. The linearly small quantities corresponding to the new variables are 
	
	\begin{align*}
	&N^i\to N^i+\beta^i\\
	&B^i\to B^i+b^i\\
	&\Theta\to \Theta+\theta\\
	&\tilde U^i\to \tilde U^i+\tilde u^i
	\end{align*}

	The eigenvalue problem of the principal symbol for the 27 variables $U=(\hat\alpha, \hat\omega,\hat\theta,\hat\kappa,\hat\beta^i,\hat b^i,\hat{\tilde u}^i,\hat{\tilde \gamma}_{ij},\hat{\tilde q}_{ij},\hat\psi,\hat a)$  can be written as follows. 
	
	\begin{subequations}\label{eig_ccz4}
		
		\begin{equation}
		i\xi_0 \hat\alpha=-2N^2\sigma (\hat\kappa-2\hat{\theta)}
		\end{equation}
		
		\begin{equation}
		i\xi_0 \hat\beta^i=f N^2 e^{-4\Omega}\hat b^i
		\end{equation}
		
		\begin{equation}
		i\xi_0\hat b^i=i\xi_0\hat{\tilde u}^i
		\end{equation}
		
		\begin{equation}
		i\xi_0\hat\theta=\frac12 N e^{-4\Omega}\left\{i\xi_k \hat{\tilde u}^k+8|\xi|_{\tilde h}^2\hat\omega-\partial_X G_3\hat{\psi}\left(-A^2|\xi|_{\tilde h}^2+e^{-4\Omega} \left(\tilde A^k \xi_k\right)^2\right)\right\}
		\end{equation}
		
		\begin{align}\nonumber
		i\xi_0 \hat \kappa-\frac14\partial_XG_3(3A^2-A_kA^k)i\xi_0\hat a=&Ne^{-4\Omega} \left\{\frac{1}{N}|\xi|_{\tilde h}^2\hat \alpha+i\xi_k\hat{\tilde u}^k+8|\xi|_{\tilde h}^2\hat\omega+\frac14\partial_XG_3(3A^2-A^kA_k)|\xi|_{\tilde h}^2\hat\psi \right.\\
		&\qquad \left. {}-\partial_XG_3\left(A\tilde A^k ~i\xi_k\hat a+\left(\xi_k\tilde A^k\right)^2\hat\psi\right)\right\}
		\end{align}
		
		\begin{equation}
		i\xi_0 \hat\omega=-\frac{N}{6}\hat\kappa+\frac16 i \xi_k\hat\beta^k
		\end{equation}
		
		\begin{equation}\label{ccz4_def_a}
		i\xi_0\hat\psi=N\hat a
		\end{equation}
		
		\begin{equation}
		i\xi_0\hat{\tilde \gamma}_{ij}=-2N\hat{\tilde q}_{ij}+2i \left(\xi_{(i}\hat\beta_{j)}\right)^\text{TF}
		\end{equation}

		\begin{align}\nonumber
		&i\xi_0\hat{\tilde q}_{ij}-\frac12e^{-4\Omega}\partial_X G_3\left(\tilde A_i\tilde A_j-\frac13\tilde h_{ij}\tilde A^k\tilde A_k\right)i\xi_0\hat a=Ne^{-4\Omega}\left\{\frac12|\xi|_{\tilde h}^2\hat{\tilde \gamma}_{ij}+i\xi_{(i}\hat{\tilde u}_{j)}-\frac13i\xi_k\hat{\tilde u}^k\tilde h_{ij}\right. \\ \nonumber
		&\qquad \left. {} +2\xi_i\xi_j\hat\omega-\frac23 |\xi|_{\tilde h}^2\hat\omega \tilde h_{ij}+\frac{1}{N}\xi_i\xi_j\hat \alpha-\frac13\frac{1}{N} |\xi|^2_{\tilde h}\hat\alpha \tilde h_{ij}-A\partial_XG_3\hat a\left(\tilde A_{(i}~i\xi_{j)}-\frac13\tilde A^k~i\xi_k\tilde h_{ij}\right)\right. \\ 
		&\qquad \left. {} +e^{-4\Omega}\partial_X G_3\hat\psi\left(\frac12(\tilde A_i\tilde A_j-\frac13\tilde h_{ij}\tilde A^k\tilde A_k)|\xi|^2_{\tilde h}-\tilde A^k \tilde A_{(i} \xi_{j)}\xi_k+\frac13\tilde A^k\tilde A^l \xi_k\xi_l\tilde h_{ij}\right)\right\}
		\end{align}
		
		\begin{align}\nonumber
		i\xi_0\hat{\tilde u}^i= &~-|\xi|_{\tilde h}^2\hat{\beta}^i-\frac13\xi^i\xi_k\hat{\beta}^k+N\left(-\frac{4}{3}i\xi^i\hat\kappa+2i\xi^i\hat\theta+\partial_XG_3A^2~i\xi^i\hat a \right.\\
		&\qquad \left. {} +e^{-4\Omega}\partial_X G_3\left(A\tilde A^k\xi_k\xi^i\hat\psi-A\tilde A^i|\xi|_{\tilde h}^2\hat\psi-\tilde A^i\tilde A^k~i\xi_k\hat a\right) \right).
		\end{align}
		
		\begin{equation}
		i\xi_0\mathcal{A}\hat a=i\mathcal{B}\hat a+\mathcal{C}\hat\psi
		\end{equation}
		
	\end{subequations}
	
	\noindent
	where, again, we do not need to deal with the precise expressions of the coefficients $\mathcal{A}$, $\mathcal{B}$ and $\mathcal{C}$, we only need to keep in mind that substituting (\ref{ev_psi}) into (\ref{ev_a}) yields
	
	\begin{equation}
	\hat\psi\left(P^\prime_{\phi\phi}\right)^{\mu\nu}\xi_\mu\xi_\nu=0,
	\end{equation}
	
	\noindent
	i.e., equation (\ref{char}).

	Our strategy is analogous to the one described in section \ref{sec:bssn}: it is easy to find 25 eigenvalues and eigenvectors of the system (\ref{eig_ccz4}) with $\hat\psi=\hat a=0$. For a more compact notation, we use $\bar\xi_0$ (see equation (\ref{xi_0_shift})) instead of $\xi_0$ to list the eigenvalues:
	
	\begin{itemize}
		\item[I.] A 12-dimensional space of modes for arbitrary $\hat {\tilde \gamma}_{ij}$ and $\hat\alpha=\hat\omega=\hat\beta^i=\hat {\tilde u}^i=0$, with eigenvalues $\bar\xi_0=\pm 1$.
		\item[II.] A 2-dimensional space of modes for arbitrary $\hat {\tilde \gamma}_{ij}$ and non-trivial $\hat\alpha$, $\hat\omega$, $\hat\beta^i$, $\hat {\tilde u}^i$, with eigenvalues $\bar\xi_0=\pm 1$.
		\item[III.] A 3-dimensional space of zero speed modes ($\bar\xi_0=0$) for arbitrary $\hat {\tilde v}^i$.
		\item[IV.] A 2-dimensional space of modes with eigenvalues $\bar{\xi_0}=\pm \sqrt{2\sigma}$.
		\item[V.] A 4-dimensional space of modes with $\bar\xi_0=\pm\sqrt{f}$, for arbitrary $e^i$ orthogonal to $\xi^i$ (w.r.t. $\tilde h_{ij}$).
		\item[VI.] A 2-dimensional space of modes with $\bar\xi_0=\pm\sqrt{\frac{4f}{3}}$.
	\end{itemize}
	
	The full expressions of the corresponding eigenvectors are given in Table \ref{tab:eig_ccz4}. Clearly, these expressions depend smoothly on $\xi_i$ if $f>0$ and $\sigma>0$.
	
	\begin{table}[H]
		
		\begin{subtable}{1\textwidth}
			\centering
			\begin{tabular}{cccccc}
				\vspace{6pt}
				&$\hat\alpha$ &$\hat\omega$ & $\hat\beta^i$  & $\hat{\tilde \gamma}_{ij}$  & $\bar \xi_0$ \\
				\hline
				\hline
				\\
				I. &$0$ & $0$ & $0$  & $\hat{\tilde\gamma}_{ij}$ &  $\pm 1$    \\
				\\
				II. & $\mp i\frac{2\sigma (4f-3)|\xi|_{\tilde h}e^{2\Omega}}{f(6\sigma+1)}$ & $\mp i\frac{ |\xi|_{\tilde h} e^{2\Omega}(f-1-2\sigma f)}{2fN(6\sigma+1)}$& $i\xi^i$ & $\hat{\tilde \gamma}_{ij}$ &    $\pm 1 $  \\
				\\
				III. & $0$ & $-\frac{1}{8|\xi|_{\tilde h}^2} i\xi_k\hat{\tilde v}^k$  & $0$ & $-\frac{2i}{|\xi|_{\tilde h}^2}\left(\xi_{(i}v_{j)}-\frac{\xi_i\xi_j}{4|\xi|_{\tilde h}^2}\xi_k\hat{\tilde v}^k\right)^\text{TF}$ &   $0$   \\
				\\
				IV. & $\pm i \frac{\sqrt{2\sigma} e^{2\Omega}|\xi|_{\tilde h}(3\sigma-2f)}{2f}$ & $\pm i \frac{\sqrt{2\sigma} e^{2\Omega}|\xi|_{\tilde h}}{8Nf}$  & $i\xi^i$ & $\pm i\frac{3\sqrt{2\sigma}e^{2\Omega}}{2fN|\xi|_{\tilde h}}\left(\xi_i\xi_j\right)^\text{TF}$ &  $\pm\sqrt{2\sigma}$   \\
				\\
				V. & $0$ & $0$  & $ie^i$ & $\pm i\frac{2 e^{2\Omega}}{N\sqrt{f}|\xi|_{\tilde h}}\xi_{(i}e_{j)}$ &  $\pm \sqrt{f}$ \\
				\\
				VI. & $0$ & $\pm i \frac{e^{2\Omega}|\xi|_{\tilde h}}{4\sqrt{3f}N}$  & $i\xi^i$ & $\pm i\frac{\sqrt{3}e^{2\Omega}}{N\sqrt{f}|\xi|_{\tilde h}}\left(\xi_i\xi_j\right)^\text{TF}$ &   $\pm \sqrt{\frac{4f}{3}} $  \\
				\\
				\hline
			\end{tabular}
			\caption{}
		\end{subtable}
		\vspace{24pt}
		\pagebreak
		
		\begin{subtable}{1\textwidth}
			\ContinuedFloat
			
			\begin{tabular}{cccccc}
				
				\vspace{6pt}
				& $\hat\theta$ & $\hat\kappa$ & $\hat b^i$  & $\hat{\tilde q}_{ij}$ &  $\hat{\tilde u}_i$  \\
				\hline
				\hline
				\\
				I. & $0$ & $0$ & $0$ & $\mp\frac12 |\xi|_{\tilde h} e^{-2\Omega}\hat{\tilde\gamma}_{ij}$ &  $0$    \\
				\\
				II. &  $\frac{|\xi|_{\tilde h}^2(4f-3)(2\sigma-1)}{2fN(6\sigma+1)}$ & $\frac{3-4f}{Nf(6\sigma+1)}|\xi|_{\tilde h}^2$ & $\mp\frac{|\xi|_{\tilde h}e^{2\Omega}}{Nf}\xi^i$ & $\mp\frac12 |\xi|_{\tilde h} e^{-2\Omega}\hat{\tilde\gamma}_{ij}-\frac{1}{N}\left(\xi_i\xi_j\right)^\text{TF}$ &  $\mp\frac{|\xi|_{\tilde h}e^{2\Omega}}{Nf}\xi_i$   \\
				\\
				III. & $0$ & $0$ & $0$ & $0$ &  $\hat{\tilde v}_i$    \\
				\\
				IV. &  $0$ & $\frac{3\sigma-2f}{2fN}|\xi|_{\tilde h}^2$ & $\mp\sqrt{2\sigma}\frac{|\xi|_{\tilde h}e^{2\Omega}}{Nf}\xi^i$ & $\pm i\frac{3\sigma-2 f}{2fN|\xi|_{\tilde h}}\left(\xi_i\xi_j\right)^\text{TF}$ & $\mp\sqrt{2\sigma}\frac{|\xi|_{\tilde h}e^{2\Omega}}{Nf}\xi_i$    \\
				\\
				V. & $0$ & $0$ & $\mp\frac{|\xi|_{\tilde h}e^{2\Omega}}{N\sqrt{f}}e^i$ & $0$ &  $\mp\frac{|\xi|_{\tilde h}e^{2\Omega}}{N\sqrt{f}}e_i$    \\
				\\
				VI. & $0$ & $0$ & $\mp\frac{2|\xi|_{\tilde h}e^{2\Omega}}{N\sqrt{3f}}\xi^i$ & $0$ &  $\mp\frac{2|\xi|_{\tilde h}e^{2\Omega}}{N\sqrt{3f}}\xi_i$    \\
				\\
				\hline
			\end{tabular}
			\caption{}
		\end{subtable}
		\caption{The list of eigenvalues and eigenvectors of the principal symbol with $\hat\psi=\hat a=0$.}
		\label{tab:eig_ccz4}
	\end{table}
	
	To show that the system consisting of equations (\ref{sys_ccz4}), (\ref{ev_def_a}), (\ref{ev_sc_bssn}) is strongly hyperbolic, it remains to be shown that (\ref{eig_ccz4}) has two eigenvectors corresponding to the eigenvalues $\xi_0^{\phi,\pm}$ (obtained by solving (\ref{char})), with smooth dependence on $\xi_i$ and with non-zero $\hat\psi$ and $\hat a^\phi\equiv \frac{i}{N}\xi_0^{\phi,\pm}\hat\psi$ (see equation (\ref{ccz4_def_a})). These eigenvectors can be found as follows. Recall the general form of the characteristic equation (\ref{gen_char}) from Section \ref{sec:setup}. When grouping the terms in the characteristic equations (\ref{eig_ccz4}) to Einstein-scalar-field and Horndeski parts (terms containing a factor $\partial_X G_3$) as in (\ref{char_split}), we see that the Horndeski terms only act on the $\hat\psi$, $\hat a$ components of $U$. In other words, the matrices $\delta \mathbb{A}$ and $\delta \mathbb{L}$ in (\ref{char_split}) are projections to the subspace associated with the scalar variables:

\begin{align*}
\delta \mathbb{A}~ U\equiv &{}\delta \mathbb{A}~(\hat\alpha, \hat\omega,\hat\theta,\hat\kappa,\hat\beta^i,\hat b^i,\hat{\tilde u}^i,\hat{\tilde \gamma}_{ij},\hat{\tilde q}_{ij},\hat\psi,\hat a)^T=\delta \mathbb{A}~(0_{25},\hat\psi,\hat a)^T,\\
\delta \mathbb{L}~U\equiv &{}\delta \mathbb{L}~(\hat\alpha, \hat\omega,\hat\theta,\hat\kappa,\hat\beta^i,\hat b^i,\hat{\tilde u}^i,\hat{\tilde \gamma}_{ij},\hat{\tilde q}_{ij},\hat\psi,\hat a)^T=\delta \mathbb{L}~(0_{25},\hat\psi,\hat a)^T.
\end{align*}

The eigenvectors $U^{\phi,\pm}$ corresponding to the eigenvalues $\xi^{\pm,\phi}$ then satisfy

\begin{equation}
\left(i\xi_0^{\pm,\phi} \mathbb{A}_0-\mathbb{L}_0(\xi_k)\right)U^{\phi,\pm}=-\left(i\xi_0^{\pm,\phi}\delta \mathbb{A}_0-\delta\mathbb{L}_0(\xi_k)\right)U^{\phi,\pm}=-\left(i\xi_0^{\pm,\phi}\delta \mathbb{A}_0-\delta\mathbb{L}_0(\xi_k)\right)\left(0_{25},\hat\psi,\hat a^\phi\right)^T.
\end{equation}

Thus we have:
	
	\begin{equation}
	U^{\phi,\pm}=-\left(i\xi_0^{\pm,\phi} \mathbb{A}_0-\mathbb{L}_0(\xi_k)\right)^{-1}\left(i\xi_0^{\pm,\phi} \delta\mathbb{A}-\delta\mathbb{L}(\xi_k)\right)\left(
	\begin{array}{c}\vspace{5pt}
	0_{25}\\ \vspace{5pt}
	\hat\psi\\ 
	\tfrac{i\xi_0^{\pm,\phi}}{N}\hat\psi
	\end{array}\right).
	\end{equation}

	A straightforward but lengthy calculation gives the following result:
	
	\begin{subequations}\label{scev_ccz4}
	\begin{align}\nonumber
	\hat{\alpha}_{\phi,\pm}={}&- \frac{\partial_X G_3 N \sigma \hat\psi}{2\left((\xi_0^{\pm,\phi})^2-2\sigma N^2 |\xi|_{\tilde{h}}^2 e^{-4\Omega}\right)}\Bigl[(3A^2-\tilde{A}^k\tilde{A}_k e^{-4\Omega})(\xi_0^{\pm,\phi})^2\\
	&+N^2 e^{-4\Omega}\Bigl(\left(A^2+\tilde{A}^k\tilde{A}_k e^{-4\Omega}\right)|\xi|_{\tilde{h}}^2-4NA\tilde{A}^k\xi_k\xi_0^{\pm,\phi}\Bigr)\Bigr]
	\end{align}

	\begin{equation}
	\hat{\theta}_{\phi,\pm}=0
	\end{equation}
	
	\begin{align}\nonumber
	\hat{\kappa}_{\phi,\pm}={}& \frac{\partial_X G_3 i\xi_0^{\pm,\phi}  \hat\psi}{4N\left((\xi_0^{\pm,\phi})^2-2\sigma N^2 |\xi|_{\tilde{h}}^2 e^{-4\Omega}\right)}\Bigl[(3A^2-\tilde{A}^k\tilde{A}_k e^{-4\Omega})(\xi_0^{\pm,\phi})^2\\
	&+N^2 e^{-4\Omega}\Bigl(\left(A^2+\tilde{A}^k\tilde{A}_k e^{-4\Omega}\right)|\xi|_{\tilde{h}}^2-4NA\tilde{A}^k\xi_k\xi_0^{\pm,\phi}\Bigr)\Bigr]
	\end{align}
	
	\begin{align}\nonumber
	\hat{\beta}^{\phi,\pm}_\xi={}&\frac{N^2 f \xi_0^{\pm,\phi}|\xi|_{\tilde{h}}^2\hat\psi\partial_X G_3 e^{-8\Omega}}{3\left((\xi_0^{\pm,\phi})^2-2\sigma N^2 |\xi|_{\tilde{h}}^2 e^{-4\Omega}\right)\left(-(\xi_0^{\pm,\phi})^2+\frac43 f N^2|\xi|_{\tilde h}^2 e^{-4\Omega}\right)}\Bigl[3\left(\tilde{A}_{\xi}^2e^{-4\Omega}-\tilde{A}^k\tilde{A}_k\right)(\xi_0^{\pm,\phi})^2\\
	&-4NA\tilde{A}_\xi \xi_0^{\pm,\phi}+N^2|\xi|_{\tilde{h}}^2\left((6\sigma+1)A^2+\tilde{A}^k\tilde{A}_k~ e^{-4\Omega}-6\sigma e^{-8\Omega}\tilde{A}_\xi^2\right)\Bigr]
	\end{align}
	
	\begin{equation}
	\hat{ \beta}_I^{\phi,\pm}=\frac{\tilde A_I  e^{-8\Omega}~f N^2\hat\psi\partial_X G_3\left(-NA|\xi|_{\tilde h}^2+\tilde A_\xi \xi_0^{\pm,\phi}\right)}{\left(-(\xi_0^{\pm,\phi})^2+ f N^2|\xi|_{\tilde h}^2 e^{-4\Omega}\right)}
	\end{equation}
	
	\begin{align}\nonumber
	\hat{\tilde u}_\xi^{\phi,\pm}=\hat{b}_\xi^{\phi,\pm}={}&\frac{ i (\xi_0^{\pm,\phi})^2|\xi|_{\tilde{h}}^2\hat\psi\partial_X G_3 e^{-4\Omega}}{3\left((\xi_0^{\pm,\phi})^2-2\sigma N^2 |\xi|_{\tilde{h}}^2 e^{-4\Omega}\right)\left(-(\xi_0^{\pm,\phi})^2+\frac43 f N^2|\xi|_{\tilde h}^2 e^{-4\Omega}\right)}\Bigl[3\left(\tilde{A}_{\xi}^2e^{-4\Omega}-\tilde{A}^k\tilde{A}_k\right)(\xi_0^{\pm,\phi})^2\\
	&-4NA\tilde{A}_\xi \xi_0^{\pm,\phi}+N^2|\xi|_{\tilde{h}}^2\left((6\sigma+1)A^2+\tilde{A}^k\tilde{A}_k~ e^{-4\Omega}-6\sigma e^{-8\Omega}\tilde{A}_\xi^2\right)\Bigr]
	\end{align}
	
	\begin{equation}
	\hat{\tilde u}_I^{\phi,\pm}=\hat{b}_I^{\phi,\pm}=\frac{\tilde A_I  e^{-4\Omega}~ i \xi_0^{\pm,\phi}\hat\psi\partial_X G_3\left(-NA|\xi|_{\tilde h}^2+\tilde A_\xi \xi_0^{\pm,\phi}\right)}{\left(-(\xi_0^{\pm,\phi})^2+ f N^2|\xi|_{\tilde h}^2 e^{-4\Omega}\right)}
	\end{equation}
	
	\begin{align}\nonumber
	\hat{\tilde q}_{ij}^{\phi,\pm}={}&\tfrac12\partial_XG_3 i \xi_0^{\pm,\phi}\left(\tilde{A}_i\tilde{A}_j-\tfrac13\tilde{A}^k\tilde{A}_k\tilde{h}_{ij}\right)\hat\psi-\partial_XG_3A\left(\tilde{A}_{(i}\xi_{j)}-\tfrac13\tilde{A}_\xi\tilde{h}_{ij}\right)\hat\psi \\
	&+\frac{\partial_XG_3\left(\xi_i\xi_j-\tfrac13 |\xi|_{\tilde{h}}^2\tilde{h}_{ij}\right)\hat\psi}{-(\xi_0^{\phi,\pm})^2+2\sigma N^2|\xi|_h^2}\left[\left(-\frac{6\sigma+1}{4}A^2+\frac{2\sigma-1}{4}\tilde{A}^k\tilde{A}_k e^{-4\Omega}\right)i\xi_0^{\pm,\phi}+2i\sigma A\tilde{A}_\xi e^{-4\Omega}\right]
	\end{align}
	
	\begin{equation}
	\hat{\tilde \gamma}_{ij}^{\phi,\pm}=\frac{1}{i\xi_0^{\pm,\phi}}\left(-2N\hat{\tilde q}_{ij}^{\phi,\pm}+2i\xi_{(i}\beta^{\phi,\pm}_{j)}-\tfrac{2i}{3}\beta_\xi^{\phi,\pm}\right)
	\end{equation}
	
	\begin{equation}
	\hat\omega_{\phi,\pm}=\frac{1}{6i\xi_0^{\pm,\phi}}\left(-N\hat\kappa_{\phi,\pm}+i\beta_\xi^{\phi,\pm}\right)
	\end{equation}
	\end{subequations}

	\noindent
	where the notation $\tilde{T}_{.\xi}$ stands for the contraction $\tilde{T}_{.i}\xi_j\tilde h^{ij}$ and $I$ is an index for the subspace orthogonal to $\xi^i$.

	Interestingly, the $\theta$ components of these two eigenvectors are $0$. However, this is not surprising at all: this variable measures the constraint violations but these two eigenvectors correspond to a physical degree of freedom and as such, they must satisfy the high frequency version of the constraints.
	
	In order to avoid singularities in the expressions (\ref{scev_ccz4}) we need to choose the parameters $\sigma$ and $f$ in such a way that the expressions appearing in the denominators
	
	$$-\left(\xi_0^{\phi,\pm}\right)^2+ \frac43f N^2|\xi|_{\tilde h}^2 e^{-4\Omega}$$
	
	$$-\left(\xi_0^{\phi,\pm}\right)^2+ f N^2|\xi|_{\tilde h}^2 e^{-4\Omega}$$
	
	$$-\left(\xi_0^{\phi,\pm}\right)^2+ 2\sigma N^2|\xi|_{\tilde h}^2 e^{-4\Omega}$$
	
	\noindent
	are non-zero for all $\xi_i\in\mathbb{S}^2$. In a generic weak field background, the null cone of $P^\prime_{\phi\phi}$ is a slightly distorted version of the null cone of the spacetime metric $g$ which means that for specific $\xi_i$ they may intersect. Comparing the critical expressions to
	
	\begin{equation}
	g^{\mu\nu}\xi_\mu\xi_\nu=-(\xi_0)^2+N^2|\xi|_h^2,
	\end{equation}
	 
	 \noindent
	 and by similar considerations as in the end of Section \ref{sec:bssn_proof}, we see that smooth dependence on $\xi_i$ might be violated for some $\xi_i\in\mathbb{S}^2$ if $\sigma=\frac12$, $f=\frac34$ and/or $f=1$. This can be easily avoided: choosing suitable $\sigma>\frac12$ and $f>1$ resolves this issue, larger values of $\sigma$ and $f$ allows stronger background fields. Therefore, we arrive at a similar conclusion as in the previous section: choosing the gauge parameters $\sigma$ and $f$ large enough ensures strong hyperbolicity of the CCZ4 system, as long as the fields are sufficiently weak. In particular, the combination of the 1+log slicing ($\sigma=\frac{1}{N}$, $N<2$) and a Gamma driver shift condition with $f>1$ appears to be a good candidate for numerical applications.

	\section{Discussion}
	
	In this paper we have provided three well-posed formulations of weak field, cubic Horndeski theories by generalizing an elliptic-hyperbolic, a BSSN-type and the CCZ4 formulation of general relativity. The elliptic-hyperbolic formulation was obtained by generalizing a combination of the constant mean curvature slicing and spatial harmonic gauge condition. In the weak field regime, on spatial slices with negative Ricci curvature, the elliptic equations can be uniquely solved for the lapse function and the shift vector. Under these assumptions, the evolution equations are strongly hyperbolic in any generalized harmonic gauge. The $2$-parameter family of BSSN-type and CCZ4-type formulations are also strongly hyperbolic when the parameters describing the slicing and shift conditions obey suitable bounds. Two important steps in the proofs of these results were noticing that fistly, the gravitational equation of motion (\ref{eom_g3}) is quasilinear and secondly, that the second derivatives of the metric disappear in a linear combination of the gravitational and scalar equations of motion.
	
	Analyzing the hyperbolicity of the evolution equations in the elliptic-hyperbolic, BSSN-type or the CCZ4-type formulations of more general Horndeski theories is more difficult. The root of the problem lies in the degeneracy of the principal symbol. In any formulation of GR the principal symbol of the equations of motion has one or more degenerate eigenspaces. This may continue to hold even when small perturbing operators are added to the equations of motion such as Horndeski terms in a weak field regime, especially if the perturbing terms have some special structure. This may lead to the failure of diagonalizability of the principal symbol and hence the failure of strong (or even weak) hyperbolicity. An upcoming paper on this issue is currently under preparation.
	
	Of course, it is possible that more general Horndeski theories do not admit a well-posed initial value formulation at all. However, these theories may still be valid as effective field theories\footnote{In fact, it was shown recently \cite{eft}, that scalar-tensor EFTs can be written in a Horndeski form up to a fairly high order in the mass dimension expansion since higher derivative operators can be removed by appropriate field redefinitions.}. A common feature of the classical equations of motion of EFTs is that they admit runaway solutions. These solutions are typically artifacts of the truncation process and cannot be considered physical. Several techniques have been developed to deal with such pathologies by modifying the equations of motion, some of them are listed in \cite{eft}. An example of such methods is the so-called reduction of order process that can be explained as follows. Given an EFT with higher order equations of motion such that the higher derivative terms are subleading in a mass dimension expansion. To any perturbative order, it is possible to derive an alternative equation of motion which is equally accurate up to the given order and contains time derivatives of the fields only up to second order\footnote{In general, it is not possible to get rid of the higher order spatial derivatives.}. The advantage of this method is that under certain assumptions (see e.g. \cite{w_f}) the modified equation no longer has runaway solutions. On the other hand, the procedure breaks Lorentz covariance. Although Horndeski theories are not higher order, the above described process may still be carried out and the “reduced order” equations of motion may have a well-posed initial value problem. Another method is based on the Israel-Stewart process \cite{israel} which is widely used in relativistic hydrodynamics to deal with the ill-posedness of the Navier-Stokes equation. Roughly speaking, consider equations of motion of the form $L(\phi)=\epsilon S(\phi)$ for a collection of fields denoted by $\phi$, \cite{fix1,fix2}. Here $L$ is a differential operator such that the zeroth order equation of motion is $L(\phi)=0$; $S(\phi)$ is another differential operator playing the role of a correction term, suppressed with a small parameter $\epsilon$.  In this procedure, one introduces an auxiliary variable $\Pi$ for $L(\phi)$ which is forced to satisfy a time evolution equation of the form $\tau\partial_t \Pi=-\Pi+\epsilon S$ where $\tau$ is a suitable timescale. The solution to this modified equation asymptotically approaches the solution to the original equation in a timescale $\tau$. This procedure also restricts solutions to the infrared and it may also fix the hyperbolicity of the equations of motion \cite{fix1,fix2}.
	
	As a final remark, we point out another (mathematical) issue which may also lead to an interesting research direction. In general, for genuinely higher order EFTs, one might have to rethink what a “well-posed initial value formulation” means. For example, classically, well-posedness is defined in a suitable function space which is typically a Sobolev space. However, for theories that are only meant to be valid up to some finite energy scale, finding a suitable function space for solutions is far from obvious.

	\section*{Acknowledgements}
	I would like to thank my supervisor, Harvey Reall, for suggesting the problem, for the many helpful	discussions and for the comments on this paper. I am also grateful to Tiago Fran\c{c}a for carefully reading and commenting on a draft version of this paper, and to Pau Figueras, Joe Keir, Jorge Santos and Ulrich Sperhake for their valuable insights. This research was supported by a George and Lilian Schiff Studentship.

	\begin{appendices}
		\section{Background information}\label{sec:back}

	In this section we provide some general information regarding pseudodifferential calculus, elliptic PDEs and hyperbolic PDEs in a self-contained manner.
	
	First, we define the norms and function spaces used in these notes. Sobolev spaces on a manifold $(\Sigma,h)$ will be denoted by $W^{s,p}(\Sigma)$ and $H^s(\Sigma)\equiv W^{s,2}(\Sigma)$. The notation $F(\mathbb{R},W^{s,p})$ stands for the space of curves of class $F$ with values in $W^{s,p}$, i.e. if $u(t,x)\in L^\infty(\mathbb{R},H^s)$ then $u(t,\cdot)\in H^s$ and $u(\cdot,x)\in L^\infty$. When discussing pseudodifferential operators, we will briefly mention Schwartz spaces, denoted by $\mathcal{S}(\mathbb{R}^n)$. This is the space of smooth functions $f$ satisfying
	
	\begin{equation}
	\sup\limits_{x\in \mathbb{R}^n}|x^\alpha \partial^\beta f |<\infty
	\end{equation} 
	
\noindent
	for any multi-indices $\alpha$, $\beta$. The Fourier transform of a function or tensor field $A$ will be denoted by $\hat A$.
	
	\subsection{Basics of pseudodifferential calculus}

	We briefly summarize the basic results from pseudodifferential calculus, based on \cite{taylor}.
	
	For $m\in \mathbb{R}$, let $S^m$ be the space of symbols, that is, the space of smooth functions $p(x,\xi)$ satisfying
	
	\begin{equation}
	|\partial_x^\beta\partial_\xi^\alpha p(x,\xi)|\leq C\langle \xi \rangle ^{m-|\alpha|},
	\end{equation}
	
\noindent
	for multi-indices $\alpha$ and $\beta$, some constant $C$ depending only on $\alpha$; $\langle \xi \rangle \equiv(1+|\xi|^2)^{1/2}$. Given a symbol $p(x,\xi)\in S^m$, one can define a corresponding \textit{pseudodifferential} operator $p(x,\partial_x)$ as an operator acting on $\mathcal{S}(\mathbb{R}^n)$ in the following way: for $u\in \mathcal{S}(\mathbb{R}^n)$
	
	\begin{equation}
	p(x,\partial_x)u=\int \frac{\mathrm{d}^n\xi}{(2\pi)^n}e^{ix\cdot \xi}p(x,\xi)\hat u(\xi).
	\end{equation}
	
	It is easy to verify that for any $u\in \mathcal{S}(\mathbb{R}^n)$, $p(x,\partial_x)u\in \mathcal{S}(\mathbb{R}^n)$, i.e. $p(x,\partial_x):\mathcal{S}(\mathbb{R}^n)\to \mathcal{S}(\mathbb{R}^n)$. The space of such operators $p(x,\partial_x)$ is denoted by $\mathcal{OP}^m$. In fact, it can be shown that if $p(x,\partial_x)\in \mathcal{OP}^m$, then $p(x,\partial_x): H^{s+m}\to H^s$.
	
	It is also possible to define the spaces $S^\infty$ and $S^{-\infty}$ by $S^\infty\equiv \cup_{m\in \mathbb{R}}S^m$ and $S^{-\infty}\equiv \cap_{m\in \mathbb{R}}S^m$. It can be shown that pseudodifferential operators with symbols in the class $S^{-\infty}$ are smoothing operators.
	
	An important subspace of $S^m$ called classical symbols and denoted by $S_\text{cl}^m$ is the one consisting of symbols $p(x,\xi)\in S^m$ which admit an asymptotic expansion in the following sense. We say that $p(x,\xi)\in S_\text{cl}^m$ (and $p(x,\partial_x)\in \mathcal{OP}_\text{cl}^m$) if there exists a sequence of symbols $p_{m-j}(x,\xi)\in S^{m-j}$, $j=0,1,2,...,\infty$ that are homogeneous in $\xi$ of degree $m-j$ for $|\xi|\geq 1$ such that
	
	\begin{equation}
	p(x,\xi)-\sum\limits_{j=0}^{N}p_{m-j}(x,\xi)\in S^{m-N}
	\end{equation} 
	
\noindent
	for all $N\geq 1$. The asymptotic expansion is denoted by
	
	\begin{equation}\label{asymp_exp}
	p(x,\xi)\sim \sum\limits_{j\geq 0}p_{m-j}(x,\xi)
	\end{equation}
	
\noindent
	where the notation $\sim$ means equality up to a smoothing operator. Indeed, by definition, the operator
	
	\begin{equation}
	p(x,\partial_x)- \sum\limits_{j\geq 0}p_{m-j}(x,\partial_x)\in\mathcal{OP}^{-\infty}
	\end{equation}
	
\noindent
	is a smoothing operator. The first term in the expansion (\ref{asymp_exp}) (i.e. the one with $j=0$) is called the \textit{principal symbol} of $p$ and will be denoted by $p^{(0)}\equiv p_{m}$.
	
	In this paper, we deal with classical pseudodifferential operators. A useful property of  operators in this class is that for $p_1\in\mathcal{OP}^{m_1}_\text{cl}$ and $p_2\in\mathcal{OP}^{m_2}_\text{cl}$, the product $q(x,\partial_x)\equiv p_1(x,\partial_x)p_2(x,\partial_x)$ is also a pseudodifferential operator in $\mathcal{OP}^{m_1+m_2}_\text{cl}$ such that $q^{(0)}(x,\xi)=p_1^{(0)}(x,\xi)p_2^{(0)}(x,\xi)$, i.e. the principal symbol of the product is the product of the principal symbols.
	
	\subsection{Elliptic equations}\label{sec:elliptic}
	
	Now we consider some results regarding elliptic equations based on \cite{taylor,bruhat,besse}. An elliptic operator $p(x,\partial_x)\in \mathcal{OP}^m_\text{cl}$ is such that $|p(x,\xi)|\geq C\langle \xi\rangle ^m$ for $|\xi|\geq 1$. Let us first consider a basic result for second order, linear elliptic operators defined by 
	
	\begin{equation}\label{2nd_order_ell}
	P(x,\partial_x) u\equiv a^{ij}(x)\partial_i\partial_j u+b^i(x)\partial_iu+c(x)u,
	\end{equation}
	
\noindent
	with coefficients $a\in H^s(\mathbb{R}^n)$, $b\in H^{s-1}(\mathbb{R}^n)$ and $c\in H^{s-2}(\mathbb{R}^n)$, $s>\frac{n}{2}$ and $a^{ij}$ is a positive definite metric. Then
	\begin{itemize}
		\item[i)]  $P$ is a map $H^s\to H^{s-2}$ with finite dimensional kernel  ker $P\subset C^\infty$. 
		\item[ii)] Furthermore, if $c\leq 0$ then ker $P$ is empty and hence $P$ is an isomorphism $H^s\to H^{s-2}$.
	\end{itemize}

	Part i) of the theorem holds for higher order elliptic operators, whereas ii) generally fails.
	The operator $q(x,\partial_x)\in \mathcal{OP}^{-m}_\text{cl}$ is said to be a \textit{parametrix} for $p$ if it satisfies $q(x,\partial_x)p(x,\partial_x)\sim p(x,\partial_x)q(x,\partial_x)\sim I$ (see notation in previous subsection). Given an equation $p(x,\partial_x)u=f$ for $u\in H^s(\mathbb{R}^n)$ and $f\in H^{s-m}(\mathbb{R}^n)$, the formal solution to this equation is $u=q(x,\partial_x)f~\text{mod}~C^\infty$. In other words, one can say that the solution map $q: f\to u$ is a pseudodifferential operator of class $\mathcal{OP}^{-2}$.  For the present purposes, it will suffice to find the principal symbol of the parametrix (or solution map) corresponding to a linear elliptic differential operator of the form (\ref{2nd_order_ell}). Using the product identity, it is easy to see that the principal symbol of the parametrix of $P$ is $Q^{(0)}(x,\xi)=(a^{ij}\xi_i\xi_j)^{-1}$.
	
	\subsection{Hyperbolic equations}\label{sec:hyp}
	
	We conclude this section by discussing the initial value problem for hyperbolic equations \cite{taylor}. Consider, firstly, a first order linear equation of the form
	
	\begin{equation}\label{hyp1}
	\partial_t u=L(t,x,\partial_x)u+g(t,x), ~~u(0)=f,
	\end{equation}
	
\noindent
	where $u$, $g\in C(\mathbb{R},H^s(\mathbb{R}^n))$ and $f\in H^s(\mathbb{R}^n)$ ($s>\frac{n}{2}+1$) are $N$-component column vectors; $L(t,x,\partial_x)\in\mathcal{OP}^1_\text{cl}$ is an $N\times N$ matrix valued function with smooth dependence on $t$, satisfying
	
	\begin{equation}
	K(t,x,\xi) L(t,x,\xi)+ L^\dagger(t,x,\xi) K(t,x,\xi)\in S^0
	\end{equation}
	
\noindent
	and
	
	\begin{equation}
	C^{-1} I\leq K(t,x,\xi) \leq C I
	\end{equation}
	
\noindent
	for $\xi\in\mathbb{S}^{n-1}$, with some constant $C>0$ and a positive definite, Hermitian matrix $K(t,x,\xi)$ that depends smoothly on its arguments. Such equations are called \textit{strongly hyperbolic} and the $N\times N$ matrix $K$ is called a \textit{symmetrizer}. It can be proved that this problem is locally well-posed in Sobolev spaces $H^s$, i.e. there exists a unique solution 
	
	$$u\in C\left([0,T),H^s\right)\cap C^1\left([0,T),H^{s-1}\right)$$
	
\noindent
	to (\ref{hyp1}) with $T>0$ depending only on $||f||_{H_s}$. 
	
	The proof of this result is based on an inequality of the form
	
	\begin{equation}
	\partial_t\left(u,u\right)_K\leq c(T)\left[\left(u,u\right)_K+\left(u,g\right)_K\right]
	\end{equation}
	
\noindent
	where $\left(\cdot,\cdot\right)_K$ is a scalar product constructed in terms of the symmetrizer $K$, that is equivalent to the $L^2$-product. This leads to an \textit{energy estimate} (after an application of the Gronwall inequality) of the form

	\begin{equation}
	|| u||_{L^2}^2\leq C(t)\left[||f||_{L^2}^2+\sup_{t\in[0,T)}||g||_{L_2}^2 \right].
	\end{equation}

	The above result can be extended to first order pseudodifferential equations, obtained by a reduction of second (or higher) order equations. This is usually done by introducing $v=\partial_t u$ (or the new variable $v$ could also be a linear combination $v\equiv\partial_t u-X^i\partial_i u$ for some $X^i$).  A second order equation can be rewritten in a form similar to (\ref{hyp1}) but the operator $L(t,x,\partial_x,\partial_x^2)$ is now a $2\times 2$ block matrix with $N\times N$ matrix blocks, acting on column vectors $U=(u,v)^T\in \mathcal{H}^s\equiv H^s\times H^{s-1}$. Then the above result holds with initial data $U(0)=F\in \mathcal{H}^s$ and the solution will be in $$U\in C\left([0,T),\mathcal{H}^s\right)\cap C^1\left([0,T),\mathcal{H}^{s-1}\right).$$
	
	It can be shown that the existence of a smooth symmetrizer implies that $L(t,x,\xi_i)$ is diagonalizable with real eigenvalues but it is not always true the other way around. However, if $S$ is the matrix whose columns are the eigenvectors of $L$, i.e. $S^{-1}LS$ is diagonal, then a positive definite symmetrizer is given by $K=\left(S^{-1}\right)^\dagger S^{-1}$. One therefore needs to check if the smoothness and boundedness conditions are also met. If the eigenvectors have a smooth dependence on these variables then so does $S$. Since the entries of $S^{-1}$ are rational functions of the entries of $S$, $S^{-1}$ and hence $K$ is also smooth in $(t,x,\xi_i)$. The boundedness condition follows straightforwardly from smoothness in a compact spacetime. ($\xi_i$ takes values in a compact set so no additional assumptions are required for the $\xi_i$ dependence). In practice, to demonstrate strong hyperbolicity, it is easiest to show that the matrix $L$ is diagonalizable, has real eigenvalues and the eigenvectors depend smoothly on $(t,x,\xi_i)$.

	The notion of strong hyperbolicity and the above well-posedness result can be extended to quasilinear or even fully non-linear PDEs. More precisely, the following theorems are proved in Chapter 5 of \cite{taylor}.

The initial value problem

	\begin{equation}\label{hyp_quasilinear}
	\partial_t u=L(t,x,\partial_x,u)u+g(t,x,u), ~~u(0)=f,
	\end{equation}
	
\noindent
	for the  quasilinear equation with $u$, $g\in C(\mathbb{R},H^s(\mathbb{R}^n))$ and $f\in H^s(\mathbb{R}^n)$ ($s>\frac{n}{2}+1$) size-$N$ column vectors; $L(t,x,\partial_x,u)\in\mathcal{OP}^1_\text{cl}$ an $N\times N$ matrix valued function that satisfies the above properties of strong (symmetrizable) hyperbolicity for any $(t,x,u,\xi_i)$, then a unique local solution $u\in C\left([0,T),H^s\right)$ exists with $T>0$.

For the nonlinear system

\begin{equation}\label{hyp_nonlinear}
\partial_t u=B(t,x,u,\partial_x u), ~~ u(0)=f
\end{equation}

\noindent
with data $f\in H^s(\mathbb{R}^n)$ ($s>\frac{n}{2}+2$), \cite{taylor} shows that if the matrix

\begin{equation}
L(t,x,u,\partial_x u,\xi)=(\partial_{\partial_i u}B)(t,x,u,\partial_x u)\xi_i
\end{equation}

\noindent
possesses a symmetrizer $K(t,x,u,\partial_x u)$ with the above properties, then the Cauchy problem (\ref{hyp_nonlinear}) is locally well-posed with a unique solution $u\in C\left([0,T),H^s\right)$, $T>0$. These results suggest that to check whether the conditions of strong hyperbolicity are met for quasilinear and nonlinear equations, it is sufficient to study the linearized equations in a generic background.

		\section{ADM and conformal decomposition}\label{app:adm}
		
		Here we provide a list of formulas used during the ADM decomposition of the equations of motion. The decomposition rules for the derivatives of the scalar field and the curvature tensors:

		\begin{align}
		\nabla_a\phi={}&-An_a+A_a\\
		\nabla_a\nabla_b\phi={}& \Phi_{ab}-2n_{(a}\Phi_{b)}+n_an_b\Phi \\
		\mathcal{R}_{abcd}={}&\rho_{abcd}+2\sigma_{ab[c}n_{d]}+2\sigma_{cd[a}n_{b]}-4n_{[a}\tau_{b][c}n_{d]}\\
		\mathcal{R}_{ab}={}&\rho_{ab}-\tau_{ab}+2\sigma_{(a}n_{b)}+\tau n_an_b\\
		\mathcal{R}={}&\rho-2\tau
		\end{align}

\noindent
where the definitions of the auxiliary variables are

\begin{align}
A\equiv{}& n^a\nabla_a\phi\\
A_a\equiv {}& D_a\phi\equiv h_a^b\nabla_b\phi\\
{\Phi}\equiv{}&\mathcal{L}_nA-{A}^a{D}_a\ln N \\
{\Phi}_a\equiv{}&{D}_aA+K_{ac}{A}^c\\
{\Phi}_{ab}\equiv{}&{D}_a{A}_b+AK_{ab}\\
{\rho}_{abcd}\equiv{}&R_{abcd}+2K_{a[c}K_{d]b}\\
{\rho}_{ab}\equiv{}&R_{ab}+KK_{ab}-K_{ac}K^{c}{}_{b}\\
{\rho}\equiv{}&R+K^2-K_{ab}K^{ab}\\
{\sigma}_{abc}\equiv{}&2D_{[a}K_{b]c}\\
{\sigma}_{a}\equiv{}&2\bigl(D^bK_{ab}-D_aK\bigr)\\
{\tau}_{ab}\equiv{}&\mathcal{L}_nK_{ab}+\tfrac{1}{N}D_aD_bN+K_{ac}K^c{}_b\\
{\tau}\equiv{}&\mathcal{L}_nK+\tfrac{1}{N}D^aD_aN-K_{ab}K^{ab}.
\end{align}

\noindent
The conformal versions of these auxiliary variables are given by the following formulas

\begin{align}
\tilde{\Phi}\equiv\Phi={}&\mathcal{L}_nA-e^{-4\Omega}\tilde{h}^{ab}\tilde{A}_a\tilde{D}_b\ln N \\
\tilde{\Phi}_a\equiv{\Phi}_a={}&\tilde{D}_aA+\tilde{Q}_{ac}\tilde{A}_b\tilde{h}^{bc}+\tfrac13 K\tilde{A}_a\\
\tilde{\Phi}_{ab}\equiv\Phi_{ab}={}&\tilde{D}_a\tilde{A}_b-2\bigl(\tilde{A}_b\tilde{D}_a\Omega+\tilde{A}_a\tilde{D}_b\Omega- \tilde{h}_{ab}\tilde{h}^{cd}\tilde{A}_c\tilde{D}_d\Omega\bigr)+Ae^{4\Omega}\bigl(\tilde{Q}_{ab}+\tfrac13 K\tilde{h}_{ab}\bigr)
\end{align}

\begin{align}
\tilde{\tau}_{kl}\equiv\tau_{kl}={}& e^{4 \Omega} \Bigl(
 \mathcal{L}_n \tilde{Q}_{kl}
+ \tfrac{1}{3}  \tilde{h}_{kl} \mathcal{L}_n K-\tfrac23 K\tilde{Q}_{kl}+\tilde{h}^{ab} \tilde{Q}_{kb} \tilde{Q}_{la}
-  \tfrac{1}{9}  \tilde{h}_{kl} K^2\Bigr)
\nonumber\\
&+\tfrac{1}{N}\Bigl(\tilde{D}_{l}\tilde{D}_{k}N+ 2 \tilde{h}_{lk} \tilde{h}^{ab} \tilde{D}_{a}N\tilde{D}_{b}\Omega
-  2 \tilde{D}_{k}\Omega \tilde{D}_{l}N -  2 \tilde{D}_{k}N \tilde{D}_{l}\Omega\Bigr).
\end{align}

\begin{align}
\tilde{\tau}\equiv\tau e^{4\Omega}={}& e^{4 \Omega} \Bigl(
 \mathcal{L}_n K- \tilde{Q}_{ac} \tilde{Q}_{bd}\tilde{h}^{ab}\tilde{h}^{cd}
-  \tfrac{1}{3}   K^2\Bigr)+\tfrac{1}{N}\Bigl(\tilde{h}^{ab}\tilde{D}_{a}\tilde{D}_{b}N+ 2 \tilde{h}^{ab} \tilde{D}_{a}N\tilde{D}_{b}\Omega\Bigr).
\end{align}

\begin{align}
\tilde{\rho}_{abcd}\equiv {\rho}_{abcd} e^{-4\Omega}={}&R_{abcd}e^{-4\Omega}+2e^{8\Omega}\Bigl(\tilde{Q}_{a[c}\tilde{Q}_{d]b} +\tfrac13K\tilde{h}_{a[c}\tilde{Q}_{d]b}+\tfrac13K\tilde{Q}_{a[c}\tilde{h}_{d]b} +\tfrac19K^2\tilde{h}_{a[c}\tilde{h}_{d]b}\Bigr)
\end{align}

\begin{align}
\tilde{\rho}_{ab}\equiv \rho_{ab}={}&R_{ab}+e^{4\Omega}\Bigl(\tfrac29K^2\tilde{h}_{ab}+ \tfrac13K\tilde{Q}_{ab}-\tilde{Q}_{ac}\tilde{Q}_{bd}\tilde{h}^{cd}\Bigr)
\end{align}

\begin{align}
\tilde{\rho}\equiv \rho e^{4\Omega}={}&e^{4\Omega}\Bigl(R+\tfrac23K^2-\tilde{Q}_{ac}\tilde{Q}_{bd} \tilde{h}^{ab}\tilde{h}^{cd}\Bigr)
\end{align}

\begin{align}
\tilde{\sigma}_{klm}\equiv\sigma_{klm} e^{-4\Omega} ={}&2\tilde{D}_{[k}\tilde{Q}_{l]m} +\tfrac23 \tilde{D}_{[k}K\tilde{h}_{l]m}+4\tilde{D}_{[k}\Omega \  \tilde{Q}_{l]m}-4\tilde{h}^{ij}\tilde{D}_i\Omega \ \tilde{Q}_{j[k}\tilde{h}_{l]m}
\end{align}

\begin{align}
\tilde{\sigma}_{k}\equiv \sigma_k={}&6 \tilde{h}^{ab} \tilde{Q}_{kb} \
\tilde{D}_{a}\Omega
+ \tilde{h}^{ab} \tilde{D}_{b}\tilde{Q}_{ka}
-  \tfrac{2}{3} \tilde{D}_{k}K.
\end{align}

\noindent
Finally, we also provide the conversion rules between curvature tensors, used in the conformal decomposition:

\begin{align}
R_{abcd}={}&e^{4\Omega}\Bigl(\tilde{R}_{abcd}+4\tilde{h}_{d[a}\tilde{D}_{b]}\tilde{D}_c\Omega -4\tilde{h}_{c[a}\tilde{D}_{b]}\tilde{D}_d\Omega \nonumber \\
&+8\tilde{D}_{[a} \ \Omega \tilde{h}_{b]d}\tilde{D}_c\Omega- 8\tilde{D}_{[a}\Omega \ \tilde{h}_{b]c}\tilde{D}_d\Omega - 8\tilde{h}_{c[a}\tilde{h}_{b]d}\tilde{h}^{ef}\tilde{D}_{e}\Omega \ \tilde{D}_c\Omega\Bigr),
\end{align}

\begin{align}
R_{ab}={}&\tilde{R}_{ab}-2\tilde{D}_{a}\tilde{D}_b\ \Omega -2\tilde{h}_{ab}\tilde{h}^{cd}\tilde{D}_{c}\tilde{D}_d\ \Omega +4\tilde{D}_{a} \ \Omega \ \tilde{D}_b\Omega - 4\tilde{h}_{ab}\tilde{h}^{cd}\tilde{D}_{c}\Omega \ \tilde{D}_d\Omega,
\end{align}

\begin{align}
R={}&e^{-4\Omega}\Bigl(\tilde{R}- 8\tilde{h}^{cd}\tilde{D}_{c}\tilde{D}_d\ \Omega -8\tilde{h}^{cd}\tilde{D}_{c}\Omega \ \tilde{D}_d\Omega\Bigr).
\end{align}

	\end{appendices}

\end{document}